\documentclass[amsmath,amssymb,aps,floats,amsfonts,notitlepage,superscriptaddress,eqsecnum,nofootinbib,prd,a4paper,twocolumn,longbibliography]{revtex4-1}

\usepackage[utf8]{inputenc}
\usepackage[]{graphicx}
\usepackage{hyperref}
\usepackage{url}
\usepackage{color}
\usepackage[usenames,dvipsnames,svgnames,table]{xcolor}
\usepackage{multirow}
\usepackage{soul}
\allowdisplaybreaks[1]

\newcommand{\eig}{\lambda_{\ell m}}

\newcommand{\diff}[2]  {\frac{d #1}{d #2}}

\newcommand{\pdiff}[2]  {\frac{\partial #1}{\partial #2}}
\newcommand{\spdiff}[2] {\frac{\partial^2 #1}{\partial #2^2}}

\newcommand{\nn}{\nonumber}

\def\beq{\begin{equation}}
\def\eeq{\end{equation}}

\begin{document}

\title{Spin-orbit precession  along eccentric orbits for extreme mass ratio black hole binaries and its effective-one-body transcription}

\author{Chris Kavanagh}
\affiliation{Institut des Hautes \'Etudes Scientifiques, 91440 Bures-sur-Yvette , France.}

\author{Donato Bini}
\affiliation{Istituto per le Applicazioni del Calcolo ``M. Picone'', CNR, 00185 Rome, Italy.}

\author{Thibault Damour}
\affiliation{Institut des Hautes \'Etudes Scientifiques, 91440 Bures-sur-Yvette , France.}

\author{Seth Hopper}
\affiliation{CENTRA, Departamento de F\'isica, Instituto Superior T\'ecnico - IST,\\
 Universidade de Lisboa, 1049 Lisboa, Portugal}

\author{Adrian C.~Ottewill}
\affiliation{School of Mathematics and Statistics and Institute for Discovery, University College Dublin, Belfield, Dublin 4, Ireland.}

\author{Barry Wardell}
\affiliation{School of Mathematics and Statistics, University College Dublin, Belfield, Dublin 4, Ireland.}
\affiliation{Department of Astronomy, Cornell University, Ithaca, NY 14853, USA}

\begin{abstract}
In this work we present an analytical gravitational self-force calculation of the spin-orbit precession along an eccentric orbit around a Schwarzschild black hole, following closely the recent prescription of Akcay, Dempsey, and Dolan. We then transcribe this quantity within the Effective-One-Body (EOB) formalism, thereby determining several new, linear-in-mass-ratio, contributions in the post-Newtonian expansion of the spin-orbit couplings entering the EOB Hamiltonian. Namely, we determine the second gyro-gravitomagnetic ratio $g_{S_*}(r,p_r,p_{\phi})$ up to order $p_r^2/r^4$ included.
\end{abstract}

\maketitle

\section{Introduction}
\label{sec:intro}

The development of accurate waveform templates for compact binaries is essential for the science of
gravitational waves. For example, the extraction of physical information from the first LIGO detections
\cite{Abbott:2016blz,Abbott:2016nmj} has made a key  use of a bank of $\sim 200, 000$
semianalytical templates ~\cite{Taracchini:2013rva,Purrer:2015tud}, describing the inspiral, merger and ringdown of two comparable-mass 
black holes, that were developed within the Effective-One-Body (EOB) formalism \cite{Buonanno:1998gg,Buonanno:2000ef,Damour:2000we,Damour:2001tu}.
For future detectors such as LISA to reach their
full potential one needs to describe systems with mass ratios varying from 1:1 to $\sim1:10^6$,
evolving over long inspirals into plunge, merger and ringdown phases. When attempting to model such
orbital evolutions one typically relies on several approximate ways of solving Einstein's equations,
valid in different asymptotic regimes:
post-Newtonian (PN) theory in the slow-motion, weak-field regime; post-Minkowskian (PM) theory in the weak-field regime;
gravitational self-force (SF) theory for small mass ratios; numerical relativity (NR) for strong-field comparable mass
binaries; and EOB theory for analytically interpolating between various regimes.

Recent years have witnessed a fruitful crossbreeding between these
various methods. Notably, the EOB formalism has provided, through its natural theoretical flexibility, a common ground for
incorporating the results of other approaches. Examples of recent works contributing to the crossbreeding
between EOB theory and other approximation methods are: $\text{EOB}\bigcup\text{PN}$ 
\cite{Damour:2015isa,Damour:2016abl,Julie:2017pkb,Messina:2017yjg};  
$\text{EOB} \bigcup \text{PM}$ \cite{Damour:2016gwp};
$\text{EOB} \bigcup \text{SF}$ \cite{Bini:2015bfb,Hopper:2015icj,Akcay:2015pjz,Bini:2016cje}; and $\text{EOB}\bigcup \text{NR}$
\cite{Nagar:2015xqa,Bohe:2016gbl,Babak:2016tgq,Dietrich:2017feu,Nagar:2017jdw}.

The primary focus of this paper is on the third of these strategies: the extraction of physical
information from SF results, and their EOB transcription. The SF approach --- in which Einstein's equations are solved
perturbatively with the mass-ratio as small parameter --- is ideally suited to describing the
motion of compact binary systems with a large discrepancy in the masses. 
An important theme (initiated in Refs. \cite{Detweiler:2008ft,Barack:2009ey})
 within the SF community over the past ten years has been the extraction
 of physically meaningful quantities through the computation of gauge-invariant SF quantities.   
 These quantities are typically defined within the conservative sector, with
dissipative effects of the self-force ignored or turned off. See Refs. \cite{Damour:2009sm,Barausse:2011dq}
for the first corresponding EOB transcriptions of gauge-invariant SF quantities.
In the literature there now exists a
wide array of gauge invariant quantities, each with varying dependencies on the perturbed metric
and its derivatives. The utility of these include: insights into the physical effects of the
self-force (see, e.g., \cite{Colleoni:2015ena}); comparisons within SF theory between calculations in differing 
gauges (e.g. \cite{Sago:2008id}); comparisons
with independent PN calculations (e.g.\cite{Blanchet:2009sd,Akcay:2015pza,vandeMeent:2016hel}) and with NR codes (e.g. \cite{LeTiec:2011bk}); 
and the extraction of high-PN-order contributions to
the potentials of EOB theory (e.g. \cite{Bini:2013zaa,Bini:2014ica,Kavanagh:2015lva}).

In a recent paper, Akcay, Dempsey and Dolan \cite{Akcay:2016dku} presented a methodology for
calculating the gauge invariant self-force correction to the spin-orbit precession of a spinning
compact body along an eccentric orbit in Schwarzschild spacetime, as well as a numerical
calculation of the precession using a Lorenz gauge code. Their presentation is the first example of
a gauge invariant quantity for an eccentric orbit binary which depends on derivatives of the metric
perturbation, and gives access for the first time to spin-orbit effects along an {\it eccentric} orbit.

The first aim of the present work is to complement the (mostly numerical) results of Ref. \cite{Akcay:2016dku} by presenting 
an {\it analytical} calculation of
their invariant as a PN expansion within SF theory. To do so we rely on a low eccentricity
assumption in a manner following closely that of Ref. \cite{Hopper:2015icj}. 
By contrast, both, with Ref. \cite{Hopper:2015icj} (which used a Regge-Wheeler gauge),
and with Ref. \cite{Akcay:2016dku} (which used a Lorenz gauge), we will work in
 a so-called radiation gauge, which unlike the
Regge-Wheeler gauge is readily extendible to a Kerr spacetime. This will provide
an independent check of the gauge invariance of the
spin-precession quantity defined in \cite{Akcay:2016dku}.

The second aim of the present work is to explicitly derive the
relationship between the spin precession invariant along eccentric orbits, and the various
potentials parametrizing spin-orbit effects within the EOB formalism. [For the corresponding relationship
in the simpler case of circular orbits see Ref. \cite{Bini:2014ica}.] We shall then use this relationship to show how the knowledge of
the $O(e^2)$ [respectively, $O(e^4)$] eccentric corrections to the spin precession translates into new information about the terms
quadratic (resp., quartic) in the radial momentum in the spin-orbit potentials of the EOB Hamiltonian.

The organisation of this paper is as follows. In Sec.~\ref{Sec:Theory} we review the formalism for calculating the eccentric spin-precession, discuss eccentric geodesics in Schwarzschild spacetime and their perturbation and review the radiation gauge approach to reconstructing the perturbed metric. In Sec.~\ref{Sec:Calculation} we describe the post-Newtonian approach we take to calculating the retarded metric perturbation, the self-force, give our regularization and metric completion and finally the eccentric spin precession. Then in Sec.~\ref{Sec:EOB}, after briefly recalling
the EOB parametrization of spin-orbit effects, we show how to transcribe the spin-precession invariant $\Delta \psi(p,e)$ into a knowledge of the
$O(\nu)$ contribution to the second gyro-gravitomagnetic ratio $g_{S_*}(u,p_r, p_{\phi}; \nu)$.

\section{Geodetic spin-precession}
\label{Sec:Theory}
\subsection{Overview}
\label{sec:SpinPrec}

We start by giving a brief summary of the prescription of Akcay, Dempsey and Dolan  to calculate the gravitational self-force (SF) correction to the spin precession. For a more detailed description we refer the reader to
\cite{Akcay:2016dku}.

We wish to calculate the amount of precession angle a test spin vector accumulates over one radial period compared to the accumulated azimuthal angle. This precession is conveniently measured by the quantity (in units where $G=c=1$)
\begin{align}
	\psi(m_2\Omega_r, m_2\Omega_\varphi; q)=\frac{\Phi-\Psi}{\Phi}  \,. \label{Eq:psiDef}
\end{align}
Here, we consider a binary system with masses $m_1$ and $m_2$ (with $q \equiv \frac{m_1}{m_2} \ll 1$); $\Phi$ is the accumulated azimuthal phase from periapsis to periapsis (i.e. during a radial period), and $\Psi$ the corresponding accumulated phase of the spin vector (both being computed along an eccentric orbit perturbed by SF effects), and, finally;  $\Omega_r$, and $\Omega_{\varphi}$ are, respectively the
radial and (mean) azimuthal angular frequencies.
The question is then how to define $\Psi$.

The spin vector $s^a$  is parallely transported along an equatorial geodesic of the (regularized \cite{Detweiler:2002mi}) $O(q)$-perturbed spacetime with four velocity $u^a$ and proper time $\tau$: $\frac{D s_a}{d\tau}=0$. Projecting this equation onto a particular (polar-type) reference frame $e^a_{\alpha}$, its spatial components satisfy
\begin{align}
	\frac{d \mathbf{s}}{d\tau}=\boldsymbol{\omega}\times\mathbf{s}
\end{align}
where 
\begin{align}
	(\mathbf{s})_i=e^a_i s_a, \quad (\boldsymbol{\omega})_i=-\tfrac{1}{2}\epsilon_{ijk}\omega^{jk},\quad \omega_{ij}=-g_{ab} e_i^a \frac{D e_j^b}{d\tau}. \label{Eq:SpininFrame}
\end{align}
The spin precession is then entirely defined by the relative motion of the tetrad throughout the orbit. Choosing the basis
so that only the ($\theta$-like) $(\boldsymbol{\omega})_2$ component is non-zero, one finds that the spin vector accumulates an angle $\Psi$ over one radial period (periapsis to periapsis)
\begin{align}
	\Psi(m_2\Omega_r, m_2\Omega_\varphi; q)=\oint \! \omega_{13}(\tau)\,d\tau. \label{Eq:PsiBackground}
\end{align}    
The aim  is then to explicitly calculate the $O(q)$, SF contributions to $\psi$, Eq. \eqref{Eq:psiDef}, and   to $\Psi$, Eq.\eqref{Eq:PsiBackground}, i.e. the quantities
\begin{align}
	\Delta \Psi=\Psi(\Omega_r,\Omega_{\varphi},q)-\Psi(\Omega_r,\Omega_{\varphi},0) \,,\label{Eq:DeltaPsiDef}
\end{align}
and
\begin{align}
	\Delta \psi=\psi(\Omega_r,\Omega_{\varphi},q)-\psi(\Omega_r,\Omega_{\varphi},0) =-\frac{\Delta\Psi}{\Phi} ,\label{Eq:DeltapsiDef}
\end{align}
where we recall that $q=m_1/m_2$ denotes the small mass ratio, and where we used the fact that 
\beq
\Phi \equiv \Omega_{\varphi} {\mathcal{T}} \equiv 2 \pi \frac{\Omega_{\varphi}}{\Omega_r} \,,
\eeq
is the same on the perturbed ($q \ne 0$) and background ($q=0$) orbits. Here, differently from Ref. \cite{Akcay:2016dku}, we use the letter $\mathcal{T}= 2\pi/\Omega_r$ to denote the coordinate-time
radial period.

In practice, we shall work below with an intermediate $O(q)$ variation (denoted $\delta$) which does not keep fixed the values
of the two frequencies  $(\Omega_r,\Omega_{\varphi})$. We can then recover the correct value of $\Delta \psi$, Eq. \eqref{Eq:DeltapsiDef}, by first `subtracting' the induced frequency shifts
\begin{align}
	\Delta\Psi=\delta\Psi-\pdiff{\Psi}{\Omega_r}\delta\Omega_r-\pdiff{\Psi}{\Omega_{\varphi}}\delta\Omega_{\varphi}\,,\label{Eq:DeltaPsi}
\end{align}
and then computing
\begin{align}
	\Delta\psi=-\frac{\Delta\Psi}{\Phi}.
	\label{Eq:Deltapsi}
\end{align}
With this broad outline in mind, the next few sections will focus on the explicit details of this calculation when using post-Newtonian (PN) expansions.

\subsection{Motion on the background ($q=0$): Schwarzschild spacetime}
\label{Sec:Background}
In the usual Schwarzschild coordinates, the unperturbed ($q=0$) metric takes the form
\begin{align}
	ds^2=-f dt^2+f^{-1}dr^2+r^2(d\theta^2+\sin^2\!\theta \,d\varphi^2)
\end{align}
where $f\equiv(1-2m_2/r)$.

\subsubsection{Background equatorial geodesics}

We start be recalling some of the defining equations for a particle undergoing bound equatorial motion in a Schwarzschild spacetime. Here and henceforth we use a subscript $p$ to denote evaluation at the position of the particle. Such motion is parameterised by two constants of motion, the specific energy and angular momentum $(\mathcal{E},\mathcal{L})$ respectively. The tangent four velocity is then given by
\begin{align}
	u_p^{\mu}=\left(\frac{\mathcal{E}}{f_p},u^r_p,0,\frac{\mathcal{L}}{r_p^2}\right).
\end{align}
Here we have made the standard restriction to equatorial motion setting $\theta_p=\tfrac{\pi}{2},u^\theta_p=0$. The radial motion can be parametrised using Darwin's relativistic anomaly $\chi$ \cite{Darw59}
\begin{align}
	r_p(\chi)=\frac{p M}{1+e\cos\!\chi} \label{Eq:rofpe}
\end{align} 
where $\chi=0$ corresponds to periapsis, $p$ is the semilatus rectum, and $e$ the eccentricity. Using these parameters, 
\begin{align}
	\mathcal{E}^2 = \frac{(p-2)^2-4e^2}{p(p-3-e^2)}, \qquad \mathcal{L}^2 = \frac{p^2 M^2}{p-3-e^2} \label{Eq:ELofpe}
\end{align}
and
\begin{align}
	\frac{dt_p}{d \chi} &= \left[ \frac{(p-2)^2 - 4 e^2}{p - 6 - 2 e \cos \chi} \right]^{1/2}
	 \nn \\
	 &\qquad\qquad \times \frac{p^2 M}{(p - 2 - 2 e \cos \chi) (1 + e \cos \chi)^2}
	, \label{Eq:dtdchi}
	\\
	\frac{d \varphi_p}{d\chi} 
	&= \left[\frac{p}{p - 6 - 2 e \cos \chi}\right]^{1/2} ,
\label{Eq:dphidchi}	\\
	\frac{d\tau_p}{d \chi} &= \frac{M p^{3/2}}{(1 + e \cos \chi)^2} 
	\left[ \frac{p - 3 - e^2}{p - 6 - 2 e \cos \chi} \right]^{1/2} .\label{Eq:dtaudchi}
\end{align}

With these one can compute the unperturbed radial period as the coordinate time taken between successive periapses
\begin{align}
	{\mathcal T}\equiv\int_0^{2\pi}\!\left(\diff{t_p}{\chi}\right)\,d\chi \,.
\end{align}
The characteristic orbital frequencies are then 
\begin{align}
	\Omega_r=\frac{2\pi}{{\mathcal T}},\qquad \Omega_\varphi=\frac{\Phi}{{\mathcal T}}, \label{Eq:backgroundOmegas}
\end{align}	
with $\Phi=\int_0^{2\pi}\!\left(\frac{d \varphi_p}{d\chi} \right)\,d\chi$.

\subsubsection{Background reference tetrad}

As recalled above, in order to define the precession of the spin vector, one needs to choose a reference frame. Following \cite{Akcay:2016dku}, a suitable polar-type tetrad is that given explicitly (when $q=0$) by Marck \cite{Marck431}:
\begin{align}
	e^a_0&=u^a=\left(\frac{\mathcal{E}}{f},u^r,0,\frac{\mathcal{L}}{r^2}\right), \\
	e^a_1&=\frac{1}{f\sqrt{1+\mathcal{L}^2/r^2}}\left(u^r,f \mathcal{E},0,0\right), \\
	e^a_2&=\left(0,0,1/r,0\right) , \\
	e^a_3&=\frac{1}{r\sqrt{1+\mathcal{L}^2/r^2}}\left(\frac{\mathcal{E L}}{f},\mathcal{L}u^r,0,1+\frac{\mathcal{L}^2}{r^2}\right).  
\end{align}
Using \eqref{Eq:SpininFrame}  the key frequency determining the precession function is
\begin{align}
	\omega_{13}=\frac{\mathcal{E L}}{r^2+\mathcal{L}^2}.
\end{align}
It is straightforward then to calculate the background spin precession using \eqref{Eq:rofpe},\eqref{Eq:ELofpe} with \eqref{Eq:psiDef},\eqref{Eq:PsiBackground}.

\subsection{Motion and spin-precession in the perturbed spacetime ($q\ne0$)}

Ref. \cite{Akcay:2016dku}, generalizing previous results by Barack and Sago \cite{Barack:2011ed}, has derived an explicit integral
expression for  the SF contribution $\delta\Psi$ to $\Psi$ (using a specific variation $\delta$ which does not fix the frequencies).
Their final result reads

\begin{align}
	\delta\Psi=\int\!\left(\frac{\delta\dot{\Psi}}{\dot{\Psi}}-\frac{\delta u^r}{u^r}\right)\dot{\Psi}\diff{\tau}{\chi}\,d\chi. \label{Eq:deltaPsi}
\end{align}
The term proportional to $\delta u^r$ appears since the proper time in \eqref{Eq:PsiBackground} also needs to be varied. 
The evaluation of this expression requires further definitions. We have introduced $\dot{\Psi}\equiv\omega_{13}=\omega_{[13]}$ to be the integrand of \eqref{Eq:PsiBackground} in favour of the $\omega$ of \cite{Akcay:2016dku} to avoid overlap with the frequency notation for the frequency domain solutions of the Teukolsky equation. Its variation is given by Akcay \textit{et al} as
\begin{align}
	\delta\dot{\Psi}=\frac{1}{2}\dot{\Psi}h_{00}+\delta\Gamma_{[31]0}+(c_{01}e_1^b+c_{03}e^b_3)e_{a[3}\nabla_b e_{1]}^a \label{Eq:deltaPsidot}
\end{align} 
where we have neglected a total derivative term which vanishes upon integration over the orbit, $\delta\Gamma_{[31]0}$ is a tetrad component of $\delta\Gamma_{\mu\nu\rho}=\tfrac{1}{2}(h_{\mu\nu,\rho}+h_{\mu\rho,\nu}-h_{\nu\rho,\mu})$ and
\begin{align}
	c_{01}&=\frac{1}{f \sqrt{1+\mathcal{L}^2/r^2}}(\mathcal{E}\delta u^r_{BS}-u^r\delta\mathcal{E}_{BS}),\\
	c_{03}&=\frac{\delta\mathcal{L}_{BS}}{r\sqrt{1+\mathcal{L}^2/r^2}}
\end{align} 
are terms arising from perturbing the tetrad. In this expression the subscript $BS$ refers to the perturbations in energy, angular momentum and radial velocity defined by Barack and Sago in \cite{Barack:2011ed}. They differ from those used by Akcay \textit{et al} by the normalisation of the 4-velocity, because Barack and Sago normalise with respect to the background metric while Akcay \textit{et al} normalise with respect to the perturbed spacetime. This leads to the following relation valid to linear order in the mass-ratio
\begin{align}
	\delta\mathcal{E}_{BS}&=\delta\mathcal{E}-\frac{1}{2}\mathcal{E}h_{00}, \\
	\delta u^r_{BS}&=\delta u^r-\frac{1}{2}u^r h_{00}, \\
	\delta\mathcal{L}_{BS}&=\delta\mathcal{L}-\frac{1}{2}\mathcal{L}h_{00}. 
\end{align}
where $h_{00} \equiv h_{\mu\nu}e^\mu_0 e^\nu_0$. The $O(q)$ perturbations $\delta\mathcal{E}_{BS}$, $\delta\mathcal{L}_{BS}$ 
(which are denoted  $\Delta E$ and $\Delta L$ by Barack and Sago)
of the energy and angular momentum $u_t\equiv -\mathcal{E}, u_\varphi \equiv\mathcal{L}$ entering the above equations  have been shown in Sec II. C of \cite{Barack:2011ed} to be determined by the
following quadratures
\begin{align}
	\delta\mathcal{E}_{BS}(\chi)&=\delta\mathcal{E}_{BS}(0)- \!\int_0^\chi \! F_t^{\text{cons}} \diff{\tau}{\chi} \,d\chi,\\
	\delta\mathcal{L}_{BS}(\chi)&=\delta\mathcal{L}_{BS}(0)+\!\int_0^\chi \! F_\varphi^{\text{cons}} \diff{\tau}{\chi} \,d\chi \,.\label{Eq:deltaEL}
\end{align} 
Here $\delta\mathcal{E}(0)$ and $\delta\mathcal{L}(0)$ are the energy and angular momentum shift at periapsis, given in Eq.~(37) and (38) of  \cite{Barack:2011ed}, and $F_\mu^{\text{cons}}$ is the conservative part of the self-force which we discuss below (see Eq. \eqref{Fmu}). [By contrast to the
Detweiler-Whiting formulation of SF we used in our presentation above, Barack and Sago use the formulation where the perturbed motion
satisfies the forced-motion equation $ \frac{D_0 u^\mu}{d\tau_0}=F^\mu(\tau_0) $.]
Using the normalisation condition of Akcay \textit{et al} the perturbed radial velocity is then calculated from the relation
\begin{align}
	\frac{1}{2}h_{00}-\frac{ \mathcal{E}\delta \mathcal{E}}{f_p}+\frac{u^r\delta u^r{}}{f_p}+\frac{\mathcal{L}\delta \mathcal{L}}{r_p^2}=0. \label{Eq:deltaurNorm}
\end{align}
Finally the gauge invariant precession function is calculated using Eqs. \eqref{Eq:DeltaPsi} and \eqref{Eq:Deltapsi}. The frequency shifts calculated by perturbing \eqref{Eq:backgroundOmegas} are given in Eq.~(75)-(76) of \cite{Barack:2011ed}. We however do not require their $\alpha$ term, which they added to ensure the asymptotic flatness of their metric perturbation, since we will work in an asymptotically flat gauge. 

\subsection{Radiation gauge metric perturbation}
\label{Sec:RgMPs}
Our strategy for computing a post-Newtonian expansion of the retarded metric perturbation in many ways follows closely that laid out in \cite{Hopper:2015icj}, in that we will use the method of extended homogeneous solutions to construct a particular solution of a particular partial differential equation whose solutions are related to the metric perturbation. The key difference however is that we shall use the tetrad formalism and radiation gauge to construct the metric, using the CCK procedure, so named after its development by Chrzanowski \cite{Chrzanowski:1975wv} and Cohen and Kegeles \cite{COHEN19755,Kegeles:1979an}. Specifically this involves building inhomogeneous solutions to the Bardeen-Press-Teukolsky (BPT) equation ($a=0$ Teukolsky equation), from this a Hertz potential and finally the metric perturbation and all its first derivatives. Since much of the details are covered by a variety of authors e.g \cite{Lousto:2002em,Keidl:2010pm,Shah:2010bi,Kavanagh:2016idg,vandeMeent:2015lxa} we will give an abridged overview of the strategy and refer the reader to the given references for details.

\subsubsection{Bardeen-Press-Teukolsky (BPT) equation}

The description of perturbations to a black hole spacetime can be reduced to a single partial differential equation for the tetrad components of the perturbed Weyl tensor $C_{\mu\nu\rho\sigma}$. In particular (essentially) all information is simultaneously contained in the two quantities $\psi_0$ and $\psi_4$ defined by:
\begin{align}
\psi_0&=-C_{\alpha\beta\gamma\delta}l^\alpha m^\beta l^\gamma m^\delta, \\
\psi_4&=-C_{\alpha\beta\gamma\delta}n^\alpha \bar{m}^\beta n^\gamma \bar{m}^\delta,
\end{align}
where $l^\mu$, $n^\nu$ $m^\mu$ and $\bar{m}^\mu$ are the Kinnersley tetrad legs given in Appendix \ref{App:Kinn}.
The dynamics of the perturbed Weyl scalars on a Schwarzschild background are described by the BPT equation:
\begin{widetext}
\begin{align}
\frac{r^4}{\Delta}\spdiff{\psi}{t}
-\frac{1}{\sin^2\theta}\spdiff{\psi}{\varphi} 
-\Delta^{-s}\pdiff{}{r}\left(\Delta^{s+1}\pdiff{\psi}{r}\right) -\frac{1}{\sin \theta}\pdiff{}{\theta}\left(\sin\theta \pdiff{\psi}{\theta}\right)  
-&2is\frac{ \cos\theta}{\sin^2\theta}\pdiff{\psi}{\varphi}
-2 s \left[\frac{m_2r^2}{\Delta}-r \right]\pdiff{\psi}{t} \nonumber\\
&+(s^2\cot^2\theta-s)\psi = 4 \pi r^2 T.
\label{Eq:BPTMaster}
\end{align} 
\end{widetext}
with
\begin{align}
\Delta \equiv r^2 - 2m_2 r.
\end{align}
Here, for $s=+2$ 
\begin{align}
	\psi=\psi_0, \qquad T=2 T_0
\end{align}
and for $s=-2$
\begin{align}
	\psi= \varrho^{-4}\psi_4, \qquad T=2 \varrho^{-4}\, T_4,
\end{align}
while the source terms are 
\begin{align}
T_0&=(\boldsymbol{\delta}+\bar{\varpi}-\bar{\alpha}-3 \beta-4 \tau) \times\nonumber\\
& \qquad \left[(\boldsymbol{D}-2\epsilon-2\bar{\varrho})T_{13} -(\boldsymbol{\delta}+\bar{\varpi}-2\bar{\alpha}-2 \beta)T_{11}\right] \nonumber \\
& +(\boldsymbol{D}-3\epsilon+\bar{\epsilon}-4\varrho-\bar{\varrho}) \times \nonumber \\
& \qquad\left[(\boldsymbol{\delta}+2 \bar{\varpi} -2\beta)T_{13}-(\boldsymbol{D}-2\epsilon+2\bar{\epsilon}-2\bar{\varrho})T_{33}\right], \\
\nonumber\\
T_4&=(\boldsymbol{\Delta}+3\gamma-\bar{\gamma}+4 \mu+\bar{\mu}) \times \nonumber \\
& \qquad \left[(\boldsymbol{\bar{\delta}}-2\bar{\tau}+2\alpha)T_{24}
  -(\boldsymbol{\Delta} +2\gamma-2\bar{\gamma}+\bar{\mu})T_{44}\right] \nonumber \\
& + (\boldsymbol{\bar{\delta}}-\bar{\tau}+\bar{\beta}+3\alpha+4\varpi) \times \nonumber \\
& \qquad \left[(\boldsymbol{\Delta}+2 \gamma +2\bar{\mu})T_{24}-(\boldsymbol{\bar{\delta}}-\bar{\tau}+2\bar{\beta}+2\alpha)T_{22}\right],
\end{align}
where $\boldsymbol{D}=l^\mu\partial_\mu$, $\boldsymbol{\Delta}=n^{\mu}\partial_\mu$ and $\boldsymbol{\delta}=m^\mu\partial_\mu $.
Here $T_{ij}$ are the Kinersley-tetrad projections of the point particle source. Now and henceforth we will focus on the $s=2$ solutions for $\psi_0$ (a similar procedure could be followed with $\psi_4$ since it contains the same information as $\psi_0$).

This equation is fully separable by means of a Fourier transform and projection over spin-weighted spherical harmonics. Due to the double-periodicity of the eccentric orbits the Fourier transform reduces to a Fourier series labelled by the discrete frequencies $\omega=m \Omega_\varphi+n\Omega_r$
\begin{align} 
	\psi_0(t,r,\theta,\varphi)&=\sum_{\ell m}\psi_0^{\ell m} (t,r) \, {}_2Y_{\ell m}(\theta,\varphi)  \,,\label{Eq:psi0}\\
	\psi_0^{\ell m}(t,r)&=\sum_{n=-\infty}^\infty \! e^{-i \omega t} {}_2R_{\ell m \omega}(r) \,.
\end{align}
The radial functions ${}_sR_{\ell m \omega}(r)$ satisfy 
\begin{align}
\bigg[\Delta^{-s}\diff{}{r}\left(\Delta^{s+1}\diff{}{r}\right)+&\frac{r^4\omega^2-2 i s (r-m_2)r^2\omega}{\Delta}\nonumber \\
+4 i s \omega r -{}_s &\eig\bigg]{}_s R_{\ell m \omega}(r)
={}_sT_{\ell m \omega}, \label{Eq:RadBPT} 
\end{align}
where ${}_s \eig=\ell(\ell+1)-s(s+1)$. Assuming a pair of homogenous solutions ${}_s \hat{R}^+_{\ell m \omega}$, ${}_s \hat{R}^-_{\ell m \omega}$ to the above, the corresponding inhomogeneous solution is written
\begin{align}
	{}_2R_{\ell m \omega}(r)=c^+{}_2 \hat{R}^-_{\ell m \omega}(r)+c^-{}_2 \hat{R}^+_{\ell m \omega}(r) \label{Eq:inhomSoln}
\end{align}
where
\begin{align}
	c^+&=\frac{1}{W_{\ell m n}}\int_{r}^{r_p}\!\Delta^2{}_2 \hat{R}^+_{\ell m \omega}(r) T_{\ell m \omega} \,dr, \\
	c^-&=\frac{1}{W_{\ell m n}}\int_{r_p}^{r}\!\Delta^2{}_2 \hat{R}^-_{\ell m \omega}(r) T_{\ell m \omega} \,dr 
\end{align}
and $W_{\ell m n}=\Delta^{s+1}\left({}_2 \hat{R}^+_{\ell m \omega} {}_2 \hat{R}^-{}'_{\ell m \omega}-{}_2 \hat{R}^-_{\ell m \omega} {}_2 \hat{R}^+{}'_{\ell m \omega}\right)$. The source term here is
\begin{align}
	T_{\ell m \omega}=\frac{1}{\mathcal T}\int_{0}^{2\pi}\int_0^{2\pi}\int_0^\pi T_0 e^{i \omega t} {}_2 Y^*_{\ell m}(\theta,\varphi) \sin\theta \diff{t}{\chi}d\theta d\varphi d\chi.
\end{align}

\subsubsection{Hertz potential and the retarded metric perturbation}

As is standard we reconstruct the metric perturbation by means of an auxiliary function known as the Hertz potential $\Psi_H$, as in for example \cite{Keidl:2010pm}. 
In the outgoing radiation gauge we work in, $\Psi_H$ satisfies the spin-2 BPT equation, as well as the angular equation
\begin{align}
	\psi_0=\frac{1}{8}\left(\mathcal{L}^4 \bar{\Psi}_H+12 m_2 \partial_t \Psi_H\right). \label{Eq:AngEq}
\end{align}
Here $\mathcal{L}^4=\mathcal{L}_1\mathcal{L}_0\mathcal{L}_{-1}\mathcal{L}_{-2}$ and $\mathcal{L}_s=-\big(\partial_\theta-s\cot\theta+i \csc\theta\partial_\varphi\big)$.
Spectrally decomposing the Hertz potential as
\begin{align}
	\Psi_H=\sum_{\ell m n} e^{-i \omega t}\Psi_{\ell m \omega} \; {}_2 Y_{\ell m}(\theta,\varphi)
\end{align}	
Eq.~\eqref{Eq:AngEq} can be algebraically inverted to give
\begin{align}
	\Psi_{\ell m \omega}=8 \frac{(-1)^m D \, {}_2\bar{R}_{\ell,-m,-\omega}+12 i m_2 \omega \, {}_2R_{\ell m \omega}}{D^2+144M^2\omega^2} \label{Eq:PsiofR}
\end{align}
where $D=\ell(\ell+1)(\ell-1)(\ell+2)$. 
The metric is then obtained from $\Psi_H$ by applying a set of differential operators:
\begin{align}
h_{\alpha\beta}&= - \varrho^{-4}\{n_\alpha n_\beta(\boldsymbol{\bar{\delta}}-3\alpha-\bar{\beta}+5\varpi)(\boldsymbol{\bar{\delta}}-4\alpha+\varpi)\nonumber\\
&+\bar{m}_\alpha \bar{m}_\beta(\boldsymbol{\Delta}+5\mu-3\gamma-\bar{\gamma})(\boldsymbol{\Delta}+\mu-4\gamma)\nonumber\\
&-n_{(\alpha}n_{\beta)}\left[(\boldsymbol{\bar{\delta}}-3\alpha+\bar{\beta}+5\varpi+\bar{\tau})(\boldsymbol{\Delta}+\mu-4\gamma)\right. \nonumber \\
&\left. +(\boldsymbol{\Delta}+5 \mu -\bar{\mu}-3\gamma -\bar{\gamma})(\boldsymbol{\bar{\delta}}-4\alpha+\varpi)\right]\}\Psi_H+\text{c.c.}, \label{Eq:hofPsi}
\end{align}
where $\text{c.c.}$ denotes complex conjugation. The first derivatives of the metric which appear in the formula for the spin precession can then be written in terms of three derivatives of the inhomogeneous solution \eqref{Eq:inhomSoln}.

Ultimately the decomposition in spin-weighted spherical harmonics of $\psi_0$, Eq. \eqref{Eq:psi0}, generates a corresponding 
decomposition of $h_{\alpha\beta}$ in {\it tensorial} spherical harmonics, with $\ell$ (together with the parity) labelling each irreducible representation of the rotation group. In turn, this generates a corresponding decomposition of the spin-precession $\psi$. We shall often
refer to the irreducible pieces of these decompositions as `` $\ell$-modes".

\section{Post-Newtonian approach}
\label{Sec:Calculation}

We now wish to proceed with the calculation laid out in the previous sections analytically, using a PN assumption that the orbital separation between the two bodies is large, i.e. $p\gg 1$. For simplicity, we will additionally assume small eccentricities ($e \ll 1$), and so our results will appear as double expansions in $e$ and $1/p$.
  
In practice, to achieve the required accuracy of the spin precession invariant we will work to $7$ orders in $1/p$ and $6$ orders in $e$. This will yield the result to 5 PN orders and accurate to order $e^2$ (since the invariant is defined as the ratio of two angles one loses two powers of $e$ throughout the calculation, i.e. $e^4$ accuracy is needed for $e^2$ results). Our extra orders are kept to reduce potential systematic errors.

\subsection{Background orbit}

The various background orbital elements of \ref{Sec:Background} can be easily calculated in the PN regime, see e.g. \cite{Hopper:2015icj}. For example the orbital period, $\mathcal T$, can be calculated by expanding \eqref{Eq:dtdchi}
\begin{align}
	&\diff{t_p}{\chi}=p^{3/2}\bigg[\big(1-2 e \cos{ \chi}
   +3 e^2 \cos ^2\chi +\mathcal{O}(e^3)\big) \nonumber\\
	&+3\big(1-e \cos \chi + e^2 \cos ^2\chi +\mathcal{O}(e^3)\big)\frac{1}{p}+\mathcal{O}\left(p^{-2}\right)\bigg]
\end{align}
and integrating order by order. This can be extended with ease to the desired orders in $1/p$ and $e$. Repeating this for $\diff{\varphi_p}{\chi}$, we can obtain the two orbital frequencies 
\begin{align}
	m_2\Omega_r&=\left(\frac{1-e^2}{p}\right)^{3/2}\bigg[1-3\frac{1-e^2}{p}+\mathcal{O}(p^{-2})\bigg] \\
	m_2\Omega_\varphi&=\left(\frac{1-e^2}{p}\right)^{3/2}\bigg[1+3\frac{e^2}{p}+\mathcal{O}(p^{-2})\bigg]  \label{Eq:rfreqPN}
\end{align}
These are equivalent in the Newtonian limit.

\subsection{PN-expanded BPT equation}

Before we compute the perturbed orbital elements we need the self-force and thus the metric perturbation and its derivatives. For our set of homogeneous solutions we use exactly those described in Ref. \cite{Kavanagh:2016idg}, with the rotation parameter $a$ limiting to zero. That is, dropping the $s=2$ subscript and translating notation,
\begin{align}
	\hat{R}^{+/-}_{\ell m \omega}=R^{\text{up}/\text{in}}_{\ell m \omega}(a\rightarrow0).
\end{align}
Note that while the computation of \cite{Kavanagh:2016idg} is aimed at circular orbits, the homogeneous solutions obtained therein are derived with only the assumptions that the orbital radius is large and that the frequency scales as $\omega\!\sim\! r^{-3/2}$. Both of the assumptions are satisfied in our current study, as can be seen explicitly by Eq.~\eqref{Eq:rfreqPN}. 

We would also like to emphasise the nature of these solutions as a function of $\ell$. Since the regularised self force is convergent in $\ell$, in numerical studies of the self force a finite number of $\ell$ values are computed, which amounts to some corresponding accuracy when computing the full sum over spherical harmonics, as in say Eq.~\eqref{Eq:psi0}. In the case of post-Newtonian expansions, the situation is somewhat different. It turns out one can compute homogeneous solutions leaving $\ell$ as a parameter, the drawback being that they typically breakdown for low $\ell$. Thus the strategy is to compute explicit PN expansions for $\ell=2\ldots6$, the rest being captured by the general expansions. Typically as one increases the order of our PN expansion more low $\ell$'s are needed.  

As a particular example we will run through the procedure with $\ell=2$. In Sec.II B of \cite{Kavanagh:2016idg}, the homogeneous solutions to the radial Teukolsky equation are computed (similarly to Ref. \cite{Bini:2013rfa}) as expansions in $\eta=\tfrac{1}{c}$ with coefficients in terms of the two variables $X_1=GM/r$, and $\sqrt{X_2}=\omega r$. These expansions are simplified by writing them as a product of an exponential factor  with radial dependence entirely contained in logarithmic terms, and a remaining series in $\eta$. After limiting $a\to0$ these are 
\begin{align}
	\hat{R}^{+}_{\ell=2 m \omega}=&-\frac{72 i X_1^4 \eta ^7}{\sqrt{X_2}}\Bigg(1+ i \sqrt{X_2} \eta+\left(5 X_1-\frac{X_2}{2}\right) \eta ^2 \nonumber \\
	&-\frac{1}{6} i X_2^{3/2} \eta ^3+\left(\frac{120 X_1^2}{7}+\frac{7 X_1 X_2}{2}+\frac{X_2^2}{24}\right) \eta ^4\Bigg) \nonumber \\
	&+\mathcal{O}(\eta^{12}) \label{Eq:R+X1X2}\\
	\hat{R}^{-}_{\ell=2, m \omega}=&\frac{3}{2}-i \sqrt{X_2} \eta-\frac{11 X_2}{28} \eta ^2+\frac{3}{28} i X_2^{3/2} \eta ^3 \nonumber\\
	&+\left(-\frac{37 X_1 X_2}{42}+\frac{23 X_2^2}{1008}\right)\eta ^4+\mathcal{O}(\eta^5)\label{Eq:R-X1X2}
\end{align}
These solutions, while coming from usual Mano-Suzuki-Takasugi (MST) expansions, have been normalised to remove certain radius independent factors that are unimportant for constructing the inhomogeneous solution. The solutions can now be converted to series expansions in $1/p$ and $e$ as a function of $\chi$ by using Eqs.~\eqref{Eq:rofpe} and \eqref{Eq:rfreqPN} with the frequency $\omega=m \Omega_\varphi+n\Omega_r$ i.e. by evaluating \eqref{Eq:R+X1X2},\eqref{Eq:R-X1X2} at the position of the particle for the relevant frequency values. In the above, the $\eta$ factors are simply an order counting tool and in practice can be dropped when converting to the expansion in $1/p$. For now we will hold off on fully switching to $p,e$ and $\chi$ and instead swap $p$ for the dimensionless frequency variable 
\begin{align}
	y \equiv (m_2\Omega_\varphi)^{2/3}. 
\end{align}
The resulting double expansion in $y$ and $e$ for our example is 
\begin{widetext}
\begin{align}
	\hat{R}^{+}_{\ell=2, m \omega}&= \left(1+5  e\cos \chi +\mathcal{O}(e^2)\right)y^5+\left(i (m+n)+4 i  (m+n) e\cos \chi +\mathcal{O}(e^2)\right)y^{11/2}+\bigg(5-\frac{1}{2} (m+n)^2 \nn \\
	&+ \left(30-\frac{3}{2} (m+n)^2\right) e\cos \chi +\mathcal{O}(e^2)\bigg)y^6-\left(3 i n+\frac{1}{6} i (m+n)^3+ \left(12 i n+\frac{1}{3} i (m+n)^3\right) e\cos \chi +\mathcal{O}(e^2)\right)y^{13/2} \nn \\
	&+\mathcal{O}(y^7)
	\\
	\hat{R}^{-}_{\ell=2, m \omega}&=\frac{3}{2}+\left(-i (m+n)+i  (m+n) e\cos \chi +\mathcal{O}(e^2)\right)y^{1/2}+\left(-\frac{11}{28} (m+n)^2+\frac{11}{14}  (m+n)^2 e\cos \chi +\mathcal{O}(e^2)\right)y \nonumber \\
	&+\left(\frac{3}{2} \left(2 i n+\frac{1}{14} i (m+n)^3\right)+\frac{3}{2}  \left(-2 i n-\frac{3}{14} i (m+n)^3\right) e\cos \chi +\mathcal{O}(e^2)\right)y^{3/2}+\mathcal{O}(y^2) 
\end{align}  
\end{widetext}
where we have removed a constant factor from  $R^{+}_{\ell m \omega}$.

\subsection{Hertz potential and retarded metric perturbation}

The modes of the Hertz potential and thus the metric perturbation are then straightforwardly given using Eqs.~\eqref{Eq:PsiofR} and \eqref{Eq:hofPsi}. The main new feature in this construction as compared to the circular case is the sum over the radial frequencies, i.e. the infinite sum over $n$. The sum is expected to be exponentially convergent for the bound geodesics we are considering. This manifests in the convergence of the small eccentricity expansion. What one finds is that for an expansion valid to $e^k$, one needs only to sum $n=-k,...,k$ with $h_{\mu\nu}^{\ell m n}=\mathcal{O}(e^{k+1})$ for $|n|>k$. In other words, to capture higher eccentricity orbits one needs more and more $n$-modes. 

Computationally speaking the $n$-sum can be time consuming and a potential bottleneck. We find it therefore more convenient to sum the $n$-modes of the Hertz potential to give $\Psi_{\ell m}(t,r)=\sum_{n=-\infty}^\infty e^{-i \omega t}\Psi_{\ell m \omega}(r)$. We must also compute the $n$-sum for each of the $t$ derivatives that then appear in \eqref{Eq:hofPsi}, for example
\begin{align}
	\partial_t \Psi_{\ell m}(t,r)=-i\sum_{n=-\infty}^\infty \omega e^{-i \omega t}\Psi_{\ell m}(r).
\end{align}
as well as the relevant $r$ derivatives (which are obtained from Eq.~\eqref{Eq:PsiofR}). With these in hand the metric perturbation as a function of $\ell$ and $m$ is more or less trivially computed. For instance, using Eq.~\eqref{Eq:hofPsi}
\begin{align}
h_{tt}^{\ell m}=&\frac{1}{8}(r-2 m_2)^2\bigg(\partial^2_{\theta}\,{}_2 Y_{\ell m}(\theta,\varphi)+2 m\partial_{\theta}\,{}_2 Y_{\ell m}(\theta,\varphi) \nn \\
&+(m^2-2)\,{}_2 Y_{\ell m}(\theta,\varphi)\bigg) \Psi_{\ell m}(t,r)
\end{align} 
(which can be recognized as being a pure scalar harmonic). Each metric component and its derivatives are in practice evaluated here at the position of the particle. At this stage we sum over $m$ to get $h_{\mu\nu}^\ell=\sum_{m=-\ell}^\ell h_{\mu\nu}^{\ell m} $.

As will be discussed in more detail below, the singular nature of $h_{\mu \nu}(t,r,\theta,\varphi)$ in the vicinity
of the source worldline requires us to separately evaluate the two different radial limits $ r\to r_p^{\pm}$, from above or below, of the modes.
These limits are indicated below by a $\pm$ subscript: e.g. $h_{\mu\nu , \, \pm}^{\ell}$, or $F_{\mu, \, \pm}^{\ell}$.

 For ease of reading we omit explicit expressions for each of the components of the metric and their first order partial derivatives.

\subsection{Perturbed geodesic and self-force}

Before calculating the perturbed orbit quantities $\delta \mathcal{E}$, $\delta\mathcal{L}$ and $\delta u^r$ we must explicitly compute the $t$ and $\varphi$ components of the conservative self-force. The self-force is given by \cite{Wardell:2015kea}
\begin{align} \label{Fmu}
	F^\mu=\mathcal{P}^{\mu\nu\lambda\rho}(2 h_{\nu\lambda;\rho}-h_{\lambda\rho;\nu}),  \\ \nonumber
	\mathcal{P}^{\mu\nu\lambda\rho}=-\frac{1}{2}(g^{\mu\nu}+u^{\mu}u^{\nu})u^{\lambda}u^{\rho}.
\end{align}
where the covariant derivatives here are taken with respect to the background metric. Formally this equation requires the regularised metric perturbation; we will however use it with the $\ell$-modes of the retarded metric and leave regularization to later. 

At linear order in the mass ratio we can uniquely define the dissipative and conservative parts of the self-force via 
\begin{align}
F^{\mu}_\text{diss}&=\frac{1}{2}(F^{\mu}[h_{\mu\nu}^\text{ret}]-F^{\mu}[h_{\mu\nu}^\text{adv}]), \\
F^{\mu}_\text{cons}&=\frac{1}{2}(F^{\mu}[h_{\mu\nu}^\text{ret}]+F^{\mu}[h_{\mu\nu}^\text{adv}]).
\end{align}
Noting the symmetry relation of Eq.~(2.80) of \cite{Hinderer:2008dm}, the authors of \cite{Barack:2010tm} rewrote these for the case of equatorial geodesics purely in terms of the retarded solution:
\begin{align}
	F^{\mu}_\text{diss}&=\frac{1}{2}(F^{\mu}_\text{ret}(\tau)-\epsilon_{(\mu)}F^{\mu}_\text{ret}(-\tau)) \\
	F^{\mu}_\text{cons}&=\frac{1}{2}(F^{\mu}_\text{ret}(\tau)+\epsilon_{(\mu)}F^{\mu}_\text{ret}(-\tau)),
\end{align}
where $\epsilon_{(\mu)}=(-1,1,1,-1)$. With our parameterization $F^{\mu}_\text{ret}(\tau\!\rightarrow\!-\tau)\equiv F^{\mu}_\text{ret}(\chi\rightarrow-\chi)$. We compute this analytically, e.g. for $\ell=2$ we have
\begin{widetext}
\begin{align}
	q^{-1} F_{t,+}^\text{cons}&=\left(3 e \sin (\chi )+3 e^2\sin (2 \chi )+\mathcal{O}(e^3)\right)p^{-5/2}+\left(-\frac{85}{14} e \sin (\chi )-\frac{127}{14} e^2 \sin (2 \chi )+\mathcal{O}(e^3)\right)p^{-7/2}+\mathcal{O}(p^{-9/2}) \, ,\\
	q^{-1} F_{\varphi,+}^\text{cons}&=	\left(-\frac{9}{7} e \sin (\chi )-\frac{9}{7} e^2 \sin (2 \chi )+\mathcal{O}(e^3)\right)p^{-2}+\left(\frac{1237}{126} e \sin (\chi )+\frac{114}{7} e^2 \sin (2 \chi )+\mathcal{O}(e^3)\right)p^{-3}+\mathcal{O}(p^{-4}) \,.
\end{align}
In this double expansion the corresponding integrals needed for Eq.~\eqref{Eq:deltaEL} are trivially computed order by order. Explicitly,
\begin{align}
	\delta\mathcal{E}_{\text{BS}}^+&=\left(\frac{3}{2}+3 e \cos (\chi )+\mathcal{O}(e^2)\right)p^{-1}+\left(\frac{1}{14}-\frac{11}{7} e \cos (\chi )+\mathcal{O}(e^2)\right)p^{-2}+\mathcal{O}(p^{-3}),\\
	\delta\mathcal{L}_{\text{BS}}^+&=\frac{3}{2}\sqrt{p}+\left(\frac{1}{14}+\frac{9}{7} e \cos (\chi )+\mathcal{O}(e^2)\right)p^{-1/2}+\mathcal{O}(p^{-3/2}),
\end{align}
and the corresponding $\delta u^{r , +}_{\text{BS}}$ is found using the normalisation condition.
\end{widetext}

Eq.~\eqref{Eq:deltaPsidot} is now straightforwardly computed here to give the $\ell=2$ values
\begin{align}
	\delta\dot{\Psi}^+=&\bigg(\frac{3}{2}+3 \cos \chi e+\mathcal{O}(e^{2})\bigg)p^{-3/2} \nn \\
	&-\bigg(\frac{19}{28}+\frac{11}{7}\cos \chi e+\mathcal{O}(e^{2} \bigg)p^{-5/2}+\mathcal{O}(p^{-3}) \,,\nn \\
	\delta\dot{\Psi}^-=&-\big(1+2 \cos \chi e+\mathcal{O}(e^{2})\big)p^{-3/2} \nn \\
	&+\frac{4}{7}\big(1+6\cos \chi e+\mathcal{O}(e^{2} \big)p^{-5/2}+\mathcal{O}(p^{-3}) \,.
\end{align}
The integration of Eq.~\eqref{Eq:deltaPsi} is as usual applied order by order. 

The only remaining perturbed quantities are the frequency shifts $\delta\Omega_r$, $\delta\Omega_\varphi$ which appear in \eqref{Eq:Deltapsi}. Using Eq.~(75)-(76) of \cite{Barack:2011ed} (without the $\alpha$ factor) we find 
\begin{align}
	m_2\delta\Omega_r^+&=\bigg(\frac{3}{2}-\frac{9}{4}e^2\bigg)p^{-3/2} \nn \\
	&+\bigg(-\frac{115}{28}+\frac{575}{56}e^2+\mathcal{O}{(e^4)}\bigg)p^{-5/2}+\mathcal{O}{(p^{-7/2})} \nn\\
	m_2\delta\Omega_\varphi^+&=\bigg(\frac{3}{2}-\frac{9}{4}e^2\bigg)p^{-3/2} \nn \\
	&+\bigg(-\frac{103}{28}+\frac{77}{8}e^2+\mathcal{O}{(e^4)}\bigg)p^{-5/2}+\mathcal{O}{(p^{-7/2})}
\end{align}
Using Eq.~\eqref{Eq:Deltapsi} we give illustrative sample expansions for the retarded $\Delta\psi$ for $\ell=2$:
\begin{align}
	\Delta\psi^+_{\ell=2}&= \frac{43}{48}p^{-1}-\bigg(\frac{2047}{1008}+\frac{1427 }{1008}e^2\bigg)p^{-2}+\mathcal{O}(p^{-3})\nn \,,\\
	\Delta\psi^-_{\ell=2}&= \frac{43}{48}p^{-1}-\bigg(\frac{5827}{1008}+\frac{1427 }{1008}e^2\bigg)p^{-2}+\mathcal{O}(p^{-3}) \,.
\end{align}
Let us also give sample expansions for the generic $\ell$ expressions. This are valid for all $\ell$ greater than a value determined by the given PN order (see Sec II B. of Ref.~\cite{Kavanagh:2015lva}). These are useful for clearly seeing the divergent behaviour discussed in the next sections
\begin{widetext}
\begin{align} \label{psigenericl}
	\Delta\psi^+_{\ell}&=\frac{3 \left(1+7 \ell +7 \ell ^2\right)}{4 (-1+2 \ell ) (3+2 \ell )}p^{-1} + \Bigg(\frac{3 \left(1920-1197 \ell -852 \ell ^2+120 \ell ^3-653 \ell ^4-402 \ell ^5+90 \ell ^6+64 \ell ^7\right)}{16 \ell  (1+\ell ) (-3+2 \ell )
   (-1+2 \ell ) (3+2 \ell ) (5+2 \ell )}\nn \\
   &-\frac{3 \left(-480-102 \ell -466 \ell ^2-583 \ell ^3+71 \ell ^4+435 \ell ^5+145 \ell ^6\right)e^2}{32 \ell  (1+\ell ) (-3+2 \ell )
   (-1+2 \ell ) (3+2 \ell ) (5+2 \ell )}\Bigg)p^{-2} +\mathcal{O}(p^{-3})  \,,\nn \\
	\Delta\psi^-_{\ell}&=\frac{3 \left(1+7 \ell +7 \ell ^2\right)}{4 (-1+2 \ell ) (3+2 \ell )}p^{-1} + \Bigg(-\frac{3 \left(-1920+1377 \ell +1104 \ell ^2-848 \ell ^3-467 \ell ^4+402 \ell ^5+358 \ell ^6+64 \ell ^7\right)}{16 \ell  (1+\ell ) (-3+2 \ell )
   (-1+2 \ell ) (3+2 \ell ) (5+2 \ell )}\nn \\
   &-\frac{3 \left(-480-102 \ell -466 \ell ^2-583 \ell ^3+71 \ell ^4+435 \ell ^5+145 \ell ^6\right)e^2}{32 \ell  (1+\ell ) (-3+2 \ell )
   (-1+2 \ell ) (3+2 \ell ) (5+2 \ell )}\Bigg)p^{-2} +\mathcal{O}(p^{-3}) \,.\nn \\
\end{align}
\end{widetext}

\subsection{Regularization and completion of the metric perturbation}

\subsubsection{Metric completion (non-radiative multipoles $\ell \leq1$)}

The reconstructed retarded metric perturbation obtained by the CCK procedure laid out in the previous section is well known to give the full metric perturbation modulo perturbations to the spacetime due to the mass and angular momentum of the small body. In Schwarzschild spacetime this amounts to the absence of the spherical harmonic $\ell=0, 1$ modes in the reconstructed metric. We include these modes using the corrected Regge-Wheeler-Zerrili low multipoles given in Appendix A of \cite{Hopper:2015icj}. Ultimately, the contribution of these modes to the spin precession is, at leading orders, \\
\\
\begin{align}
	\Delta\psi^+_{\ell=0}&=\frac{-\frac{3}{4}+\frac{3 e^2}{4}}{p^2}+\frac{-\frac{73}{16}+\frac{45 e^2}{8}+\mathcal{O}(e^4)}{p^3}+\mathcal{O}(p^{-4}) \,,\nn \\
	\Delta\psi^-_{\ell=0}&=\frac{-\frac{3}{4}+\frac{3 e^2}{2}}{p^3}+\frac{-\frac{97}{16}+\frac{199 e^2}{16}+\mathcal{O}\left(e^4\right)}{p^4}+\mathcal{O}(p^{-5})  \,,
\end{align}
\begin{align}
	\Delta\psi^+_{\ell=1}&=\frac{1}{p}+\frac{\frac{9}{2}-e^2}{p^2}+\frac{\frac{163}{8}-6 e^2+\mathcal{O}(e^4)}{p^3}+\mathcal{O}(p^{-4}) \,,\nn \\
	\Delta\psi^-_{\ell=1}&=\frac{1}{p}+\frac{\frac{9}{4}+\frac{5 e^2}{4}}{p^2}+\frac{\frac{143}{16}+\frac{51 e^2}{8}+\mathcal{O}(e^4)}{p^3}+\mathcal{O}(p^{-4}) \,.
\end{align}

\subsubsection{Regularization}

So far in the calculation we have been constructing the $\ell$-modes of the spin precession
invariant from the retarded metric perturbation.  The full retarded metric perturbation is, however, a singular quantity due
to the point particle delta function source term. This singular nature manifests itself as a {\it direction-dependent}
divergent sum over $\ell$-modes (each individual $\ell$-mode being, however, finite), which for our
spin-precession invariant takes the following form for large values of $\ell$
\begin{align} \label{psisinglargel}
\Delta\psi^{\ell\to\infty}_{\pm}=\pm A_{\infty} \, (2\ell+1) +B_{\infty}+\mathcal{O}(\ell^{-2}).
\end{align}
Here, the sign of the $A_{\infty}$ term is dependent on whether one takes the limit $r\to r_p^\pm$ from above or
below. By definition, the two coefficients $A_{\infty}$ and $B_{\infty}$ parametrizing the large-$\ell$ singular behavior Eq. \eqref{psisinglargel}
of $\Delta \psi$ are independent of $\ell$.
[We have shown above (see, e.g., Eq. \eqref{psigenericl}) how our  PN-expanded method allows us to analytically
extract the values of $A_{\infty}$ and $B_{\infty}$.]

Calculating the regularized finite value for $\Delta\psi$ involves calculating, and subtracting out,  the 
$\ell$-modes of the corresponding singular piece $\Delta\psi^{\text{S}}$ of $\Delta\psi$, say
\begin{align} \label{psising}
\Delta\psi^{\text{S}}_{\ell , \pm}=\pm A_{\text{S}} \, (2\ell+1) +B_{\text{S}}.
\end{align}
Detweiler and Whiting \cite{Detweiler:2002mi} have shown how to define the singular pieces of $h_{\mu \nu}^{\text{S}}$,
$h_{\mu \nu , \lambda}^{\text{S}}$, and thereby $\Delta\psi^{\text{S}}$, in the Lorenz gauge.
Recently, Ref. \cite{Wardell:2015ada} emphasized that, when using (as we do here, following most of the
analytical work on SF corrections to gauge-invariant quantities starting with Ref. \cite{Detweiler:2008ft})
a decomposition of  $h_{\mu\nu}$ in {\it tensor-harmonics} $\ell$-modes, there are subtleties concerning
the value of the coefficient $A_{\text{S}}$ of $(2\ell+1)$ in Eq. \eqref{psising}. Indeed, while the  value of $B_{\text{S}}$ 
was found to be independent of $\ell$ (and therefore equal to the large-$\ell$ value $B_{\infty}$ entering Eq. \eqref{psisinglargel}), 
Ref. \cite{Wardell:2015ada} found that the  value of $A_{\text{S}}$ stabilized to its asymptotic value $A_{\infty}$
only when when $\ell \geq 2$. The (nonradiative) modes $\ell=0,1$ involve  $A_{\text{S}}$ coefficients that differ from $A_{\infty}$.
This difference comes from the peculiar low-$\ell$ dependence of certain radial derivatives of $h_{\mu\nu}$, such as $h_{t \varphi ,r}$ or $h_{\varphi \varphi ,r}$, see notably Eqs. (6.11j) and (6.11l) in Ref. \cite{Wardell:2015ada}. [Although the mentioned equations deal
with the case of circular orbits, nothing fundamentally changes in the eccentric orbit case, and
the same conclusions hold.]

 In the Lorenz gauge, the correct regularized spin-precession invariant would then be given by
\begin{align}
	\Delta\psi=\sum_{\ell=0}^{\infty}\left(\Delta\psi^\ell_\pm - \Delta\psi^{\ell, \pm}_{\text{S}} \right)=\sum_{\ell=0}^{\infty}\left(\Delta\psi^\ell_\pm    \mp A_{\text{S}} \, (2\ell+1) -B_{\text{S}} \right) \label{Eq:ModeSumLor}.
\end{align}   
This result can be simplified by working with the average over the  limits $r\to r_p$ from both sides,
thereby avoiding the need to keep track of the low $\ell$ dependence of $A_{\text{S}}$, and being able to use
the value of $B_{\infty}$ extracted from the large $\ell$ behavior of  $\Delta \psi$, Eq. \eqref{psisinglargel}:
\begin{align}
	\Delta\psi=\sum_{\ell=0}^{\infty}\left(\frac{1}{2}\big(\Delta\psi^{\ell,+}+\Delta\psi^{\ell,-}\big)-B_{\infty}\right). \label{Eq:ModeSumRad}
\end{align}
There is one remaining subtlety, which is that the regularization procedure explained above was derived in the
Lorenz gauge while our calculation of the tensorial $\ell$ modes was done in a radiation gauge. 
There are two ways to deal with this additional subtlety. 
On the one hand, as was shown by Barack and Ori in the case of the gravitational self-force \cite{Barack:2001ph}, 
the procedure should also work in a gauge which can be
reached from Lorenz gauge by a gauge transformation of a smooth enough nature. The radiation gauge
we work in does not, however, fall within this class (for a discussion of the singular structure of
the various radiation gauges we refer the reader to \cite{Pound:2013faa}). Much work has been
focussed recently on formally defining the correct regularization procedure in radiation gauge. The
strategy employed is to construct a gauge transformation $\xi^\mu$ which is defined locally in the
vicinity of the worldline, and which can be used to transform from the radiation gauge to a
``locally Lorenz'' gauge. It is then argued that this generates a correction to the mode-sum
formula, Eq.~\eqref{Eq:ModeSumLor}, which has an overall sign that differs on either side of the
$r\to r_p$ limit. A straightforward solution, then, is to take the average of both limits (as we
are doing), thus eliminating the troublesome singular gauge contribution.

On the other hand,  it is likely that a much simpler solution resolves the issue in our case. The quantity we are
computing is a gauge invariant, and so we are free to use the regularization procedure derived in
the Lorenz gauge (with averaging to compensate for the fact that we are not decomposing into
scalar spherical harmonics) and apply it to any other gauge in the same invariance class. We
stress the point, however, for two reasons: (i) while we have partially checked the correctness of this fact
(notably by checking the continuity  of $\Delta\psi^\ell_\pm - \Delta\psi^{\ell, \pm}_{\text{S}}$ across $r=r_p$), 
we have not analytically shown that
$\Delta\psi$ is invariant under the transformation from Lorenz to radiation gauge; and (ii) things
would not be so straightforward if one were interested in computing non-gauge-invariant quantities
such as the self-force (in which case a more careful analysis along the lines of
Ref.~\cite{Pound:2013faa} may be required).

Finally, taking the large-$\ell$ limit of our expression \eqref{psigenericl} (to all orders we computed) we can read off the $A_\infty$ and $B_\infty$ coefficients:
\begin{widetext} 
\begin{align}
	A_{\infty}=& \frac{3}{8}p^{-2}+\left(\frac{49}{32}-\frac{27 e^2}{32}+\mathcal{O}(e^4)\right)p^{-3}+\left(\frac{1007}{128}-\frac{221 e^2}{32}+\mathcal{O}(e^4)\right)p^{-4}+\left(\frac{23441}{512}-\frac{25437 e^2}{512}+\mathcal{O}(e^4)\right)p^{-5} \nn \\
	&+\left(\frac{575983}{2048}-\frac{354297 e^2}{1024}+\mathcal{O}(e^4)\right)p^{-6}+\mathcal{O}(p^{-7})\\
	B_{\infty}=&\frac{21}{16} p^{-1}+ \left(-\frac{201}{128}-\frac{435 e^2}{512}+\mathcal{O}(e^4)\right)p^{-2}+\left(\frac{529}{1024}-\frac{1155 e^2}{1024}+\mathcal{O}(e^4)\right)p^{-3}+\left(\frac{152197}{16384}-\frac{352849 e^2}{65536}+\mathcal{O}(e^4)\right)p^{-4} \nn \\
	&+\left(\frac{17145445}{262144}-\frac{5100243 e^2}{131072}+\mathcal{O}(e^4)\right)p^{-5}+\left(\frac{886692225}{2097152}-\frac{2456459237
   e^2}{8388608}+\mathcal{O}(e^4)\right)p^{-6}+\mathcal{O}(p^{-7}).
\end{align}
\end{widetext}

Although we do not use the $A_\infty$ parameter due to our averaging, we have checked that both of these expressions agree with independently calculated $A_S$ and $B_S$. These were computed from expansions in $p$ and $e$ of the singular metric perturbation calculated using the methods described in \cite{Barack:1999wf,Heffernan:2012su,Wardell:2015ada} with a projection onto scalar spherical harmonics. This provides a valuable check that our radiative modes ($\ell\geq2$) are demonstrating the correct singular behaviour.

\subsection{Spin-precession invariant}

Upon regularization, our generic-$\ell$ terms can be seen analytically to converge as $\ell^{-2}$.
This allows us to explicitly compute the infinite series over $\ell$, giving one of our main results
\newpage
\begin{widetext}
\begin{align}
	\Delta\psi=&-p^{-1}+\left(\frac{9}{4}+e^2\right)p^{-2}+\left(\left(\frac{739}{16}-\frac{123 \pi ^2}{64}\right)+\left(\frac{341}{16}-\frac{123 \pi
   ^2}{256}\right) \mathit{e}^2+\mathcal{O}(e^4)\right)p^{-3}+\Bigg(\bigg(-\frac{587831}{2880}+\frac{1256 \gamma }{15} \nonumber\\
   &+\frac{31697 \pi ^2}{6144}+\frac{296 \log
   (2)}{15}+\frac{729 \log (3)}{5}-\frac{628 \log (p)}{15}\bigg)+\bigg(-\frac{164123}{480}+\frac{536 \gamma }{5}-\frac{23729 \pi ^2}{4096}+\frac{11720 \log
   (2)}{3} \nonumber\\
   &-\frac{10206 \log (3)}{5}-\frac{268 \log (p)}{5}\bigg)e^2+\mathcal{O}(e^4)\Bigg)p^{-4} +\Bigg(\bigg(-\frac{48221551}{19200}-\frac{22306 \gamma }{35}+\frac{2479221 \pi ^2}{8192}+\frac{22058
   \log (2)}{105} \nonumber\\
   &-\frac{31347 \log (3)}{28}+\frac{11153 \log (p)}{35}\bigg)+\bigg(-\frac{164123}{480}+\frac{536 \gamma }{5}-\frac{23729 \pi ^2}{4096}+\frac{11720 \log
   (2)}{3}-\frac{10206 \log (3)}{5} \nonumber \\
   &-\frac{268 \log (p)}{5}\bigg)e^2+\mathcal{O}(e^4)\Bigg)p^{-5}+\Bigg(\frac{49969 \pi }{315}+\frac{319609 \pi }{630}e^2+\mathcal{O}(e^4)\Bigg)p^{-11/2}+\Bigg(\bigg(-\frac{1900873914203}{101606400}-\frac{344021 \gamma }{1890} \nonumber \\
   & +\frac{7230119267 \pi
   ^2}{2359296}-\frac{7335303 \pi ^4}{131072}-\frac{2514427 \log (2)}{270}+\frac{234009
   \log (3)}{70}+\frac{9765625 \log (5)}{9072}+\frac{344021 \log (p)}{3780}\bigg) \nonumber \\
   &+\bigg(-\frac{464068669129}{5080320}-\frac{2508913 \gamma }{945}+\frac{32088966503 \pi
   ^2}{2359296}-\frac{146026515 \pi ^4}{1048576}+\frac{273329813 \log
   (2)}{945}-\frac{159335343 \log (3)}{8960} \nonumber \\
   &-\frac{17193359375 \log
   (5)}{145152}+\frac{2508913 \log (p)}{1890}\bigg)e^2+\mathcal{O}(e^4)\Bigg)p^{-6}+\mathcal{O}\left(p^{-13/2}\right). \label{Eq:DeltaPsiPN}
\end{align}
\end{widetext}

We could perform several checks on our result Eq. \eqref{Eq:DeltaPsiPN}. First, the first three terms of our expansion can be seen immediately to agree with the coefficients of $p^{-1}, p^{-2}, p^{-3}$ derived in Ref. \cite{Akcay:2016dku}. 

Second, we have compared the numerical values predicted by our
analytical result Eq. \eqref{Eq:DeltaPsiPN} 
to the numerical estimates of $\Delta \psi$ obtained in Ref.
\cite{Akcay:2016dku}. More precisely, Table II in Ref.
\cite{Akcay:2016dku} lists  a sample of numerical estimates  $\Delta
\psi^{\rm num}(p_i, e_k)$  of  $\Delta \psi(p,e)$, together with  estimates of the 
corresponding numerical error $\sigma^{\rm num}(p_i, e_k)$. This sample
includes nineteen values, $10 \leq p_i \leq 100$, of the semi-latus rectum $p$,
and, for each value of $p$, nine values of $e$, namely: $e_k = 0.050+0.025k\,, k=0,1,\ldots,8$.
As our main aim was to compare the eccentricity dependence of these numerical data to the
one predicted by our analytical result  Eq. \eqref{Eq:DeltaPsiPN}, we extracted a numerical estimate
of the $\mathcal{O}(e^2)$ contribution, say $e^2 \Delta \psi^{(2)}(p)$, in  $\Delta \psi(p,e)$, 
\beq
\Delta \psi(p,e)=\Delta \psi^{(0)}(p)+e^2 \Delta \psi^{(2)}(p)+e^4 \Delta \psi^{(4)}(p)+\mathcal{O}(e^6)\,,
\eeq
in the following way. First, we used  the analytical knowledge of the leading-order
$\mathcal{O}(e^4)$ contribution, $e^4 \Delta \psi^{e^4}_{\rm LO}(p) = - \frac12 e^4
p^{-3}$ (derived in Ref. \cite{Akcay:2016dku}) to work with the $\mathcal{O}(e^4)$-corrected\footnote{We have checked 
that taking into account a next-to-leading-order  $\mathcal{O}(e^4)$ contribution, $e^4 \Delta \psi^{e^4}_{\rm NLO}(p) = c_{\rm NLO}\, e^4
p^{-4}$, with $ |c_{\rm NLO}| \leq 3$ did not significantly change our results.}
inclusion of  numerical data 
$\Delta \psi^{\rm num'}(p_i, e_k) \equiv \Delta \psi^{\rm num}(p_i, e_k) - e_k^4 \Delta \psi^{e^4}_{\rm LO}(p_i) $.
Then, we fitted, for each value of $p_i$ the nine numerical data $\Delta \psi^{\rm num'}(p_i, e_k); \,  k=0,1,\ldots,8$
to a linear function of $e^2$, say
\beq
\Delta \psi^{\rm fit}(p_i, e_k) =  m_i \, e_k^2 +q_i \,.
\eeq
We used a least-squares fit, weighted by the (inverse squares of the) corresponding numerical errors $\sigma^{\rm num}(p_i, e_k)$ listed in 
Table II  of \cite{Akcay:2016dku}. This fitting procedure gave us (for each $p_i$) estimates of $m_i (\approx \Delta \psi^{(2)}(p_i))$
and $q_i (\approx \Delta \psi^{(0)}(p_i))$, together with corresponding fitting errors (obtained from the covariance matrix). In addition,
the goodness of each fit is measured by the corresponding reduced 
$\chi^2_{\rm red}(p_i)= \chi^2_{\rm min}/(N_{\rm data} - N_{\rm param})$, with $N_{\rm data}=9$ and $N_{\rm param}=2$.

The results for our estimates of the $e^2$ slope $m_i \approx \Delta \psi^{(2)}(p_i)$ are given in Table I below.

\begin{table}[t]
  \caption{\label{tab:num_analysis} Comparison between the  values $m^{\rm thy}(p)$, predicted by our
 analytical result Eq. \eqref{Eq:DeltaPsiPN}, of the coefficient of $e^2$  in $\Delta \psi(p,e)= q(p) + m(p) e^2+ \mathcal{O}(e^4)$,
to the numerical estimates $m^{\rm num}(p)$ of $m(p)$ obtained by least-squares fitting the numerical data for 
$\Delta \psi(p,e)$ obtained in Ref.\cite{Akcay:2016dku}. The last entry is the ratio \eqref{rm} between the difference
$m^{\rm num} -  m^{\rm thy}$ and our estimate of the total error $\sigma^{\rm tot}_{m} \equiv \sqrt{ (\sigma^{\rm num}_{m})^2 +   (\sigma^{\rm thy}_{m})^2}$ on $m$.
 See text for more details (notably about the estimates of the errors $\sigma^{\rm thy}_{m}$ and $\sigma^{\rm num}_{m}$).}
  \begin{center}
    \begin{ruledtabular}
      \begin{tabular}{lcccc}
$p$& $\chi^2_{\rm red}$& $m^{\rm num}(\sigma^{\rm num}_m)$& $m^{\rm thy}(\sigma^{\rm thy}_{m})$& $r_m$\\
\hline
10&  0.221&   2.83892(11)$\times 10^{-2}$   &     3.9(2.1)$\times 10^{-2}$  &  -0.50 \cr
15&  4.119&   9.12787(61)$\times 10^{-3}$    &   10.0(1.9)$\times 10^{-3}$  & -0.45 \cr
20&  2.591&   4.40237(32)$\times 10^{-3}$ &    4.54(34)$\times 10^{-3}$ & -0.41 \cr
25& 0.0357&   2.561664(35)$\times 10^{-3}$   &    2.956(92)$\times 10^{-3}$ &  -0.38 \cr
30& 0.867&    1.66508(23)$\times 10^{-3}$      &     1.677(31)$\times 10^{-3}$ & -0.37 \cr
35& 0.646&    1.16553(20)$\times 10^{-3}$    &     1.170(12)$\times 10^{-3}$  &  -0.33 \cr
40& 0.372&    8.5943(13)$\times 10^{-4}$     &     8.611(56)$\times 10^{-4}$   & -0.30 \cr
45& 0.102&    6.58803(62)$\times 10^{-4}$    &     6.597(28)$\times 10^{-4}$    &  -0.32 \cr
50& 0.227&    5.20356(82)$\times 10^{-4}$    &     5.211(15)$\times 10^{-4}$   &   -0.48 \cr
55& 0.481&    4.2124(14)$\times 10^{-4}$    &     4.2171(85)$\times 10^{-4}$ &   -0.55 \cr
60& 1.249&    3.4798(15)$\times 10^{-4}$    &     3.4813(50)$\times 10^{-4}$    &   -0.29 \cr
65& 3.630&    2.9178(37)$\times 10^{-4}$      &     2.9215(31)$\times 10^{-4}$  &   -0.75 \cr
70& 1.800&    2.4793(43)$\times 10^{-4}$     &     2.4859(20)$\times 10^{-4}$    &   -1.40 \cr
75& 3.113&    2.1347(49)$\times 10^{-4}$     &     2.1405(13)$\times 10^{-4}$    &   -1.14 \cr
80& 6.313&    1.8583(90)$\times 10^{-4}$   &    1.86200(91)$\times 10^{-4}$   &  -0.41 \cr
85& 4.832&    1.6289(97)$\times 10^{-4}$    &    1.63430(64)$\times 10^{-4}$  &   -0.56 \cr
90& 1.707&    1.4552(39)$\times 10^{-4}$    &    1.44578(45)$\times 10^{-4}$  &  2.42 \cr
95& 3.687&    1.292(10)$\times 10^{-4}$      &    1.28797(33)$\times 10^{-4}$  &  0.37 \cr
100& 1.882&   1.130(14)$\times 10^{-4}$   &    1.15456(24)$\times 10^{-4}$ &   -1.70\cr
   \end{tabular}
  \end{ruledtabular}
\end{center}
\end{table}

Namely, the first four entries of Table I are: $p_i$, $\chi^2_{\rm red}(p_i)$, the numerical estimate $m_i^{\rm num}$ of the slope 
$m_i \approx \Delta \psi^{(2)}(p_i)$ obtained from our fit, and the corresponding theoretical prediction 
$m_i^{\rm thy} \equiv \Delta \psi^{(2) \rm PN}(p_i)$, as obtained from our PN-expanded analytical result Eq. \eqref{Eq:DeltaPsiPN}.
In addition, we have indicated in parentheses, both for the numerical estimates of $m_i$, and their theoretical ones, estimates of the 
corresponding uncertainties in their values. The estimates of the numerical uncertainty on $m_i^{\rm num}$ was obtained by 
renormalizing the fitting error by a factor $\sqrt{\chi^2_{\rm red}(p_i)}$. Indeed, though
the goodness-of-fit parameters $\chi^2_{\rm red}(p_i)$ were always (as tabulated) of order unity, they were not always numerically close to 1.
In view of the difficulty, in numerical SF computations, to accurately estimate the numerical error, we considered that a value of $\chi^2_{\rm red}(p_i)$
different from 1 indicated an inaccurate estimate of the numerical errors $\sigma^{\rm num}(p_i, e_k)$, which we (coarsely) corrected by
multiplying $\sigma^{\rm num}_m(p_i, e_k)$ (and correlatively $\sigma^{\rm fit}_{m_i}$) by a factor $\sqrt{\chi^2_{\rm red}(p_i)}$.
Concerning the theoretical error $\sigma^{\rm thy}_{m_i}$, it was estimated by the value of the last analytically computed contribution to
$\Delta \psi^{(2)}(p_i)$ in Eq. \eqref{Eq:DeltaPsiPN}, i.e. 
$\sigma^{\rm thy}_m(p) =  |(18101.8418+1327.467196\ln(p)) 1/p^6|$.

Finally, the last entry in Table I displays the values of the ratios
\beq \label{rm}
r_{m_i}\equiv \frac{ m_i^{\rm num} -  m_i^{\rm thy} }{\sigma^{\rm tot}_{m_i}},
\eeq
where $\sigma^{\rm tot}_i \equiv \sqrt{ (\sigma^{\rm num}_{m_i})^2 +   (\sigma^{\rm thy}_{m_i})^2}$ is an estimate of the combined
numerical-analytical error on $m_i \approx \Delta \psi^{(2)}(p_i)$. The most significant fact (for our purpose) in Table I is that the values
of the latter ratios are all of order unity. This is a valuable, independent check on our analytical computations.

Our fitting procedure has
also given use numerical estimates of  $q_i \approx \Delta \psi^{(0)}(p_i)$ that we have satisfactorily compared (with corresponding ratios
$r_{q_i}\equiv  (q_i^{\rm num} -  q_i^{\rm thy} )/\sigma^{\rm tot}_{q_i}$ found to be of order unity)
to the 9.5PN current analytical knowledge of $\Delta \psi^{(0)}(p)$ (given in Appendix A of \cite{Akcay:2016dku}). 
[We recall (see the end of section IV for more details) that 
$ \Delta \psi^{(0)}(p) =\lim_{e\to0} \Delta \psi(p, e)$ differs from $\Delta \psi^{\rm circ}(p)$, and that the difference  $\Delta \psi(p, e\to0) -\Delta \psi^{\rm circ}(p)$ was related in Section IIIB of \cite{Akcay:2016dku} to the EOB function $\rho(x)$ measuring periastron precession. 
While $\Delta \psi^{\rm circ}(p)$ is known to very high PN orders \cite{Kavanagh:2015lva,online},  $\rho(x)$ is currently known to the 9.5-PN level \cite{Bini:2016qtx}.]

\section{Improving the spin-orbit  sector of the EOB Hamiltonian}
\label{Sec:EOB}
In this section, we shall show how to transcribe the new SF results contained in Eq. \eqref{Eq:DeltaPsiPN} above into an improved knowledge
of  the spin-orbit sector of the EOB Hamiltonian.
The inclusion of spin couplings in the EOB Hamiltonian was initiated in Ref. \cite{Damour:2001tu} and  developed in Refs. \cite{Damour:2008qf,Barausse:2009aa,Barausse:2009xi,Nagar:2011fx,Barausse:2011ys,Taracchini:2012ig,Taracchini:2013rva,Damour:2014sva,Bini:2014ica,Balmelli:2015lva,
Bini:2015mza,Balmelli:2015zsa,Bini:2015xua,Bohe:2016gbl,Babak:2016tgq}. Here, we will focus on the case of non-precessing spins (parallel or antiparallel to the orbital angular momentum), and  only consider effects {\it linear} in spins. Following the formulation of Refs. \cite{Damour:2008qf,Damour:2014sva}, the spin-orbit
couplings are described by two phase-space-dependent gyrogravitomagnetic ratios $g_S$ and $g_{S*}$.

We would like to emphasise that throughout this section all coordinate variables will be referring to EOB variables which, despite overlapping labelling, are not to be confused with those of the previous section. For example we will encounter an EOB eccentricity $e$, which is distinct to that used in Eq. \eqref{Eq:DeltaPsiPN} to parameterize the noncircular dependence of the SF spin precession invariant. To avoid issues when relating the EOB spin precession function to its  previous SF version, we shall henceforth relabel all independent variables from the previous section with an additional overbar, i.e. the variables entering Eq. \eqref{Eq:DeltaPsiPN} will be now written as $\bar{p}$ and $\bar{e}$.
In addition, in order to better explicate the introduction of various dimensionless quantities in the EOB formalism, we shall often return,
in this section, to the use of physical units where $G$ and $c$ are not set to unity.

\subsection{EOB notation and reminders}

Let us first recall the standard EOB results and notation, which we shall follow here. The total Hamiltonian of the system is expressed as 
\beq
\label{H_Heff}
H({\mathbf R},{\mathbf P},{\mathbf S}_1,{\mathbf S}_2)=Mc^2\sqrt{1+2\nu \left(\frac{H_{\rm eff}}{\mu c^2}-1\right)}\equiv Mc^2 h\,,
\eeq
where (with the convention $m_1<m_2$, and $m_1\ll m_2$ in  the extreme-mass-ratio limit)
\beq
M=m_1+m_2\,,\mu=\frac{m_1m_2}{(m_1+m_2)}\,, \nu=\frac{\mu}{M}=\frac{m_1m_2}{(m_1+m_2)^2},
\eeq
and where the {\it effective EOB Hamiltonian} $H_{\rm eff}$ is decomposed as 
\beq
H_{\rm eff}=H^{\rm O}_{\rm eff}+H^{\rm SO}_{\rm eff}\,.
\eeq
Here
\beq
\label{orb_ham}
H^{\rm O}_{\rm eff}=c^2 \sqrt{A \left(\mu^2c^2 +{\mathbf P}^2+\left(\frac{1}{B}-1\right)P_R^2+ Q\right)}\,,
\eeq
denotes the {\it orbital} part of the effective Hamiltonian, expressed in terms of the squared linear momentum 
\beq
{\mathbf P}^2=\frac{P_R^2}{B}+\frac{{\mathbf L}^2}{R^2}=\frac{P_R^2}{B}+\frac{P_\phi^2}{R^2}\,,
\eeq
with ${\mathbf L}={\mathbf R}\times {\mathbf P}$ denoting the orbital angular momentum (with magnitude $L\equiv P_\phi$), and in terms of the EOB radial potentials parametrizing the effective metric 
(specialized here to equatorial motions) 
\beq
\label{eff_met}
ds^2_{\rm (eff)}=g^{\rm (eff)}_{\mu\nu}dX^\mu dX^\nu=-A c^2 dT_{\rm eff}^2 + B dR^2 +R^2 d\phi^2\,.  
\eeq
The last (quartic in momenta) contribution $Q$ on the right-hand-side (rhs) of Eq. \eqref{orb_ham} will be defined below.
In addition  
\beq
H^{\rm SO}_{\rm eff}= G_S^{\rm phys} {\mathbf L}\cdot {\mathbf S}+G_{S*}^{\rm phys} {\mathbf L}\cdot {\mathbf S}_* 
\eeq
denotes the {\it spin-orbit} part of the effective Hamiltonian, expressed in terms of the following two symmetric combination of the spin vectors ${\mathbf S}_1$ and ${\mathbf S}_2$ of the system
\beq
{\mathbf S}={\mathbf S}_1+{\mathbf S}_2\,,\qquad {\mathbf S}_*=\frac{m_2}{m_1} {\mathbf S}_1+\frac{m_1}{m_2}{\mathbf S}_2\,.
\eeq
In the parallel-spin case that we consider here ${\mathbf L}\cdot {\mathbf S}=LS=P_\phi S$ and ${\mathbf L}\cdot {\mathbf S}_*=LS_*=P_\phi S_*$.
It is convenient to work with the following dimensionless, rescaled variables 
\beq
\label{rescal}
r=\frac{c^2 R}{GM}\,, u=\frac{GM}{c^2 R}\equiv\frac{1}{r}\,, j\equiv p_\phi=\frac{c P_\phi}{GM\mu}\,, p_r=\frac{P_R}{\mu c}\,,
\eeq
and quantities
\begin{eqnarray}
\label{rescal2}
\hat H_{\rm (eff)}&=&\frac{H_{\rm (eff)}}{\mu c^2}\equiv \hat H^{\rm O}_{\rm (eff)}+\hat H^{\rm SO}_{\rm (eff)}\nonumber\\
g_S &=& R^3 G_S^{\rm phys}\,,\quad g_{S*}= R R_c^2 G_{S*}^{\rm phys}
\end{eqnarray}
where $R_c^2=R^2+\mathcal{O}({\rm spin}^2)$ \cite{Damour:2014sva}. Here, as we work linearly in spins, we can neglect the spin quadratic contribution to $R_c^2$.
In the following, we shall sometimes set, for simplicity, the velocity of light to 1.

\subsection{Present knowledge of the EOB  gyrogravitomagnetic ratios $g_{S}$ and $g_{S*}$}

Let us describe the present knowledge of the two (phase-space-dependent) dimensionless  gyrogravitomagnetic ratios $g_{S}$ and $g_{S*}$.
First, from the PN-expanded point of view, $g_{S}$ and $g_{S*}$ are known  at the next-to-next-to-leading-order  (NNLO) level \cite{Nagar:2011fx,Barausse:2011ys}
\begin{widetext}
\begin{eqnarray}
\label{gssstar_PN}
g_S^{\rm PN}(u,p_r,p_\phi) &=&2 +\eta^2 \left[-\frac58 \nu u-\frac{27}{8}\nu p_r^2   \right]+\eta^4 \left[\nu \left(-\frac{51}{4}u^2-\frac{21}{2}up_r^2+\frac58 p_r^4 \right)+\nu^2 \left( -\frac18 u^2 +\frac{23}{8}u p_r^2+\frac{35}{8}p_r^4 \right) \right] +\mathcal{O}(\eta^6) \,,\nonumber\\
g_{S*}^{\rm PN}(u,p_r,p_\phi) &=& \frac32 +\eta^2 \left[-\frac98 u -\frac{15}{8}p_r^2  +\nu \left(-\frac34 u-\frac94 p_r^2  \right)\right]+\eta^4 \left[-\frac{27}{16}u^2+\frac{69}{16}up_r^2+\frac{35}{16}p_r^4 \right. \nonumber\\
&& \left. +\nu \left(-\frac{39}{4}u^2-\frac{9}{4}u p_r^2+\frac52 p_r^4 \right)+\nu^2 \left( -\frac{3}{16}u^2 +\frac{57}{16}u p_r^2+\frac{45}{16}p_r^4 \right) \right]+\mathcal{O}(\eta^6)\,. 
\end{eqnarray}
\end{widetext}
Here  $\eta\sim 1/c$ is a place-holder for keeping track of the PN order, which we  shall generally ignore in the following.
The values of $g_{S}$ and $g_{S*}$ cited above have been expressed in the Damour-Jaranowski-Schaefer (DJS) spin gauge \cite{Damour:2007nc,Damour:2008qf}, which is defined so that these quantities do not actually depend on $p_\phi$.

Note that, at the PN order indicated above, $g_{S}$ and $g_{S*}$ depend on the symmetric mass ratio $\nu$ in the following way
\begin{eqnarray}
g_S(u,p_r,p_\phi)&=& 2 +\nu g_{S}^{(\nu^1)}(u,p_r,p_\phi)+\nu^2 g_{S}^{(\nu^2)}(u,p_r,p_\phi) \nn \\
&&+\mathcal{O}(\nu^3) \,,\nonumber\\
g_{S*}(u,p_r,p_\phi)&=& g_{S*}^{(\nu^0)}(u,p_r,p_\phi) +\nu g_{S*}^{(\nu^1)}(u,p_r,p_\phi) \nn \\
&&+\nu^2 g_{S*}^{(\nu^2)}(u,p_r,p_\phi)+\mathcal{O}(\nu^3)\,.
\end{eqnarray}

Analytical gravitational self-force theory allowed one to improve the knowledge on the first gyrogravitomagnetic ratio $g_S$, to linear order in $\nu$ and for circular orbits. Namely, Ref. \cite{Bini:2015xua} derived (along circular orbits) the PN  expansion of $g_S^{\rm (circ)}(u)= 2-\frac{5}{8} \nu u \delta G_S^{\rm resc}+\mathcal{O}(\nu^2)$  to the 7.5PN level, see Eq. (4.3) there.  
For concreteness, let us quote here only the first few terms of this expansion
\begin{align}
g_S^{\rm (circ)}(u)=&2-\frac{5}{8} \nu u\bigg[1+\frac{102}{5}u+\left( \frac{80399}{720}-\frac{241}{120}\pi^2 \right)u^2 \nn \\
& +\mathcal{O}(u^3)  \bigg]+\mathcal{O}(\nu^2)\,.
\end{align}

Concerning the second gyrogravitomagnetic ratio $g_{S*}$, it was emphasized in Ref.\cite{Barausse:2009xi} that the $\nu$-independent piece of $g_{S*}$, $g_{S*}^{(\nu^0)}$,  could be exactly determined  from considering 
a spinning particle in an external background \cite{Barausse:2009aa}. Taking as external background a Schwarzschild metric (consistently with our working linearly in spins) this leads to the explicit expression
\begin{align}
\label{gsstar_expl}
g_{S*}^{ (\nu^0)}(u,p_r,p_\phi)=\frac{1}{1+\sqrt{ 1+p_\phi^2 u^2 +(1-2u)p_r^2}} \nn \\+\frac{1}{\sqrt{ 1+p_\phi^2 u^2 +(1-2u)p_r^2}}\, \frac{2}{1+\frac{1}{\sqrt{1-2u}}}\,.
\end{align}

This exact expression for $g_{S*}^{ (\nu^0)}$ has introduced an explicit dependence on $p_\phi$, corresponding to being in a different spin gauge than the DJS one used in the PN-expanded expressions \eqref{gssstar_PN}. 

Gravitational self-force theory allowed, starting in 2014, to acquire new knowledge on spin precession in extreme-mass-ratio binaries \cite{Dolan:2013roa,Bini:2014ica,Bini:2015mza,Kavanagh:2015lva}. The knowledge acquired in the latter references was limited to the case of circular orbits and was  transcribed within the EOB formalism in \cite{Bini:2014ica}. 
When considering circular orbits, it is natural 
to decompose $g_{S*}$  (within self-force theory) in the following way
\beq
\label{g_star_circ_n}
g_{S*}^{\rm  circ }(u,\nu)= g_{S*}^{{\rm circ }\, (\nu^0)}(u)+\nu  g_{S*}^{{\rm circ }\, (\nu^1) }(u)+\mathcal{O}(\nu^2)\,,
\eeq
where
\beq
\label{0_circ}
g_{S*}^{{\rm circ }\, (\nu^0)}(u) \equiv\frac{3}{1+\frac{1}{\sqrt{1-3u}}}\,,
\eeq
is defined by replacing $p_r$ by 0 and $p_\phi$ by $p_\phi^{\rm (circ)}(u)=[u(1-3u)]^{-1/2}$ in \eqref{gsstar_expl}.
The additional $\mathcal{O}(\nu)$ correction $\nu  g_{S*}^{{\rm circ }\, (\nu^1) }(u)$ in Eq. \eqref{g_star_circ_n} has been analytically determined as a PN expansion, up
to the 9.5PN level in \cite{Bini:2014ica,Bini:2015mza}. [From the analytical results of \cite{Kavanagh:2015lva} one could further determine $g_{S*}^{{\rm circ }\, (\nu^1) }(u)$  to the 23PN level]. Moreover, Ref. \cite{Bini:2014ica}, combining analytical knowledge with a fit to numerical SF data from Ref. \cite{Dolan:2013roa} [together with numerical SF data from \cite{Akcay:2012ea}] derived a simple representation of $g_{S*}^{{\rm circ }\, (\nu^1) }(u)$ as a rational function of $u$ in the interval $0\le u <\frac13$, see Eqs. (6.39)--(6.40) in Ref. \cite{Bini:2014ica}.
Let us quote here only the first  terms of the PN expansion of  $g_{S*}^{{\rm circ }\, (\nu^1) }(u)$

\begin{align}
\label{gsstar_circ_expl3}
g_{S*}^{{\rm circ }\, (\nu^1) }(u)=& -\frac34 u -\frac{39}{4} u^2+\left(\frac{41}{32}\pi^2-\frac{7627}{192}\right) u^3\nonumber\\
&+\bigg(-24\ln(u)-\frac{1017}{20}-\frac{1456}{15}\ln(2)-48\gamma \nn \\
&+\frac{23663}{2048}\pi^2\bigg)u^4+\mathcal{O}(u^5).
\end{align}
See Eq. (A1) in the Appendix of Ref. \cite{Bini:2015mza} for the 9.5PN accurate extension of this expression.

\subsection{Improving the analytical knowledge of $g_{S}$ and $g_{S*}$}

In the present paper we shall improve the SF knowledge of $g_{S*}$ by 
computing the $\mathcal{O}(p_r^2)$ corrections to Eq. \eqref{g_star_circ_n}.
To this end, let us decompose $g_{S*}$ in the following way
\begin{align}
\label{gsstar_param}
g_{S*}(u,p_r,p_\phi; \nu) &=g_{S*}^{ (\nu^0)}(u,p_r,p_\phi)+\nu \big[g_{S*}^{\rm 1SF 0}(u)\nn \\
&+p_r^2 g_{S*}^{\rm 1SF 2}(u)+\mathcal{O}(p_r^4)\big]+\nu^2 g_{S*}^{(\nu^2)}(u) \nn \\
&+\mathcal{O}(\nu^2p_r^2)+\mathcal{O}(\nu^3)\,.
\end{align}
Here $ g_{S*}^{ (\nu^0)}(u,p_r,p_\phi)$ is the phase-space function defined in Eq. \eqref{gsstar_expl} above, while we have written the additional $\mathcal{O}(\nu)$ and $\mathcal{O}(\nu^2)$ contributions
as $p_\phi$-independent functions of $u$ and $p_r$, expanded in powers of $p_r^2$. [We use here the freedom allowed by DJS-type spin gauge to eliminate any $p_\phi$ dependence in the $\mathcal{O}(\nu)$ terms.]
In the following we shall use the SF result Eq. \eqref{Eq:DeltaPsiPN} to determine the PN expansions of $g_{S*}^{\rm 1SF 0}(u) $ and $g_{S*}^{\rm 1SF 2}(u)$, namely
\begin{align}
\label{coeff_undet}
g_{S*}^{\rm 1SF 0}(u) =&  g_{*01}u+g_{*02}u^2+g_{*03}u^3+\left(g_{*04}^c+g_{*04}^{\ln{}}\ln u\right)u^4 \nn\\
&+\ldots \nonumber\\
g_{S*}^{\rm 1SF 2}(u)=& g_{*20}+ g_{*21}u+g_{*22}u^2+\left(g_{*23}^c+g_{*23}^{\ln{}}\ln u\right)u^3 \nn \\
&+\left(g_{*24}^c+g_{*24}^{\ln{}}\ln u\right)u^4+\ldots\nonumber\\
g_{S*}^{\rm 1SF 4}(u)=& g_{*40}+ g_{*41}u+g_{*42}u^2+\left(g_{*43}^c+g_{*43}^{\ln{}}\ln u\right)u^3 \nn \\
&+\left(g_{*44}^c+g_{*44}^{\ln{}}\ln u\right)u^4+\ldots\,.
\end{align}

The first step towards the determination of the coefficients in Eq. \eqref{coeff_undet} is to relate the circular limit of Eq. \eqref{gsstar_param} to  the previous  circular result, Eq. \eqref{g_star_circ_n}.
Indeed, the circular limit of \eqref{gsstar_param} reads
\begin{align}
\label{lim_circ_2}
g_{S*}(u,p_r,p_\phi)\bigg|_{\rm circ} =& g_{S*}^{ (\nu^0)}(u,0,p_\phi^{\rm (circ)}(u,\nu))+\nu \left[g_{S*}^{\rm 1SF 0}(u) \right] \nn \\
&+\mathcal{O}(\nu^2)\,.
\end{align}
where the expression of the {\it $\nu$-dependent} value of the square of $p_\phi^{\rm (circ)}(u,\nu)$, which is well known from EOB theory \cite{Buonanno:1998gg,Damour:2009sm},
reads
\begin{eqnarray}
[p_\phi^{\rm (circ)}(u,\nu)]^2&=& -\frac{\partial_u A(u; \nu)}{\partial_u (u^2A(u; \nu))} \nn \\
&=&\frac{1}{u(1-3u)}\left[ 1-\nu \frac{2a(u)+(1-2u)a'(u)}{2(1-3u)} \right] \nn \\
&&+\mathcal{O}(\nu^2)\,.
\end{eqnarray}
Here $A(u; \nu)=1-2u+\nu a(u)+\mathcal{O}(\nu^2)$ is the $\nu$-expansion of the main radial EOB potential $A(u;\nu)=-g_{00}^{\rm (eff)}$, see Eq. \eqref{eff_met}.
This yields 
\beq
p_\phi^{\rm (circ)}(u,\nu)=j_{\rm (circ)}(u)+\nu \delta j(u)+\mathcal{O}(\nu^2)\,,
\eeq
with
\beq
j_{\rm (circ)}(u)=\frac{1}{\sqrt{u(1-3u)}}\,, \delta j(u)=-  \frac{2a(u)+(1-2u)a'(u)}{4\sqrt{u}(1-3u)^{3/2}}\,.
\eeq
Inserting this result in  Eq. \eqref{lim_circ_2}, we see that the first term on the rhs contributes an additional $\mathcal{O}(\nu)$ contribution, namely
\begin{align}
g_{S*}^{ (\nu^0)}&(u,0,p_\phi^{\rm (circ)}(u,\nu))= \nn \\ &\frac{3}{1+\frac{1}{\sqrt{1-3u}}}
+\nu \frac{\partial  g_{S*}^{ (\nu^0)}(u,0,p_\phi)}{\partial p_\phi}\bigg|_{p_\phi=j_{\rm (circ)}} \delta j(u) +\mathcal{O}(\nu^2).
\end{align}
This implies the following link
\beq
g_{S*}^{{\rm circ }\, (\nu^1) }(u)=\frac{\partial  g_{S*}^{ (\nu^0)}(u,0,p_\phi)}{\partial p_\phi} \bigg|_{p_\phi=j_{\rm (circ)}} \delta j(u) +  g_{S*}^{\rm 1SF 0}(u)\,,
\eeq
which determines the value of $g_{S*}^{\rm 1SF 0}(u)$ from the previously known results on $g_{S*}^{{\rm circ }\, (\nu^1) }(u)$, given to 4PN fractional accuracy in Eq. \eqref{gsstar_circ_expl3} and to 9.5PN accuracy in Eq. (A1) of Ref. \cite{Bini:2015mza}.
Let us cite explicitly here only the first coefficients in the PN expansion \eqref{coeff_undet} of  $g_{S*}^{\rm 1SF 0}(u)$
\begin{eqnarray}
g_{*01}&=& -\frac34 \nonumber\\
g_{*02}&=& -\frac{39}{4} \nonumber\\
g_{*03}&=& \frac{41}{32}\pi^2-\frac{7987}{192}\nonumber\\
g_{*04}^c&=&  -\frac{11447}{120}-\frac{1456}{15}\ln(2)-48\gamma+\frac{26943}{2048}\pi^2\nonumber\\
g_{*04}^{\ln{}}&=&  -24 \,,
\end{eqnarray}
leading to
\begin{eqnarray}
g_{S*}^{\rm 1SF 0}(u)&=&  
-\frac34 u -\frac{39}{4}u^2+\left(\frac{41}{32}\pi^2-\frac{7987}{192}\right) u^3 \nonumber\\
&&  +\bigg(-24\ln(u)-\frac{11447}{120}-\frac{1456}{15}\ln(2)-48\gamma \nn \\
&&+\frac{26943}{2048}\pi^2\bigg) u^4 +\mathcal{O}(u^5)\,.
\end{eqnarray}

In a second step, we can determine the coefficients $g_{*20},\ldots ,g_{*24}^{\ln{}}$ in the PN expansion, Eq.\eqref{coeff_undet}, of the $\mathcal{O}(p_r^2)$  contribution to $g_{S*}(u,p_r,p_\phi)$, Eq. \eqref{gsstar_param}. The technicalities of this determination will be explained in the next subsection. 
For clarity, let us quote in advance the result we shall obtain
\begin{eqnarray}
g_{*20}&=& -\frac94 \nonumber\\
g_{*21}&=&-\frac94 \nonumber\\
g_{*22}^c&=& -\frac{717}{32} \nonumber\\
g_{*22}^{\ln{}}&=& 0\nonumber\\
g_{*23}^c&=& \frac{1447441}{960}-\frac{4829}{256}\pi^2-\frac{16038}{5}\ln(3)+\frac{46976}{15}\ln(2) \nn \\
&&-\frac{512}{5}\gamma\nonumber\\
g_{*23}^{\ln{}}&=&-\frac{256}{5} \nonumber\\
g_{*24}^c&=&-\frac{185195453}{38400}+\frac{19162}{35}\gamma+\frac{2097479}{8192}\pi^2 \nn \\
&&+\frac{454167}{20}\ln(3)-\frac{1081966}{35}\ln(2) \nonumber\\
g_{*24}^{\ln{}}&=&+\frac{9581}{35} \,.
\end{eqnarray}
Note that only the values of  the first two coefficients were known before, see Eq. \eqref{gssstar_PN}. The results for  $g_{*22}, g_{*23}^c, g_{*23}^{\ln{}}, g_{*24}^c$ and  $g_{*24}^{\ln{}}$ are new with this work. At the level $\mathcal{O}(\nu,p_r^4)$, instead, the only known coefficient \cite{Nagar:2011fx,Barausse:2011ys} is 
\beq
g_{*40}=\frac52\,,
\eeq
as shown in Eqs. \eqref{gssstar_PN}.

\subsection{EOB computation of $\Omega_r$ and  $\Omega_\phi$ as functions of energy and angular momentum}
We have seen above how SF theory led to a determination of the functional link between the spin precession quantity $\psi$
 and the two gauge-invariant frequencies of 
the orbital motion, $\Omega_r$  and  $\Omega_\phi$:
\begin{align}
\psi&(Gm_2 \Omega_r,  G m_2\Omega_\phi; q)=\nn \\
&\psi_0(G m_2\Omega_r, G m_2 \Omega_\phi )+q\Delta\psi(G m_2\Omega_r, G m_2\Omega_\phi)+\mathcal{O}(q^2)\,,
\end{align}
where $q=m_1/m_2=\nu+\mathcal{O}(\nu^2)$.
In order to relate the SF result, Eq. \eqref{Eq:DeltaPsiPN}, on $\Delta\psi(G m_2\Omega_r,  G m_2\Omega_\phi)$ to the PN expansion of the EOB gyrogravitomagnetic ratio $g_{S*}(u,p_r,p_\phi; \nu)$, Eq.  
\eqref{gsstar_param}, we need, as a first task, to compute the functional link predicted by EOB theory between $\Omega_r$  and  $\Omega_\phi$ and the (gauge-invariant) total energy $E_{\rm tot}$ and orbital  angular momentum $L=P_\phi$ of the corresponding motion of the binary system.

Having in mind the link, Eq. \eqref{H_Heff}, between $E_{\rm tot}=H$ and the effective energy ${\mathcal E}_{\rm eff}=H_{\rm eff}$ (or, equivalently, 
$\hat {\mathcal E}_{\rm eff}={\mathcal E}_{\rm eff}/(\mu c^2)$ and $\hat H_{\rm eff}=H_{\rm eff}/(\mu c^2)$), together with the definition \eqref{rescal} of the rescaled angular momentum $j$,  our first task will be to compute $\Omega_r$ and  $\Omega_\phi$ as functions  of $\hat {\mathcal E}_{\rm eff}$ and $j$.

The two frequencies we are interested in can be written as
\beq
\Omega_r=\frac{2\pi}{{\mathcal T}}\,,\qquad {\rm with}\qquad {\mathcal T}=\oint dT\,,
\eeq
and
\beq
\Omega_\phi=\frac{\Phi}{{\mathcal T}}=\frac{\oint d\phi}{\oint dT}=\frac{\oint \dot \phi\, dT}{{\mathcal T}}\,.
\eeq
Here $T$ denotes the physical\footnote{$T$ is the standard Schwarzschild-like coordinate time, as observed at infinity.} time (to be distinguished from the effective time $T_{\rm eff}$ entering Eq. \eqref{eff_met}), while $\oint$ denotes a periapsis-to-periapsis integral.

Using the rescaled quantities \eqref{rescal} and \eqref{rescal2}, together with the dimensionless, rescaled physical time 
\beq
t=\frac{c^3T}{GM}\,,
\eeq
the above frequencies become
\beq
GM \Omega_r =\frac{2\pi}{\oint dt}\,,\qquad GM\Omega_\phi=\frac{\oint d\phi}{\oint dt}\,.
\eeq

The time integration in these integrals can be replaced by radial integration  using Hamilton's equations for the rescaled radial variable
\beq
\label{r_dot}
\frac{dr}{dt}=\frac{1}{\nu}\frac{\partial h}{\partial p_r}\,,
\eeq
together with Hamilton's equation for the azimuthal variable
\beq
\label{phi_dot}
\frac{d\phi}{dt}=\frac{1}{\nu}\frac{\partial h}{\partial j}\,.
\eeq
In order to turn Eq. \eqref{r_dot} into a relation of the type
$dt=f(r)dr$ we need the explicit expression of the (rescaled) radial momentum $p_r$ as a function of $r$.
The latter relation is obtained by writing the law of conservation of energy
\begin{widetext}
\begin{align}
\label{energy_cons}
\hat {\mathcal E}_{\rm eff}&=\hat H_{\rm eff}(u,p_r,j; S,S_*) =\sqrt{A(u) \left( 1+j^2u^2+ A(u)\bar D(u) p_r^2+\hat Q\right)}+\frac{ju^3}{M^2} (g_S S+g_{S*}S_*)
\end{align}
\end{widetext}
where we replaced the metric potential $B=g_{RR}^{\rm eff}$, Eq. \eqref{eff_met}, by $\bar D\equiv (AB)^{-1}$.

As our aim here is to compute the coupling coefficients $g_S$, $g_{S*}$ parametrizing effects linear in spins, it is easily seen that it is enough to compute $\Omega_r$ and $\Omega_\phi$ to {\it zeroth} order in spins. In other words, we can neglect the spin-dependent terms in the energy conservation law \eqref{energy_cons} and work with the 
simplified mass-shell condition
\beq
\hat {\mathcal E}_{\rm eff}^2=A(u) \left( 1+j^2u^2+ A(u)\bar D(u) p_r^2+\hat Q\right)+\mathcal{O}({\rm spin})\,.
\eeq

The SF expansions of the EOB potential $A$, $\bar D$ and $\hat Q=Q/\mu^2$ read
\begin{align}
\label{SFEOB_funct}
A(u)&= 1-2u+\nu a(u) +\mathcal{O}(\nu^2) \,, \nonumber\\
\bar D(u) &= 1+\nu \bar d(u)  +\mathcal{O}(\nu^2) \,, \nonumber\\
\hat Q(u) &= \nu q_4 (u) p_r^4 + \nu q_6 (u) p_r^6+\nu q_8 (u) p_r^8+ \mathcal{O}(\nu^2,p_r^{10})\,. 
\end{align}
The PN expansions of the first SF-order  radial functions $a(u)$, $\bar d(u)$, $q_4(u)$, $\ldots$ entering the latter equations have been determined by SF theory to very high PN-orders: see Refs. \cite{Blanchet:2010zd,Barack:2010ny,Bini:2013zaa,Bini:2013rfa,Bini:2014nfa,Bini:2015bla,Kavanagh:2015lva} for $a(u)$, Refs. \cite{Bini:2015bfb} for $\bar d(u)$ and Refs. \cite{Bini:2015bfb,Hopper:2015icj,Bini:2016qtx} for  $q_4(u)$, $q_6(u)$ etc...

For concreteness, let us quote here the beginning of these expansions

\begin{widetext}
\begin{eqnarray}
\label{eq_39}
A(u)&=& 1-2 u+2\nu u^3+\left(\frac{94}{3}-\frac{41}{32}\pi^2\right)\nu u^4\nonumber\\
&+& \left[\left(\frac{2275}{512}\pi^2-\frac{4237}{60}+\frac{128}{5}\gamma+\frac{256}{5}\ln(2)\right)\nu+\left(\frac{41}{32}\pi^2-\frac{221}{6}\right)\nu^2+\frac{64}{5}\nu\ln(u)\right] u^5 +\mathcal{O}(u^6)\nonumber\\
\bar D(u)&=& 1+6\nu u^2 +(52\nu-6\nu^2)u^3 +\left[ \left(-\frac{533}{45}-\frac{23761}{1536}\pi^2 +\frac{1184}{15}\gamma-\frac{6496}{15}\ln 2 +\frac{2916}{5}\ln 3 \right)\nu \right. \nonumber\\
&+& \left. \left( \frac{123}{16}\pi^2 -260\right)\nu^2 +\frac{592}{15}\nu \ln(u) \right]u^4+\mathcal{O}(u^5)\nonumber\\
\hat Q(u,p_r) &=& \left[ 2\left(4-3\nu \right) \nu u^2+\left(\left(-\frac{5308}{15}+\frac{496256}{45}\ln 2 -\frac{33048}{5}\ln 3\right)\nu-83 \nu^2 + 10\nu^3  \right)u^3 +\mathcal{O}(u^4)\right]p_r^4 \nonumber\\
&+& \left[ \left( -\frac{827}{3}-\frac{2358912}{25}\ln 2+\frac{1399437}{50}\ln 3 +\frac{390625}{18}\ln 5 \right) \nu -\frac{27}{5}\nu^2 +6 \nu^3\right]u^2 p_r^6 +\mathcal{O}(u^3,p_r^8)\,.
\end{eqnarray}
\end{widetext}

Inserting the SF expansion, Eqs. \eqref{SFEOB_funct}, into the mass-shell condition allows us to compute the functional dependence of  $p_r$ 
on $u$, $\hat {\mathcal E}_{\rm eff}$ and $j$. To the first SF-order, i.e., 
\beq
\label{pr_exp}
p_r(u,\hat {\mathcal E}_{\rm eff},j; S,S_*)=p_r^{(0)}(u,\hat {\mathcal E}_{\rm eff},j)+\nu p_r^{(1)}(u,\hat {\mathcal E}_{\rm eff},j)+\mathcal{O}(\nu^2,{\rm spin}) 
\eeq
the explicit expressions of $p_r^{(0)}$ and $p_r^{(1)}$ read (when neglecting spins)
\begin{widetext}
\begin{align}
\label{pr_expressions}
p_r^{(0)}(u,&\hat {\mathcal E}_{\rm eff},j)= \frac{\sqrt{\hat {\mathcal E}_{\rm eff}^2-(1-2u)(1+u^2j^2)}}{1-2u}\nonumber\\
p_r^{(1)}(u,&\hat {\mathcal E}_{\rm eff},j)= -\frac{\hat {\mathcal E}_{\rm eff}^2}{p_r^{(0)}}\frac{a(u)}{(1-2u)^3} -p_r^{(0)} \left[\bar d(u)+\frac{a(u)}{(1-2u)}+(p_r^{(0)})^2 \frac{q_4(u)+q_6(u)(p_r^{(0)})^2+ q_8(u)(p_r^{(0)})^4+\ldots}{1-2u}\right]\,.
\end{align}
\end{widetext}
We can then obtain an explicit expression for replacing time integration by radial integration 
by inserting Eqs. \eqref{pr_exp} and \eqref{pr_expressions} in the relation
\begin{align}
\label{dt_dr}
dt
&=  \frac{\sqrt{1+2\nu(\hat {\mathcal E}_{\rm eff}-1)} \, \hat {\mathcal E}_{\rm eff} }{Ap_r(A\bar D+2p_r^2 \nu q_4 +3p_r^4 \nu q_6+4 p_r^6 \nu q_8+\ldots)}\, dr\,,
\end{align}
which follows from the radial equation of motion \eqref{r_dot}.
Inserting Eq. \eqref{dt_dr} in the $t$-integral expressions for $\Omega_r$ and $\Omega_\phi$, Eqs. \eqref{r_dot} and \eqref{phi_dot}, and formally expanding $p_r$ in powers of $\nu$ according to Eqs. \eqref{pr_exp} and \eqref{pr_expressions}, leads to $r$-integral expressions for the frequencies of the type
\beq
\label{end_points}
GM \Omega_{r, \phi}\sim \sum_n \int_{u_{\rm (min)}}^{u_{\rm (max)}} du  \frac{f_n(u)}{\left(\hat {\mathcal E}_{\rm eff}^2-(1-2u)(1+u^2j^2)\right)^{n+\frac12}}\,
\eeq
where $f_n(u)$ involves a combination of $(1-2u)^k$, $a(u)$, $\bar d(u)$, etc.
Note that these radial integrals are divergent for $n\ge 1$.
The appearance of singular integral is due to our formal replacement of the expansion  \eqref{pr_exp} in originally convergent integrals of the type
$\oint  dr  f(r)/p_r$. As shown in Ref. \cite{Damour:1988mr}, the correct result for these expansions is obtained simply by taking Hadamard's partie finie (Pf) of the singular  integrals, Eq. \eqref{end_points}.

We are interested here in computing the PN-expansions of the frequencies. 
These PN-expansions can be conveniently obtained: i) by replacing the various first SF-order  EOB potentials $a(u)$, $\bar d(u)$, etc., by their PN expansion illustrated in Eq. \eqref{eq_39}; and,   using again the general result of Ref. \cite{Damour:1988mr}, ii) by formally expanding the radicals entering the denominators above, namely
\beq
\hat {\mathcal E}_{\rm eff}^2-(1-2u)(1+u^2j^2)=\hat {\mathcal E}_{\rm eff}^2-1 +2u -j^2u^2+2j^2u^3\,,
\eeq
around the Newtonian-like, quadratic radical
\beq
\label{R_0_Def}
{\mathcal R}_0(u)=\hat {\mathcal E}_{\rm eff}^2-1 +2u -j^2u^2\,.
\eeq
In other words, the combination of the $\nu$-expansion and the PN-expansion leads to integral expressions for the frequencies of the type
\beq
\label{Newt_like_int}
GM \Omega_{r, \phi}\sim \sum_n {\rm Pf}\,\int_{u^{\rm (min)}_0}^{u^{\rm (max)}_0} du  \frac{\tilde f_n(u)}{{\mathcal R}_0^{n+\frac12}}(u)\,.
\eeq
Here the symbol Pf denotes Hadamard's partie finie. Note that the values of the end points of the $u$-integrations are different from the ones in Eq. \eqref{end_points} above and now denote the two roots of the Newtonian-like quadratic radical ${\mathcal R}_0(u)$, Eq. \eqref{R_0_Def}.
Finally, we are left with evaluating Newtonian-like  radial integrals of the type \eqref{Newt_like_int}.
When evaluating the partie finie of these singular integrals it is convenient to replace $\hat {\mathcal E}_{\rm eff}$ and $j$ by 
the quantities\footnote{As said above, the so-defined EOB variables $p\equiv 1/u_p$ and $e$ should be distinguished from the
SF-defined variables $\bar{p}$ and $\bar{e}$ used as independent variables in Eq. \eqref{Eq:DeltaPsiPN} above.} $u_p$ and $e$ defined
so that the two roots of ${\mathcal R}_0(u)$ are $u_p(1\pm e)$, namely
\beq
\label{ec_and_up}
\hat {\mathcal E}_{\rm eff} ^2-1=-u_p(1-e^2) \,,\qquad j^2=\frac{1}{u_p}\,.
\eeq
As we are interested in slightly eccentric \footnote{The parameter $e$ defined in Eq. \eqref{ec_and_up} does not measure the exact eccentricity of the corresponding EOB orbits that would vanish when $p_r$ vanishes. Nevertheless, when used in PN-expansion, it will allow us to correctly evaluate the gauge-invariant quantities we are interested in.} motions, we can further expand the latter (singular) Newtonian-like integrals in powers of $e$.

When doing so, we use the following Newtonian-like parametrization of the inverse radius $u=1/r$ 
\beq
\label{u_pdef}
u=u_p (1+e\cos \chi_0)\,,
\eeq
so that
\beq
du=-u_p e \sin \chi_0 d\chi_0\,,\quad {\mathcal R}_0(u)=u_p e^2 \sin^2 \chi_0\,.
\eeq
We have then shown that the $e$-expansion of the partie finie of the integrals \eqref{Newt_like_int}
is correctly obtained by taking the $\epsilon^0$ term in the Laurent expansion in $\epsilon$ of the  integrals of the type
\beq
\int_{\epsilon}^{\pi-\epsilon}  \frac{g_n(\chi_0)}{\sin^{2n} \chi_0} d\chi_0  \,,
\eeq
that are generated by the expansions \eqref{Newt_like_int}.

Finally, the combined PN-, SF- and eccentricity-expansions of the frequencies yield 
\beq
\label{eq_ecc_Omega_r}
GM\Omega_r = \Omega_r^{(0)}(u_p,\nu)+e^2  \Omega_r^{(2)}(u_p,\nu)+ e^4  \Omega_r^{(4)}(u_p,\nu)+\mathcal{O}(e^6)\,,
\eeq
with
\begin{widetext}
\begin{eqnarray}
\label{exp_om_r}
\Omega_r^{(0)}(u_,\nu)&=& u_p^{3/2}+\left(\frac12 \nu-\frac32\right) u_p^{5/2} +\left(\frac{19}{8}\nu-\frac{33}{8}\right)u_p^{7/2} +\left[-\frac{523}{16}+\left(-\frac{41}{64}\pi^2+\frac{2591}{48}\right)\nu  \right] u_p^{9/2} \nonumber\\
&+& \left[-\frac{40389}{128}+\left(\frac{692771}{1152}-\frac{26735}{1536}\pi^2+\frac{1168}{15}\gamma+\frac{584}{15}\ln(u_p)+\frac{1458}{5}\ln(3)-\frac{2096}{15}\ln(2)\right)\nu\right] u_p^{11/2} \nonumber\\
&+&\left[-\frac{849573}{256}+\left(\frac{50524}{105}\gamma-\frac{611631}{70}\ln(3)+\frac{13140979}{2304}+\frac{25262}{105}\ln(u_p)+\frac{255236}{15}\ln(2) +\frac{481}{96}\pi^2\right)\nu\right] u_p^{13/2} \nonumber\\
&+& \frac{11770}{63}u_p^7\pi\nu +\mathcal{O}(u^{15/2},\nu^2)\nonumber\\
\Omega_r^{(2)}(u_p,\nu)&=& -\frac32 u_p^{3/2}+\left(-\frac{5}{4}\nu+\frac{15}{4}\right) u_p^{5/2} +\left(-\frac{109}{16}\nu+\frac{171}{16}\right) u_p^{7/2}+\left[\frac{2607}{32}+\left(\frac{123}{64}\pi^2-\frac{4609}{32}\right)\nu\right] u_p^{9/2} \nonumber\\
&+& \left[\frac{186639}{256}+\left(-\frac{3208}{15}\gamma-\frac{1604}{15}\ln(u_p)-\frac{28431}{5}\ln(3)+\frac{128728}{15}\ln(2)+\frac{273443}{6144}\pi^2-\frac{18601901}{11520}  \right)\nu\right] u_p^{11/2} \nonumber\\
&+& \left[\frac{3564729}{512}+\left(\frac{26578611}{224}\ln(3)-\frac{1657307}{6144}\pi^2-\frac{92786}{105}\gamma-\frac{2905678}{7}\ln(2)-\frac{46393}{105}\ln(u_p)\right.\right. \nonumber\\
&& \left.\left.+\frac{21484375}{224}\ln(5)-\frac{8195349553}{806400}\right)\nu\right] u_p^{13/2} -\frac{92341}{315}u_p^7\pi\nu+\mathcal{O}(u_p^{15/2},\nu^2)\,\nonumber\\
\Omega_r^{(4)}(u_p,\nu)&=& \frac38 u_p^{3/2}+\left(-\frac{45}{16}+\frac{15}{16}\nu  \right) u_p^{5/2} +\left(\frac{401}{64}\nu-\frac{495}{64}\right) u_p^{7/2}+\left[-\frac{6489}{128}+\left(-\frac{123}{64}\pi^2+\frac{13927}{128}\right)\nu\right]u_p^{9/2}\nonumber\\
&+&\left[-\frac{333831}{1024}+\left(\frac{5218813}{5120}+\frac{436}{5}\ln(u_p)-\frac{566832}{5}\ln(2)+\frac{872}{5}\gamma -\frac{59563}{2048}\pi^2+\frac{1953125}{96}\ln(5)\right.\right.\nonumber\\
&& \left.\left.+\frac{6647751}{160}\ln(3)\right)\nu\right] u_p^{11/2}\nonumber\\
&+&\left[-\frac{3194019}{2048}+\left(\frac{451817433}{4480}\ln(3)-\frac{30997}{420}\ln(u_p)-\frac{4017578125}{2688}\ln(5)+\frac{99386957}{30}\ln(2)\right.\right.\nonumber\\
&& \left.\left. -\frac{30997}{210}\gamma+\frac{18324577}{24576}\pi^2-\frac{24403514197}{3225600}\right)\nu\right] u_p^{13/2}-\frac{11660111}{53760} u_p^7\pi\nu +\mathcal{O}(u^{15/2},\nu^2) \,.
\end{eqnarray}
Note that the dependence on $\ln u_p$ starts at $\mathcal{O}(u_p^{11/2})$.
Similarly, the azimuthal frequency reads
\beq
GM \Omega_\phi = \Omega_\phi^{(0)}(u_p,\nu)+e^2  \Omega_\phi^{(2)}(u_p,\nu)+ e^4  \Omega_\phi^{(4)}(u_p,\nu)+\mathcal{O}(e^6)\,,
\eeq
with, for example,  
\begin{eqnarray}
\label{exp_om_phi}
\Omega_\phi^{(0)}(u_p,\nu)&=&  u_p^{3/2}+\left(\frac32+\frac12 \nu\right)u_p^{5/2} 
+\left(-\frac{17}{8}\nu+\frac{111}{8}\right) u_p^{7/2} 
+\left[\frac{2099}{16}+\left(-\frac{5935}{48}+\frac{205}{64}\pi^2\right)\nu\right] u_p^{9/2} \nonumber\\
&+&\left[\frac{172059}{128}+\left(-\frac{3557387}{1920}+\frac{38195}{1024}\pi^2-\frac{448}{5}\gamma-\frac{224}{5}\ln(u_p)-\frac{896}{5}\ln(2)\right)\nu \right] u_p^{11/2}\nonumber\\
&+&\left[\frac{3720501}{256}+\left(-\frac{33940}{21}\gamma-\frac{174231}{70}\ln(3)-\frac{1332163499}{57600}-\frac{16970}{21}\ln(u_p)-\frac{74468}{105}\ln(2)\right.\right.\nonumber\\
&& \left.\left. +\frac{291883}{1536}\pi^2\right)\nu\right] u_p^{13/2}-\frac{13696}{105} u_p^7\pi \nu +\mathcal{O}(u_p^{15/2},\nu^2)\nonumber\\
\Omega_\phi^{(2)}(u_p,\nu)&=&  -\frac32  u_p^{3/2}+\left(-\frac34-\frac54\nu  \right)u_p^{5/2)}+\left(-\frac{129}{16}-\frac{49}{16}\nu\right) u_p^{7/2}+\left[-\frac{1407}{32}+\left(-\frac{369}{128}\pi^2-\frac{303}{32}\right)\nu\right] u_p^{9/2}\nonumber\\
&+&\left[-\frac{20793}{256}+\left(-\frac{2656}{15}\gamma-\frac{1328}{15}\ln(u_p)-\frac{5832}{5}\ln(3)+\frac{12416}{15}\ln(2)+\frac{148949}{6144}\pi^2-\frac{6765029}{11520}\right)\nu\right] u_p^{11/2}\nonumber\\
&+&\left[\frac{1631139}{512}+\left(\frac{788049}{20}\ln(3)-\frac{15875}{1536}\pi^2-\frac{100046}{105}\gamma-\frac{1566262}{21}\ln(2)-\frac{50023}{105}\ln(u_p)\right.\right. \nonumber\\
&-&\left.\left.\frac{7980766891}{806400}\right)\nu\right] u_p^{13/2}-\frac{262043}{315}\pi \nu u_p^7 +\mathcal{O}(u_p^{15/2},\nu^2)\nonumber\\
\Omega_\phi^{(4)}(u_p,\nu)&=& \frac38 u_p^{3/2}+\left(-\frac{27}{16}+\frac{15}{16}\nu\right) u_p^{5/2}+\left(\frac{581}{64}\nu-\frac{855}{64}\right) u_p^{7/2}+\left[-\frac{20799}{128}+\left(-\frac{123}{64}\pi^2+\frac{30193}{128}\right)\nu\right] u_p^{9/2}\nonumber\\
&+&\left[-\frac{2037927}{1024}+\left(\frac{4235449}{1024}+\frac{1046}{5}\ln(u_p)-19324\ln(2)+\frac{2092}{5}\gamma-\frac{395017}{4096}\pi^2+\frac{63423}{5}\ln(3)\right)\nu\right] u_p^{11/2}\nonumber\\
&+&\left[-\frac{49225221}{2048}+\left(-\frac{1153657809}{4480}\ln(3)+\frac{655871}{420}\ln(u_p)-\frac{654296875}{2688}\ln(5)+\frac{29552521}{30}\ln(2)\right.\right. \nonumber\\
&& \left.\left. +\frac{655871}{210}\gamma+\frac{4564433}{12288}\pi^2+\frac{130198554749}{3225600}\right)\nu\right] u_p^{13/2)}+\frac{21293}{28} u_p^7\pi\nu +\mathcal{O}(u_p^{15/2},\nu^2)\,.
\end{eqnarray}
\end{widetext}
Note again that the dependence on $\ln u_p$ starts at $\mathcal{O}(u_p^{11/2})$. 

In the expansions  \eqref{exp_om_r} and \eqref{exp_om_phi} above we have only displayed the beginning of the PN expansions of $\Omega_r^{(0,2,4)}(u_p,\nu)$ and $\Omega_\phi^{(0,2,4)}(u_p,\nu)$. Actually we have computed them to the highest PN order currently known from SF calculations. Similarly, we have also computed the eccentricity expansions \eqref{eq_ecc_Omega_r} up to $\mathcal{O}(e^{10})$.

It will be convenient in the following to trade the two dimensionless frequencies $GM \Omega_r$ and $GM \Omega_\phi$ by two other related dimensionless (equally gauge-invariant) parameters, namely
the (fractional) periastron advance per orbit
\beq
k\equiv \frac{\Omega_\phi}{\Omega_r}-1=\frac{\Phi}{2\pi}-1\,,
\eeq
and the dimensionless azimuthal frequency variable
\beq
y\equiv (G m_2 \Omega_\phi )^{2/3}=\left(1-\frac23 \nu+\mathcal{O}(\nu^2)\right) (GM \Omega_\phi )^{2/3}\,.
\eeq
The combined eccentricity-, PN- and SF-expansions of these quantities read
\beq
k(u_p,e;\nu)= k^{(0)}(u_p,\nu)+e^2 k^{(2)}(u_p,\nu)+ e^4 k^{(4)}(u_p,\nu)+\mathcal{O}(e^6)\,,
\eeq
with
\begin{widetext}
\begin{eqnarray}
k^{(0)}(u_p,\nu)&=& 3 u_p +\left(\frac{45}{2}-6\nu\right) u_p^2 +\left[210+\left(-205+\frac{123}{32}\pi^2\right)\nu\right] u_p^3\nonumber\\ 
&+&
\left[\frac{17325}{8}+\left(-\frac{2512}{15}\gamma+\frac{191671}{3072}\pi^2-\frac{592}{15}\ln(2)-\frac{1458}{5}\ln(3)-\frac{1256}{15}\ln(u_p)-\frac{559223}{180}\right)\nu\right] u_p^4\nonumber\\
&+&\left[\frac{189189}{8}+\left(-\frac{364664}{21}\ln(2)+\frac{739519}{2048}\pi^2-\frac{45188}{35}\ln(u_p)+\frac{172773}{35}\ln(3)-\frac{90376}{35}\gamma-\frac{7826703}{200} \right)\nu\right] u_p^5 \nonumber\\
&-& \frac{99938}{315}\pi\nu u_p^{11/2} +\mathcal{O}(u_p^6,\nu^2)\nonumber\\
\nonumber\\ 
k^{(2)}(u_p,\nu)&=&\left(-\frac32 \nu+\frac{15}{4}\right) u_p^2 +\left[\frac{315}{4}+\left(-106+\frac{123}{128}\pi^2\right)\nu\right]u_p^3\nonumber\\
&+&
\left[\frac{10395}{8}+\left(-\frac{519697}{240}-\frac{23440}{3}\ln(2)-\frac{1072}{5}\gamma +\frac{111797}{2048}\pi^2+\frac{20412}{5}\ln(3)-\frac{536}{5}\ln(u_p)\right)\nu\right] u_p^4 \nonumber\\
&+& \left[\frac{315315}{16}+\left(\frac{1791937}{4096}\pi^2-\frac{7486}{5}\ln(u_p) -\frac{14661947}{400}+\frac{29100532}{105}\ln(2)-\frac{14972}{5}\gamma-\frac{7611489}{160}\ln(3)\right.\right.\nonumber\\
&& \left.\left. -\frac{21484375}{224}\ln(5)\right) \nu\right] u_p^5\nonumber\\
& -& \frac{319609}{315}\pi u_p^{11/2}\nu  +\mathcal{O}(u_p^6,\nu^2)\nonumber\\
k^{(4)}(u_p,\nu)&=& 
-\frac32 u_p^3\nu\nonumber\\
&+& \left[\frac{3465}{64}+\left(\frac{411686}{5}\ln(2)-\frac{3620943}{160}\ln(3)-\frac{37}{5}\ln(u_p)-\frac{74}{5}\gamma+\frac{13529}{48}-\frac{1953125}{96}\ln(5)+\frac{33601}{8192}\pi^2\right)\nu\right]u_p^4\nonumber\\
& +& \left[\frac{135135}{64}+\left(\frac{455078125}{448}\ln(5)-\frac{16271}{70}\ln(u_p)-\frac{16271}{35}\gamma -\frac{1358100027}{2240}\ln(3)+\frac{4363277}{16384}\pi^2\right.\right.\nonumber\\
&& \left.\left. -\frac{148342613}{105}\ln(2)+\frac{10223249}{16800}\right)\nu\right] u_p^5\nonumber\\
& -& \frac{1089581}{2560}\pi \nu u_p^{11/2}  +\mathcal{O}(u_p^6,\nu^2)\,.
\end{eqnarray}
\end{widetext}
In the following  it will be convenient to work with the rescaled periastron advance per orbit $\hat k$, defined by  
\beq
\hat k \equiv \frac{k}{3}\,.
\eeq

Similarly 
\beq
y(u_p,e;\nu)= y^{(0)}(u_p,\nu)+e^2 y^{(2)}(u_p,\nu)+ e^4 y^{(4)}(u_p,\nu)+\mathcal{O}(e^6)\,,
\eeq
with   
\begin{widetext}
\begin{eqnarray}
y^{(0)}(u_p,\nu)&=&\left(1-\frac23 \nu\right) u_p+\left(-\frac13 \nu+1\right) u_p^2 +\left(9 -\frac{91}{12}\nu  \right)u_p^3 +\left[83+\left(-\frac{9967}{72}+\frac{205}{96}\pi^2\right)\nu  \right] u_p^4\nonumber\\
&+& \left[\frac{1671}{2}+\left(-\frac{1792}{15}\ln(2)+\frac{12185}{512}\pi^2-\frac{896}{15}\gamma-\frac{1686559}{960}-\frac{448}{15}\ln(u_p)\right)\nu  \right] u_p^5\nonumber\\
&+& \left[\frac{17835}{2}+\left(-\frac{164996}{315}\ln(u_p)+\frac{971767}{9216}\pi^2-\frac{26024}{63}\ln(2)-\frac{329992}{315}\gamma-\frac{58077}{35}\ln(3)\right.\right. \nonumber\\
&-& \left.\left. \frac{1768379519}{86400}\right)\nu\right]u_p^6-\frac{27392}{315}\pi\nu  u_p^{13/2}
+\mathcal{O}(u_p^6,\nu^2) \nonumber\\
y^{(2)}(u_p,\nu)&=&  
\left(\frac23 \nu-1\right) u_p-\frac23  u_p^2\nu +\left(-1-\frac{23}{12}\nu\right) u_p^3 +\left[\frac{21}{2}+\left(-\frac{41}{48}\pi^2-\frac{457}{9}\right)\nu\right] u_p^4\nonumber\\
& +& \left[\frac{667}{2}+\left(-\frac{1967801}{1728}+\frac{128815}{4608}\pi^2-\frac{3328}{45}\ln(u_p)+\frac{22144}{45}\ln(2)-\frac{6656}{45}\gamma-\frac{3888}{5}\ln(3) \right)\nu\right] u_p^5\nonumber\\
&+& \left[ 6043+\left(-\frac{4996270991}{302400}+\frac{180792}{7}\ln(3)-\frac{168544}{315}\ln(u_p)-\frac{15795776}{315}\ln(2)-\frac{337088}{315}\gamma \right.\right. \nonumber\\
&+& \left.\left. \frac{121213}{4608}\pi^2\right)\nu\right] u_p^6-\frac{565174}{945}\pi\nu u_p^{13/2} 
 +\mathcal{O}(u_p^6,\nu^2)\,\nonumber\\
y^{(4)}(u_p,\nu)&=&  (-1+\nu) u_p^2 +\left(-\frac{63}{8}+\frac{119}{12}\nu\right) u_p^3 +\left[-\frac{343}{4}+\left(\frac{1443}{8}-\frac{369}{256}\pi^2\right) \nu\right] u_p^4 \nonumber\\
&+& \left[-\frac{7687}{8}+\left(\frac{11677919}{4320}+\frac{8888}{45}\gamma+\frac{4444}{45}\ln(u_p)+\frac{40338}{5}\ln(3)-\frac{1686137}{36864}\pi^2-\frac{113864}{9}\ln(2)\right)\nu\right] u0^5 \nonumber\\
&+& \left[-\frac{42707}{4}+\left(\frac{1620873859}{67200}-\frac{72938151}{448}\ln(3)+\frac{7624295}{24576}\pi^2-\frac{654296875}{4032}\ln(5)+\frac{200895602}{315}\ln(2)\right.\right. \nonumber\\
&& \left.\left. +\frac{70123}{105}\ln(u_p)+\frac{140246}{105}\gamma\right)\nu\right] u_p^6+\frac{372467}{1890}\pi\nu u_p^{13/2} +\mathcal{O}(u_p^6,\nu^2)\,.
\end{eqnarray}
\end{widetext}
Finally, we will need in the following to invert the functional link between $(u_p,e)$ and $(\hat k,y)$, i.e., to compute the functions
\beq
u_p=f_{u_p}(\hat k,y) \,,\qquad e^2=f_{e^2}(\hat k,y)\,.
\eeq
This inversion requires some care, because the Jacobian $\partial (\hat k,y)/\partial(u_p, e^2)$ is of order $u_p$ near the origin of the $u_p$, $e^2$ plane.
If we provisionally introduce the quantity $\epsilon\equiv e^2 u_p$, the Jacobian $\partial (\hat k,y)/\partial(u_p, \epsilon)$ will be of order unity  near the origin of the $u_p$, $\epsilon$ plane. This shows that we can invert the link  $X^i=(u_p,\epsilon)\to  Y^i=(\hat k,y)$ by standard Taylor expansions of the symbolic type 
\beq
X^i=A^{i}{}_{j}Y^j+A^{i}{}_{jk}Y^jY^k+\ldots\,.
\eeq
When going back from the pair $(u_p,\epsilon\equiv e^2 u_p)$ to the original pair $(u_p,e^2)$ we obtain the following transformation
\begin{widetext}
\begin{eqnarray}
\label{up_and_e2_vs_hatk_and_y}
u_p&=& \hat k+   \left[\left(\frac{5}{2}+\frac13  \frac{y}{\hat k}\right)\nu-\frac{35}{4}+\frac54  \frac{y}{\hat k}\right] \hat k^2\nonumber\\
&& + \left[\left(\frac{41}{128}\pi^2 \frac{y}{\hat k}-\frac{151}{8} \frac{y}{\hat k}+\frac{25}{12}\left(\frac{y}{\hat k}\right)^2+\frac{50}{3}-\frac{205}{128}\pi^2\right)\nu-\frac{145}{16} \frac{y}{\hat k}+\frac{45}{16} \left(\frac{y}{\hat k}\right)^2+\frac{455}{8}\right]\hat k^3+\mathcal{O}(\hat k^4, y^3)\nonumber\\
e^2 &=& 1-\left(1+\frac23 \nu \right)\frac{y}{\hat k}+  \left[-\frac{27}{4} \frac{y}{\hat k}+\frac14  \left(\frac{y}{\hat k}\right)^2+\left(-2 \frac{y}{\hat k}+\frac16  \left(\frac{y}{\hat k}\right)^2\right)\nu\right] \hat k\nonumber\\
&& +\left[\left(-\frac{45}{8} \frac{y}{\hat k}+\frac78 \left(\frac{y}{\hat k}\right)^2+\frac{15}{16} \left(\frac{y}{\hat k}\right)^3-\frac{5}{16}\right) 
\right. \nonumber\\
&&\left.+\left(-\frac{205}{128}\pi^2 \frac{y}{\hat k}+\frac{41}{128} \pi^2 \left(\frac{y}{\hat k}\right)^2+\frac{1225}{24} \frac{y}{\hat k}-\frac{473}{24} \left(\frac{y}{\hat k}\right)^2+\frac94  \left(\frac{y}{\hat k}\right)^3+\frac18 \right) \nu\right]\hat k^2 +\mathcal{O}(k^3,y^4)\,.
\end{eqnarray}
\end{widetext}
Again we have only indicated, for concreteness, the beginning of these expansions.

\subsection{EOB computation of the spin precession frequency}
When considering, as we do here, spin couplings to linear order, i.e., an Hamiltonian of the form $H=H_{\rm orbital}+{\mathbf \Omega_{S_1}} \cdot {\mathbf S_1}+{\mathbf \Omega_{S_2}} \cdot {\mathbf S_2}$, Hamilton's equations of motion for the spins ($a=1,2$)
\beq
\frac{d{\mathbf S}_a}{dt}=\{{\mathbf S}_a,H\}
\eeq
yields
\beq
\frac{d{\mathbf S}_a}{dt}={\mathbf \Omega}_{S_a} \times {\mathbf S}_a\,,
\eeq
showing that ${\mathbf \Omega}_{S_a}=\frac{\partial H}{\partial {\mathbf S}_a}$ is the vectorial precession frequency of ${\mathbf S}_a$.
Restricting to the case of interest of parallel spins we conclude that the (algebraic) magnitude of the spin frequency of body 1 is given by

\beq
\Omega_{S_1}= \frac{\partial H}{\partial S_1}=\frac{M\nu}{h}\frac{\partial \hat H_{\rm eff}}{\partial S_1}= \left( \frac{1}{G M }\right)  \frac{\nu }{h} j u^3 \left(g_S  +g_{S*} \frac{m_2}{m_1}\right)\,.
\eeq
At first order in $\nu$, this reads
\begin{align}
GM& \Omega_{S_1}(u,p_r,j; \nu) \nn \\
&= j(1-\nu)[1-\nu (\hat {\mathcal E}_{\rm eff}-1)] u^3[\nu g_S+(1-\nu)g_{S*}]+\mathcal{O}(\nu^2) \nonumber\\
&= j(1-\nu  \hat {\mathcal E}_{\rm eff}) u^3 [\nu g_S+(1-\nu)g_{S*}]+\mathcal{O}(\nu^2)\,.
\end{align}
The averaged spin frequency is then given by
\beq
\langle  \Omega_{S_1}\rangle (\hat {\mathcal E}_{\rm (eff)},j; \nu)= \frac{1}{{\mathcal T}}\oint  \Omega_{S_1} dT
=
\frac{\Omega_r}{2\pi}\oint   \Omega_{S_1} dT \,.
\eeq
As indicated, the averaged spin frequency $\langle  \Omega_{S_1}\rangle$ is a function of the conserved dynamical quantities, $\hat {\mathcal E}_{\rm (eff)}$ and $j$.
Replacing as above the $T$-integration by a radial integration, say $dT=f(r)dr=\tilde f(u) du$, see Eq. \eqref{dt_dr}, we see that the computation of 
$\langle  \Omega_{S_1}\rangle$ amounts to computing radial integrals of the type
\beq
\oint du \tilde f(u) u^3 [\nu g_S+(1-\nu)g_{S*}(u,p_r,j; \nu)]\,.
\eeq
In the radial integral involving $g_S$ we can simply replace $g_S=2$ because of the $\nu$ prefactor. By contrast, in the radial integral involving $g_{S*}$ we need to insert the expression \eqref{gsstar_param} which involves  undetermined coefficients at the $p_r^2$ and $p_r^4$ levels, such as $g_{*22},g_{*23}^c$, etc.
This radial integral can be computed by the same technique we used above for computing $\Omega_r$ and $\Omega_\phi$. Replacing  as above $\hat {\mathcal E}_{\rm eff}$
and $j$ by the Newtonian-like quantities $u_p$ and $e^2$, defined by Eqs. \eqref{u_pdef}, we found

\begin{align}
GM &\langle  \Omega_{S_1}\rangle (u_p,e^2;\nu) =\nn \\
& \Omega_{S_1}^{(0)}(u_p,\nu)+e^2 \Omega_{S_1}^{(2)}(u_p,\nu)+ e^4  \Omega_{S_1}^{(4)}(u_p,\nu)+\mathcal{O}(e^6)\,,
\end{align}
with, for example,  
\begin{widetext}
\begin{eqnarray}
\Omega_{S_1}^{(0)}(u_p,\nu)&=& \left(\frac32 -\nu\right) u_p^{5/2}+\left(\frac{63}{8}-\frac{15}{4}\nu  \right) u_p^{7/2}+\left(-\frac{207}{4}\nu+\frac{1131}{16}\right) u_p^{9/2} \nonumber\\
&& +\left(-\frac{174979}{192}\nu+\frac{1271}{128}\nu\pi^2+\frac{90939}{128}  \right) u_p^{11/2} \nonumber\\
&& +\left(\frac{387023}{2048}\nu\pi^2-\frac{8368}{15}\nu\ln(2)-\frac{1392}{5}\nu\gamma-\frac{544259}{40}\nu+\nu g_{*22}+\frac{1946997}{256}-\frac{696}{5}\nu\ln(u_p)\right) u_p^{13/2}   \nonumber\\
&+& \mathcal{O}(u_p^7,\nu^2)\nonumber\\
\Omega_{S_1}^{(2)}(u_p,\nu)&=&   \left(-\frac94 +\frac32\nu  \right) u_p^{5/2}+\left(-\frac{105}{16}+\frac{15}{8}\nu\right) u_p^{7/2}+\left(\frac{135}{16}\nu-\frac{1167}{32}\right) u_p^{9/2}\nonumber\\
&& +\left( \frac12 \nu g_{*22} -\frac{38361}{256}-\frac{21319}{128}\nu-\frac{615}{256}\nu\pi^2\right) u_p^{11/2}\nonumber\\
&&+\left(8\nu g_{*22}+\frac{636407}{4096}\nu\pi^2-\frac{1708303}{256}\nu-696\nu\gamma+\frac32 \nu  g_{*41}+\frac12 \nu g_{*23}^c -348\nu\ln(u_p)+\frac12 \nu g_{*23}^{\ln{}}\ln(u_p)\right. \nonumber\\
&& \left. +\frac{331875}{512}-\frac{13122}{5}\nu\ln(3)+\frac{18952}{15}\nu\ln(2)\right) u_p^{13/2}+ \mathcal{O}(u_p^7,\nu^2)\,\nonumber\\
\Omega_{S_1}^{(4)}(u_p,\nu)&=&
\left(\frac{9}{16}-\frac{3}{8}\nu  \right) u_p^{5/2}+\left(-\frac{315}{64}+\frac{135}{32}\nu  \right) u_p^{7/2}+\left(\frac{1281}{16}\nu-\frac{9255}{128}\right) u_p^{9/2} \nonumber\\
&+& \left[-\frac{971463}{1024}+\left(\frac{437543}{256}-\frac{3813}{256}\pi^2-\frac38 g_{*22}\right)\nu\right]u_p^{11/2} \nonumber\\
&+&\left[-\frac{24272901}{2048}+\left(\frac{224683}{8}-\frac72 g_{*22}+\frac92  g_{*41}+\frac38 g_{*42}^{\ln{}} \ln(u_p)+\frac{6642}{5}\gamma-\frac{187742}{5}\ln(2)-\frac{3855321}{8192}\pi^2\right.\right. \nonumber\\
&+&\left.\left.\frac{3321}{5}\ln(u_p)+\frac{505197}{20}\ln(3)+\frac38 g_{*42}^c+\frac{7}{16} g_{*23}^{\ln{}} \right)\nu\right]u_p^{13/2}  + \mathcal{O}(u_p^7,\nu^2) \,.
\end{eqnarray}
\end{widetext}
As above, we have only indicated here the first terms in the PN-expansions we computed.

Let us now consider the ratio  
\beq
\psi\equiv \frac{\langle  \Omega_{S_1}\rangle}{\Omega_\phi }\,.
\eeq
It is easily checked (see, e.g., Refs. \cite{Dolan:2013roa,Bini:2014ica}) that the so-defined ratio is identical to the spin-precession measure
$\psi$ introduced in Eq. \eqref{Eq:psiDef} above. 
The combined eccentricity-, PN-  and SF-expansions of the function $\psi(u_p,e;\nu)$ is then found to be of the type
\beq
\psi(u_p,e)=\psi^{(0)}(u_p,\nu)+e^2 \psi^{(2)}(u_p,\nu)+e^4\psi^{(4)}(u_p,\nu)+\mathcal{O}(e^6)\,,
\eeq
where the beginnings of the expansions of $\psi^{(0)}(u_p,\nu)$ and $\psi^{(2)}(u_p,\nu)$ are given by
\begin{widetext}
\begin{eqnarray}
\psi^{(0)}(u_p,\nu)&=&\left(\frac32  -\nu\right)u_p+(\frac{45}{8}-3\nu) u_p^2+\left(-33\nu+\frac{663}{16}\right) u_p^3+\left(-\frac{1537}{3}\nu+\frac{41}{8}\nu\pi^2+\frac{47805}{128}\right) u_p^4\nonumber\\
&& +\left(\frac{951309}{256}-\frac{4336}{15}\nu\ln(2)+\nu g_{*22}-144\nu\gamma+\frac{109897}{1024}\nu\pi^2-\frac{1162607}{160}\nu-72\nu\ln(u_p)\right)u_p^5+\mathcal{O}(u_p^6,\nu^2)\nonumber\\
\psi^{(2)}(u_p,\nu)&=& \left(3-\frac32 \nu\right)u_p^2+\left(-39\nu+\frac{75}{2}\right)u_p^3+\left[\left(-\frac{56179}{64}+\frac{615}{64}\pi^2+\frac12  g_{*22}\right)\nu+\frac{3639}{8}\right] u_p^4\nonumber\\
&&+\left[\frac{89751}{16}+\left(\frac32  g_{*41}+\frac12  g_{*23}^c -\frac{6176}{15} \ln(2)+\frac{35}{4}  g_{*22}-\frac{3232}{5} \gamma+\frac{282651}{1024} \pi^2\right.\right.\nonumber\\
&& \left.\left.-\frac{5618027}{384}  +\frac12  g_{*23}^{\ln{}}\ln(u_p) -\frac{1616}{5} \ln(u_p)-\frac{4374}{5} \ln(3)\right)\nu\right] u_p^5+\mathcal{O}(u_p^6,\nu^2)\nonumber\\
\psi^{(4)}(u_p,\nu)&=& \left(-\frac34 \nu+\frac{3}{16}\right) u_p^3 +\left[\frac{1101}{32}+\left(-\frac{52051}{512}+\frac{123}{256}\pi^2+\frac38 g_{*22}+\frac38 g_{*41}\right)\nu\right] u_p^4\nonumber\\ 
&+&\left[ \frac{124533}{128}+\left(-\frac{1074}{5}\gamma-\frac{45358}{5}\ln(2)+\frac{675243}{8192}\pi^2-\frac{537}{5}\ln(u_p)+\frac{19683}{4}\ln(3)+\frac{145}{16} g_{*22}\right.\right. \nonumber\\
&+&\left.\left.\frac{99}{16}g_{*41}+\frac34 g_{*23}^{\ln{}} \ln(u_p)+\frac38 g_{*42}^{\ln{}}\ln(u_p)+\frac34  g_{*23}^c+\frac38 g_{*42}^c +\frac{7}{16} g_{*23}^{\ln{}} -\frac{17185101}{5120}\right)\nu\right] u_p^5\nonumber\\
& +& \mathcal{O}(u_p^6,\nu^2)\,.
\end{eqnarray}
\end{widetext}
  
Though the  so obtained function $\psi(u_p,e;\nu)$ is gauge-invariant ($u_p$ and $e$ being functions of the gauge-invariant dynamical quantities $\hat{ \mathcal E}_{\rm eff}$ and $j$), it cannot be directly compared with the (equally gauge-invariant) function $\psi(m_2\Omega_r,m_2\Omega_\phi;q)$ obtained in the SF computation above.

In order to compare our EOB-derived result with the SF result we first need to transform the dependence on $u_p$ and $e$ into a dependence on $Gm_2\Omega_r$ and $Gm_2\Omega_\phi$, or equivalently, on $y=(Gm_2 \Omega_\phi)^{2/3}$ and $\hat k=\frac13 (\frac{\Omega_\phi}{\Omega_r}-1)$.

Using Eqs. \eqref{up_and_e2_vs_hatk_and_y} above we then find that the EOB-derived functional dependence of $\psi$ on $(\hat k ,y)$ is given, at first order in $\nu$, by
\beq
\psi(\hat k,y)=\psi_0(\hat k,y)+\nu \Delta \psi (\hat k,y)+\mathcal{O}(\nu^2)\,,
\eeq
where the beginnings of the expansions of $\psi_0(\hat k,y)$ and $\Delta \psi (\hat k,y)$ read
\begin{eqnarray}
\label{delta_psi_ky}
\psi_0(\hat k,y)&=& \frac32 \hat k-\left(\frac98\hat k y+\frac{9}{2}\hat k^2\right) +\left(\frac{75}{32}\hat k^2 y-\frac{75}{32}y^2\hat k+\frac{27}{2}\hat k^3\right) \nonumber\\
&+& \left(-\frac{135}{128}\hat k^3 y-\frac{45}{128}\hat k^2 y^2-\frac{45}{8}\hat k y^3-\frac{5229}{128}\hat k^4\right)+\ldots \nonumber\\
\Delta \psi (\hat k,y) &=& -\hat k+\left(8\hat k^2-\frac54 \hat k y\right)+\bigg(\frac{123}{256}\hat k^2\pi^2 y-\frac{615}{256}\hat k^3\pi^2 \nn \\
&-&\frac{93}{8}\hat k^2 y-\frac{53}{16}y^2\hat k+\frac{69}{4}\hat k^3\bigg)
+\left(c_6^c+c_6^{\ln{}}\ln (\hat k) \right) \hat k^4  \nn \\
&+&\ldots 
\end{eqnarray}

 with
\begin{widetext}
\begin{eqnarray}
c_6^c&=& -\frac{36735959}{23040}+\left(-\frac{1}{16}g_{*41}+\frac{177067}{1440}-\frac{3995649}{320}\ln(3)-\frac{11518508}{45}\ln(2)+\frac{68359375}{576}\ln(5)\right) \left(\frac{y}{\hat k}\right)^3\nonumber\\
&& +\left(-\frac{5137157}{7680}+\frac{9}{16} g_{*41}+\frac{10900979}{15}\ln(2)+\frac{37}{5}\gamma-\frac{33203125}{96}\ln(5)+\frac{3}{8}g_{*22}-\frac{4081}{16384}\pi^2+\frac{1560789}{32}\ln(3)\right) \left(\frac{y}{\hat k}\right)^2\nonumber\\
&& +\left(-\frac{15}{16}g_{*41}-\frac54 g_{*22}+\frac{1111433}{768}+\frac{11195}{8192}\pi^2-689470\ln(2)+\frac{21484375}{64}\ln(5)-122\gamma-\frac{18575649}{320}\ln(3)\right) \frac{y}{\hat k}\nonumber\\
&& -\frac{1953125}{18}\ln(5)+\frac78 g_{*22}+\frac{7}{16}g_{*41}+\frac{9842609}{45}\ln(2)+\frac{595}{3}\gamma+\frac{438129}{20}\ln(3)+\frac{1580185}{49152}\pi^2\nonumber\\
c_6^{\ln{}}&=& \frac{595}{6}  -61 \frac{y}{\hat k}+\frac{37}{10} \left(\frac{y}{\hat k}\right)^2 \,.
\end{eqnarray}
\end{widetext}
Of most interest for our present work is the function $\Delta \psi (\hat k,y)$, Eq. \eqref{delta_psi_ky}. We can directly compare the latter function with the one computed with SF theory above in Eq.~\eqref{Eq:DeltaPsiPN}, modulo the fact that the SF computed one above was expressed not in terms of $\hat k$ and $y$, but instead in terms of   eccentricity and semi-latus rectum parameters, $\bar e$ and $\bar{p} =1/\bar u_p$, different from the ones, $e$ and $u_p$, used here (recalling the notational change discussed at the beginning of this section).

The transformation between $(\bar e,\bar u_p)$ and $(e,u_p)$ is only needed at order $\nu^0$ and can be obtained by identifying 
the $\mu$-rescaled Schwarzschild energy and angular momentum used to define $\bar e$ and $\bar u_p$ to the EOB quantities $\hat {\mathcal E}_{\rm eff}$ and $j$.
In other words, while  $(\bar e,\bar u_p)$  were defined in Eqs. \eqref{Eq:rofpe}, \eqref{Eq:ELofpe}, i.e., equivalently, by  writing
\begin{align}
\hat {\mathcal E}_{\rm eff}&=\sqrt{\frac{(1-2\bar u_p)^2-4\bar u_p^2\bar e^2}{1-3\bar u_p-\bar u_p\bar e^2}}\,,\\
j&= \frac{1}{\sqrt{\bar u_p(1-3\bar u_p-\bar u_p\bar e^2)}}\,,
\end{align}
the other pair $(e,u_p)$, used in our EOB computation, was defined by writing
\beq
\hat {\mathcal E}_{\rm eff}= \sqrt{1-u_p(1-e^2)}\,,\qquad \quad j= \frac{1}{\sqrt{u_p}}. 
\eeq
The  comparison between these two expressions implies the following transformation law
\begin{eqnarray}
u_p &=& \bar u_p (1-3\bar u_p-\bar u_p\bar e^2) \,,\nonumber\\
1-e^2 &=& (1-\bar e^2)  \, \frac{1-4\bar u_p}{(1-3\bar u_p-\bar u_p\bar e^2)^2}\,.
\end{eqnarray}
Using either this transformation (together with intermediate equations given above) or directly the well known elliptic-integrals expressions giving $\Omega_\phi$ and $k$ in a Schwarzschild background, consistently with Eqs. (2.13)-(2.16) in Ref. \cite{Akcay:2016dku}, we obtain
\\
\begin{widetext}
\begin{eqnarray}
\hat k &=& \bar u_p+\left(\frac92 +\frac14 \bar e^2\right)\bar u_p^2 +\left(\frac{45}{2}+\frac{15}{4} \bar e^2\right) \bar u_p^3 
+\left(\frac{35}{64} \bar e^4+\frac{945}{8}+\frac{315}{8} \bar e^2\right) \bar u_p^4 +\left(\frac{2835}{8} \bar e^2+\frac{945}{64} \bar e^4+\frac{5103}{8}\right) \bar u_p^5 \nonumber\\
&+&\left(\frac{31185}{128} \bar e^4+\frac{93555}{32} \bar e^2+\frac{385}{256} \bar e^6+\frac{56133}{16}\right) \bar u_p^6 +
\left(\frac{405405}{128} \bar e^4+\frac{729729}{32} \bar e^2+\frac{312741}{16}+\frac{15015}{256} \bar e^6\right) \bar u_p^7 \nonumber\\
&+&\left(\frac{18243225}{512} \bar e^4+\frac{10945935}{64} \bar e^2+\frac{14073345}{128}+\frac{675675}{512} \bar e^6\right) \bar u_p^8 +\mathcal{O}(\bar u_p^9)\nonumber\\
y&=& (1-\bar e^2) \bar u_p+(2 \bar e^2-2 \bar e^4) \bar u_p^2 +\left(-\frac{23}{8} \bar e^4-\frac{55}{16} \bar e^6+6 \bar e^2\right) \bar u_p^3 
+\left(-\frac{13}{4} \bar e^4+24 \bar e^2-\frac{209}{12} \bar e^6\right) \bar u_p^4 \nonumber\\
&+& \left(-\frac{1237}{16} \bar e^6-\frac14 \bar e^4+120 \bar e^2\right) \bar u_p^5 
+\left(-\frac{3069}{8} \bar e^6+672 \bar e^2+51 \bar e^4\right) \bar u_p^6 
+\left(730 \bar e^4-\frac{52265}{24} \bar e^6+3936 \bar e^2\right) \bar u_p^7 \nonumber\\
& +&\left(-\frac{26279}{2} \bar e^6+8032 \bar e^4+23424 \bar e^2\right) \bar u_p^8  +\mathcal{O}(\bar u_p^9)\,.
\end{eqnarray}
\end{widetext}

Inserting these expressions in the EOB-derived function  $\Delta\psi(\hat k,y)$ obtained above gives
\beq
\Delta\psi(\bar u_p, \bar e) =\Delta\psi^{(0)}(\bar u_p)+\bar e^2 \Delta\psi^{(2)}(\bar u_p)+\bar e^4 \Delta\psi^{(4)}(\bar u_p)+\ldots
\eeq
where $\Delta\psi^{(0)}(\bar u_p)$ will be discussed below, and where the $\mathcal{O}(\bar e^2)$ contribution (which is new with this work) reads
\begin{widetext}
\begin{eqnarray}
\Delta\psi^{(2)}(\bar u_p) &=& \bar u_p^2 +\left(\frac{341}{16} -\frac{123}{256}\pi^2\right)\bar u_p^3 \nonumber\\
&& +\left(-\frac{317491}{960} +\frac{268}{5}\ln(\bar u_p) -\frac{23729}{4096}\pi^2 +\frac12 g_{*22}  +\frac{536}{5}\gamma+\frac{11720}{3}\ln(2)-\frac{10206}{5}\ln(3)\right)\bar u_p^4\nonumber\\
&& +\left(\frac{21797069}{49152}\pi^2  -\frac{131386901}{57600} -\frac{10957}{15} \ln(\bar u_p) +\frac54 g_{*22}+\frac{9765625}{1344}\ln(5)+\frac12g_{*23}^c  +\frac12   g_{*23}^{\ln{}}\ln(\bar u_p)\right. \nonumber\\
&& \left.-\frac{21914}{15}\gamma-\frac{333378}{7}\ln(2)+\frac{4943349}{320}\ln(3)\right)\bar u_p^5  \nonumber\\
&& + \frac{319609}{630}\pi \bar u_p^{11/2} \nonumber\\
&& +\left(\frac54 g_{*23}^c  -\frac{9232551768011}{101606400} -\frac{33}{16} g_{*22}-\frac{225143631}{8960}\ln(3)+\frac12  g_{*24}^c+\frac12 \ln(\bar u_p) g_{*24}^{\ln{}} -\frac{33}{16}\right. \nonumber\\
&& +\frac54  g_{*23}^{\ln{}} \ln(\bar u_p)-\frac{17193359375}{145152}\ln(5)+\frac{284236994}{945}\ln(2)-\frac{529328}{189}\gamma-\frac{146026515}{1048576}\pi^4+\frac{31842559607}{2359296}\pi^2 \nonumber\\
&& \left.-\frac{264664}{189}\ln(\bar u_p)\right)\bar u_p^6 +\mathcal{O}(\bar u_p^{13/2})\,,
\end{eqnarray}
\end{widetext}
Here the terms up to $\mathcal{O}(u_p^3)$ coincide with those given in Eq. (5.4) of \cite{Akcay:2016dku}, and
the undetermined coefficients $g_{*22}, g_{*23}^c,$ etc.   which parametrized the unknown $\mathcal{O}(p_r^2)$ dependence of the EOB gyrogravitomagnetic ratio $g_{S*}(u,p_r,p_\phi)$ first enter $\Delta\psi^{(2)}(\bar u_p)$ at order $\mathcal{O}(u^4)$.

Comparing with Eq.~\eqref{Eq:DeltaPsiPN} above we then find unique values for the so far undetermined EOB coefficients, namely
\begin{widetext}
\begin{eqnarray}
g_{*22}&=& -\frac{717}{32} \nonumber\\
g_{*23}^c&=& \frac{1447441}{960}-\frac{4829}{256}\pi^2-\frac{16038}{5}\ln(3)+\frac{46976}{15}\ln(2)-\frac{512}{5}\gamma\nonumber\\
g_{*23}^{\ln{}}&=&-\frac{256}{5} \nonumber\\
g_{*24}^c&=&-\frac{185195453}{38400}+\frac{19162}{35}\gamma+\frac{2097479}{8192}\pi^2+\frac{454167}{20}\ln(3)-\frac{1081966}{35}\ln(2) \nonumber\\
g_{*24}^{\ln{}}&=&+\frac{9581}{35} \,.
\end{eqnarray}

We also computed  the $\mathcal{O}(\bar e^4)$ contribution to $\Delta\psi(\bar u_p, \bar e)$ which we give below in its parametrized form

\begin{eqnarray}
\Delta\psi^{(4)}(\bar u_p) &=&
-\frac12 \bar u_p^3\nonumber\\
&+& \left(-\frac{366857}{1536}+\frac38  g_{*41}+\frac{1953125}{192} \ln(5)+\frac{37}{10}\ln(\bar u_p)-\frac{23761}{16384}\pi^2+\frac38 g_{*22}+\frac{37}{5}\gamma-\frac{205843}{5}\ln(2)\right. \nonumber\\
&& \left. +\frac{3620943}{320}\ln(3)\right) \bar u_p^4\nonumber\\
&+& \left(-\frac{524787}{65536}\pi^2+\frac{720900871}{537600}-\frac{7143}{56}\ln(\bar u_p)+\frac38 g_{*42}^c +\frac{21}{16}g_{*41}+\frac{57}{16}g_{*22}+\frac{72265625}{5376}\ln(5)\right. \nonumber\\
&& \left. +\frac{7}{16} g_{*23}^{\ln{}} +\frac34 g_{*23}^c+\frac34 g_{*23}^{\ln{}} \ln(\bar u_p)+\frac38 g_{*42}^{\ln{}} \ln(\bar u_p)-\frac{7143}{28}\gamma-\frac{30009569}{420}\ln(2)+\frac{210887307}{8960}\ln(3)\right)\bar u_p^5\nonumber\\
&+& \frac{1089581}{5120}\pi \bar u_p^{11/2} \nonumber\\
&+&\left(\frac{55}{32} g_{*23}^{\ln{}} -\frac{255}{64}g_{*41}+\frac{41}{8}  g_{*23}^c-\frac{797597168303}{16934400}+\frac{661}{64}g_{*22}-\frac{678223072849}{153600}\ln(7)+\frac{4018699514169}{358400}\ln(3)\right. \nonumber\\
&+& \frac{9}{16}g_{*24}^{\ln{}} +\frac54 g_{*24}^c+\frac{21}{16} g_{*42}^c  +\frac38 g_{*43}^c  +\frac{5}{4}\ln(\bar u_p) g_{*24}^{\ln{}} +\frac38 g_{*43}^{\ln{}} \ln(\bar u_p) +\frac{21}{16} g_{*42}^{\ln{}} \ln(\bar u_p)\nonumber\\
&+&\frac{41}{8} g_{*23}^{\ln{}}\ln(\bar u_p)-\frac{17944462890625}{1161216} \ln(5)+\frac{2304030315373}{75600}\ln(2)-\frac{48044911}{15120}\gamma-\frac{17998485}{524288}\pi^4\nonumber\\
&+&\left. \frac{151169660653}{18874368}\pi^2-\frac{48044911}{30240}\ln(\bar u_p)\right) \bar u_p^6 +\mathcal{O}(\bar u_p^{13/2})\,.
\end{eqnarray}
\end{widetext}
As soon as some SF data on $\psi$ at order $\mathcal{O}(e^4)$  become available, this result will allow  one to determine the coefficients  parametrizing the $\mathcal{O}(p_r^4)$ EOB gyrogravitomagnetic ratio $g_{S*}$.

Concerning the quantity $ \Delta \psi (\bar u_p,\bar e \to 0)$, it can be obtained  (as pointed out in Ref. \cite{Akcay:2016dku}) by combining the previous SF results on 
$\Delta\psi_{\rm (circ)}(\bar u_p)$ \cite{Bini:2014ica,Bini:2015mza,Kavanagh:2015lva}
\begin{align}
\Delta\psi_{\rm (circ)}(\bar u_p)&=\bar u_p^2-3\bar u_p^3-\frac{15}{2}\bar u_p^4+\bigg(-\frac{6277}{30}-\frac{496}{15}\ln(2) \nn \\
&-16\gamma-8\ln(\bar u_p)+\frac{20471}{1024}\pi^2\bigg)\bar u_p^5+\ldots\,,
\end{align}
with the knowledge of the EOB function $\rho(x)$ measuring periastron precession at the 1SF-level \cite{Damour:2009sm,Bini:2016qtx}, namely
\begin{align}
 \Delta \psi &(\bar u_p,\bar e \to 0)=\nn \\
 &\Delta\psi_{\rm (circ)}(\bar u_p)-\frac12 \frac{(1-3\bar u_p)^{1/2}(1-6 \bar u_p)^{5/2}}{1-\frac{39}{4}\bar u_p+\frac{43}{2}\bar u_p^2} k_{\rm (circ)}^{\rm 1SF}(\bar u_p)\,,
\end{align}
where
\beq
k_{\rm (circ)}^{\rm 1SF}(y)=- \frac{\rho(y)-4 y}{2(1-6  y)^{3/2}}\,,
\eeq
as obtained from the definition $(1+k_{\rm (circ)}(x,\nu))^{-2}=1-6x+\nu \rho(x)+\mathcal{O}(\nu^2)$, i.e.,  using the link $x=(GM\Omega_\phi)^{2/3}=(1+\frac23 \nu)y+\mathcal{O}(\nu^2)$ and
\begin{align}
k_{\rm (circ)}(y,\nu)
&=-1+\frac{1}{\sqrt{1-6(1+\frac23 \nu)y+\nu \rho(y)}}\nn \\
&=
k_{\rm (circ)}^{(0)}(y)+\nu k_{\rm (circ)}^{\rm 1SF}(y)+\mathcal{O}(\nu^2)\,,
\end{align}
with
\beq
k_{\rm (circ)}^{(0)}(y)=-1+\frac{1}{\sqrt{1-6 y}}\,.
\eeq
\\

\section{Resummation of $g_{S*}$}

SF technology allows one to reach high PN orders when working linearly in the symmetric mass ratio $\nu$ \cite{Blanchet:2010zd,Barack:2010ny,Bini:2013zaa,Bini:2013rfa,Bini:2014nfa,Bini:2015bla,Kavanagh:2015lva,Bini:2015bfb,Hopper:2015icj,Bini:2016qtx}. 
We have given here one more example of this feature by deriving the $\mathcal{O}(e^2)$ (respectively, $\mathcal{O}(p_r^2)$) piece in the spin precession function $\psi$ (respectively, gyrogravitomagnetic ratio  $g_{S*}$) to many more PN-orders than was previously known. Past work has shown that an efficient way of using the high-PN information obtained by SF computations is to incorporate it, together with known PN contributions which are higher order in $\nu$, within some suitably resummed coupling functions of the EOB formalism. Let us show how such an approach can be applied to the $g_{S*}$ EOB coupling function.

To blend and resum SF and PN information concerning $g_{S*}$ let us start from the structure of the test-mass value of $g_{S*}$, as written (when neglecting effects quadratic in spins) in Eq. \eqref{gsstar_expl}. First, we note that this structure can be rewritten as 
\begin{align}
g_{S*}^{(\nu^0)}&(u,p_r,p_\phi)=\nn \\
&\frac{\sqrt{A_{\rm schw}(u)}}{\hat {\mathcal E}_{\rm schw}+ \sqrt{A_{\rm schw}(u)}}+\frac{2 A_{\rm schw}(u)}{\hat {\mathcal E}_{\rm schw}(1+\sqrt{A_{\rm schw}(u)})}\,,
\end{align}
where $A_{\rm schw}(u)=1-2u$ and where $\hat {\mathcal E}_{\rm schw}=\sqrt{(1-2u)(1+p_\phi^2u^2 +(1-2u)p_r^2)}$ is the conserved energy of a test particle in a Schwarzschild background.
Taking this reformulation as a model, let us define the following EOB-compatible, $\nu$-deformed $g_{S*}$ coupling
\begin{align}
\label{eq_3PN}
g&_{S*}^{\rm (EOB-like)}(u,p_r,p_\phi;\nu)=\nn \\
&\frac{\sqrt{A_{3PN}(u;\nu)}}{\hat H^{3PN}_{\rm (orb,eff)}+ \sqrt{A_{3PN}(u;\nu)}}+\frac{2 A_{3PN}(u;\nu)}{\hat H^{3PN}_{\rm (orb,eff)}(1+\sqrt{A_{3PN}(u;\nu)})}\,,
\end{align}
where $A_{3PN}(u;\nu)$ is defined as
\beq
A_{3PN}(u;\nu)\equiv  1-2 u+2\nu u^3+\left(\frac{94}{3}-\frac{41}{32}\pi^2\right)\nu u^4\,,
\eeq
and where
\begin{align}
\hat H&^{3PN}_{\rm (orb,eff)}\equiv \sqrt{A_{3PN}(u;\nu)} \nn\\
&\times\sqrt{   \left(1+p_\phi^2u^2 + A_{3PN}(u;\nu) \bar D_{3PN}(u;\nu) p_r^2+\hat Q_{3PN}\right)}\,,
\end{align}
with 
\begin{align}
\bar D_{3PN}(u;\nu)&= 1+6\nu u^2 +(52\nu-6\nu^2)u^3\,,\nn\\
\hat Q_{3PN}(u;\nu)&=2(4-3\nu)\nu u^2 p_r^4\,.
\end{align}

We propose to use the $\nu$-deformed ratio $g_{S*}^{\rm (EOB-like)}$, Eq. \eqref{eq_3PN}, as a reference value for the (unknown) exact $g_{S*}$, and to incorporate the current PN and SF information of $g_{S*}$ in the form of a {\it correcting factor} $r_{S*}(u,p_r,p_\phi; \nu)$ of the type $r_{S*}=1+\mathcal{O}(\nu)$.
[We have also explored the possibility of translating $g_{S*}$ in terms of the gyrogravitomagnetic ratio of a test-mass in some spin-effective metric. However, this led to an effective metric whose $\nu$-deformation was drastically different (starting at the 1PN-order) from the one of the usual orbital EOB effective metric, and which did not seem to offer a good starting point for resumming $g_{S*}$.]
In other words, we propose to define a resummed $g_{S*}$ of the form
\beq
\label{gsstar_resum}
g_{S*}^{\rm resum}(u,p_r,p_\phi;\nu)=g_{S*}^{\rm (EOB-like)}(u,p_r,p_\phi;\nu)r_{S*}(u,p_r,p_\phi;\nu) \,.
\eeq
When considering fast-spinning black holes, one can still use the factorized form \eqref{gsstar_resum}, but with suitably spin-quadratic-extended values of $A_{3PN}$ and $\hat H^{3PN}_{\rm (orb,eff)}$ in the definition \eqref{eq_3PN} of $g_{S*}^{\rm (EOB-like)}$ (e.g., as defined in Ref. \cite{Damour:2014sva}).

We will suggest several possible estimates of the correcting factor $r_{S*}$ in Eq. \eqref{gsstar_resum}. All these estimates will be constructed from the PN expansion of the correcting factor $r_{S*}=g_{S*}^{SF+PN}/g_{S*}^{\rm (EOB-like)}$, which is of the form
\beq
\label{rstar_PN}
r_{S*}^{\rm PN}(u,p_r,p_\phi;\nu)=1+\nu \hat g_1(u,p_r,p_\phi)+\nu^2 \hat g_2(u,p_r,p_\phi)\,,
\eeq
where
\begin{widetext}
\begin{eqnarray}
\label{g1}
\hat g_1(u,p_r,p_\phi)& =& 
\eta^2\left(-\frac12  u-\frac32 p_r^2\right)+\eta^4 \left[\frac{25}{24} p_r^4+\left(-\frac58 p_\phi^2 u^2-\frac{53}{24} u\right) p_r^2-\frac{5}{24} p_\phi^2 u^3-\frac{20}{3} u^2\right]\nonumber\\
&+& \eta^6 \left[ (-\frac{2297}{144} u^2+\frac{17}{96} p_\phi^4 u^4-\frac{97}{144} p_\phi^2 u^3) p_r^2+\frac{17}{288} p_\phi^4 u^5-\frac{8771}{288} u^3-\frac{199}{72} p_\phi^2 u^4+\frac{41}{48} u^3\pi^2\right]\nonumber\\
&+& \eta^8 \left[ \left(-\frac{1483}{192} p_\phi^2 u^4-\frac{10692}{5} u^3\ln(3)-\frac{1024}{15} u^3\gamma-\frac{14077}{1152} u^3\pi^2+\frac{93952}{45} u^3\ln(2)-\frac{55}{576} p_\phi^6 u^6\right.\right. \nonumber\\
&+&\left. \frac{77}{576} p_\phi^4 u^5+\frac{17436377}{17280} u^3-\frac{512}{15} u^3\ln(u)\right) p_r^2-\frac{354523}{4320} u^4-\frac{42575}{3456} p_\phi^2 u^5+\frac{113}{144} p_\phi^4 u^6\nonumber\\
&-& \left. \frac{55}{1728} p_\phi^6 u^7-\frac{2912}{45} u^4\ln(2)+\frac{85421}{9216} u^4\pi^2+\frac{205}{576} p_\phi^2 u^5\pi^2-32 u^4\gamma-16 u^4\ln(u)\right]\nonumber\\
&+& \eta^{10} \left[\left(\frac{144261}{10} u^4\ln(3)-\frac{256}{9} u^5\gamma p_\phi^2-\frac{433}{10368} p_\phi^6 u^7-891 u^5\ln(3) p_\phi^2+\frac{23488}{27} u^5 \ln(2) p_\phi^2\right.\right. \nonumber\\
&-& \frac{128}{9} u^5\ln(u) p_\phi^2-\frac{6281068}{315} u^4 \ln(2) +\frac{103604}{315} u^4\gamma+\frac{869}{13824} p_\phi^8 u^8\nonumber\\
&-& \left. \frac{505706219}{172800} u^4+\frac{22577}{6912} p_\phi^4 u^6+\frac{8367161}{20736} p_\phi^2 u^5+\frac{146005}{864} u^4\pi^2-\frac{75223}{13824} p_\phi^2 u^5\pi^2+\frac{51802}{315} u^4\ln(u)\right) p_r^2\nonumber\\
&+& 
\frac{5248}{35} u^5\gamma-\frac{2887}{96} p_\phi^2 u^6+\frac{145379}{41472} p_\phi^4 u^7-\frac{2197}{5184} p_\phi^6 u^8+\frac{869}{41472} p_\phi^8 u^9-\frac{4577190091}{3628800} u^5\nonumber\\
&+&\frac{13059635}{110592} u^5\pi^2-\frac{40}{3} u^6 \gamma p_\phi^2+\frac{2624}{35} u^5\ln(u)-\frac{20}{3} u^6\ln(u) p_\phi^2\nonumber\\
&+& \left. \frac{353392}{945} u^5\ln(2)-\frac{697}{6912} p_\phi^4 u^7\pi^2-\frac{728}{27} u^6\ln(2) p_\phi^2-\frac{486}{7} u^5\ln(3)+\frac{412673}{110592} p_\phi^2 u^6\pi^2\right]+\mathcal{O}(u^6)\,,
\end{eqnarray}
\end{widetext}
and
\beq
\hat g_2(u,p_r,p_\phi)=\left(\frac{19}{8}p_r^2 u-\frac18 u^2+\frac{15}{8}p_r^4 \right)\eta^4\,.
\eeq
In Eq. \eqref{g1} we have displayed only the beginning of the PN expansion of $\hat g_1$, including in particular all the $\mathcal{O}(p_r^2)$ terms that have been derived in the present paper. We kept in Eq. \eqref{g1} only the terms that are fully known, i.e., the terms that do not involve any yet undetermined coefficient such as $g_{*41}up_r^4 \eta^6$, etc.
[More precisely, in $\hat g_1$ we see that the $p_r^4$ contribution  is fully known only at $\mathcal{O}(\eta^4)$; by contrast, the $p_r^2$ one is now known from $\mathcal{O}(\eta^2)$ up to $\mathcal{O}(\eta^{10})$; in $\hat g_2$ only terms at $\mathcal{O}(\eta^4)$ are known.] Beyond the contributions indicated, only the circular 
limit of $\hat g_1$ is known and it can be straightforwardly computed (in the DJS gauge) by PN-expanding the prescription given above, using for instance either the explicit 9.5PN accurate $g_{S*}^{\rm circ}$ result of Ref. \cite{Bini:2015mza} or the implicit 22PN-accurate result of Ref. \cite{Kavanagh:2015lva} on $\psi_{\rm circ}$.

In the rest of this subsection we will motivate various ways (namely,  Taylor-like or inverse-Taylor-like, at various PN-approximations) of defining the correcting factor $r_{S*}$ in Eq. \eqref{rstar_PN}
by studying the circular limit of $g_{S*}^{\rm resum}$, Eq. \eqref{gsstar_resum}.
For concreteness, let us exhibit the explicit 4PN-accurate value of $r_{S*}$ when setting $p_r\to 0$ 
\begin{widetext}
\begin{eqnarray}
\label{eq_rSstar4PN}
r_{S*}^{4PN}(u,p_\phi;\nu) &=& 1-\frac12u \nu \eta^2 +\left(-\frac{20}{3}\nu u^2-\frac18 \nu^2 u^2-\frac{5}{24}\nu p_\phi^2 u^3\right)\eta^4\nonumber\\
&+& \left(-\frac{8771}{288} u^3+\frac{41}{48} u^3\pi^2-\frac{199}{72} p_\phi^2 u^4+\frac{17}{288} p_\phi^4 u^5\right)\nu\eta^6\nonumber\\
&+& \left(-\frac{354523}{4320}  u^4-\frac{42575}{3456} u^5 p_\phi^2-\frac{55}{1728}  p_\phi^6 u^7+ \frac{113}{144}  p_\phi^4 u^6-16  u^4\ln(u\eta^2)\right. \nonumber\\
&-&\left.\frac{2912}{45}  u^4\ln(2)-32  u^4\gamma+\frac{85421}{9216}  u^4\pi^2+\frac{205}{576} u^5\pi^2 p_\phi^2\right)\nu\eta^8\,.
\end{eqnarray}
\end{widetext}
Note the explicit appearance of $p_\phi$ in $r_{S*}^{4PN}$. We wish to study an estimate of $g_{S*}$ that would be approximately valid when considering   (circularized)  inspiralling and coalescing binary black holes. Following the results of EOB theory \cite{Buonanno:1998gg,Buonanno:2000ef} we can approximately replace $p_\phi$ by a function of $u$ having the following properties.

\begin{figure}
\begin{center}
\includegraphics[scale=0.35]{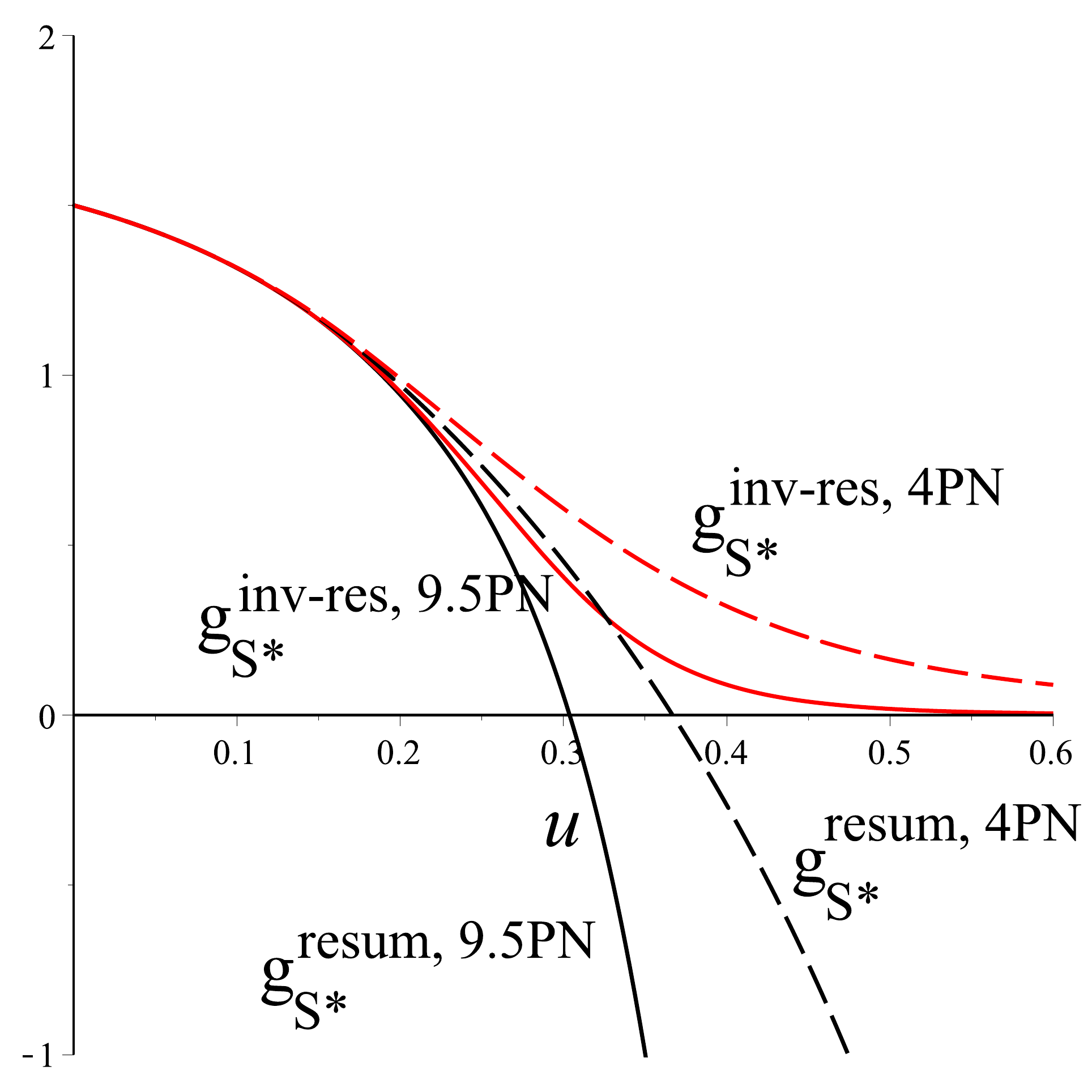}
\caption{\label{fig:1BD} 
Comparison of various estimates of the EOB gyrogravitomagnetic ratio $g_{S*}$ in the equal-mass case ($\nu=1/4$), and in the circularized inspiralling and coalescing approximation. The resummed $g_{S*}$ defined by Eq. \eqref{gsstar_resum} is plotted as a function of $u=GM/(c^2R)$ for the four cases explained in text. 
}
\end{center}
\end{figure}

Above the last stable orbit we estimate $p_\phi\approx p_\phi^{\rm (circ)}(u)=\sqrt{-{\partial_u A(u; \nu)}/{\partial_u (u^2A(u; \nu))}}$, which we simply approximate in our present qualitative study by $p_\phi\approx[u(1-3u)]^{-1/2}$. Beyond the last stable orbit we instead approximate $p_\phi$ by a constant equals to its value at the last stable orbit (say $p_\phi\approx \sqrt{12}$).
This approximation defines the evolution of $r_{S*}$, and correlatively of  $g_{S*}$, during coalescence, as a function of $u$.
The results of this exploratory study of the evolution of $g_{S*}$  during coalescence are illustrated in Fig. \ref{fig:1BD}, which considers the equal-mass case, $\nu=1/4$. This figure compares the $u$-evolution of  various approximations to $g_{S*}=g_{S*}^{\rm (EOB-like)}\, r_{S*} $: 1) the one defined by using the Taylor-like 4PN approximant for $r_{S*}$,  Eq. \eqref{eq_rSstar4PN} together with the just explained $p_\phi$ replacements; 2) the one defined by using inverse resummation of $r_{S*}^{4PN}$, i.e.,  by taking the $P^{[1,4]}$ Pad\'e approximant of the PN-expansion of Eq. \eqref{eq_rSstar4PN}; 3) {\it ditto} with the 9.5 PN-accurate generalization of \eqref{eq_rSstar4PN}; and finally 4) {\it ditto} with the inverse resummation of the latter  $r_{S*}^{9.5PN}$ expansion.

The main message of Fig. \ref{fig:1BD} is that, in confirmation of what had been already found at the next-to-leading order \cite{Damour:2008qf}, and whatever be the approximation used, $g_{S*}$ significantly decreases as the separation $R=GM/u$ between the two bodies decreases.  This decrease becomes more pronounced when one uses higher PN-approximants. When using Taylor-approximants the decrease of   $g_{S*}$ is so extreme that it formally changes a sign below a certain separation $R_{\rm (crit)}$. For instance, at 4PN we have $R_{\rm (crit)}\approx GM/0.37\approx 2.7 GM$, while at 9.5PN we have $R_{\rm (crit)}\approx GM/0.31\approx 3.2 GM$.
As this sign change is physically unwarranted,  we advise, when using Taylor-approximants, to replace $g_{S*}$ by zero beyond the critical separation $R_{\rm (crit)}$ (i.e., to replace $g_{S*}\to \frac12 (g_{S*}+|g_{S*}|)$).

If one considers that such a vanishing  of $g_{S*}$ is too extreme a behaviour, one might consider using one of the inverse-resummed estimates of $r_{S*}$.
At this stage the spread between the curves in Fig.  \ref{fig:1BD} is a measure of our uncertainty on the true value of $g_{S*}$ in the strong-field domain. One will need comparisons between numerical relativity simulations and EOB computations using various $g_{S*}$ functions (possibly including some free parameter parametrizing strong-field effects) to learn more about the exact extent to which $g_{S*}$ decreases in the strong-field domain.

\section*{Acknowledgments}
CK thanks the authors of \cite{Akcay:2016dku} for generously providing an early draft of their work, and in particular Sarp Akcay and Sam Dolan for many informative discussions.
DB thanks ICRANet and the italian INFN for partial support and IHES for warm hospitality at various stages during the development of the present project. 
SH acknowledges financial support provided under the European Union's H2020 ERC Consolidator Grant ``Matter and strong-field gravity: New frontiers in Einstein's theory'' grant agreement no.~MaGRaTh-646597.

\appendix

\section{Kinnersley tetrad}
\label{App:Kinn}
Using Boyer-Lindquist coordinates, the Kinnersley tetrad upon which the Weyl tensor is projected is given by
\begin{align*}
e^\alpha_1&=l^\alpha=\frac{1}{\Delta}(r^2+a^2,\Delta,0,a), \\
e^\alpha_2&=n^\alpha=\frac{1}{2 \Sigma}(r^2+a^2,-\Delta,0,a), \\
e^\alpha_3&=m^\alpha=-\frac{\bar{\varrho}}{\sqrt{2}}(i a \sin \theta,0,1,\frac{i}{\sin \theta}), \\
e^\alpha_4&=\bar{m}^\alpha=-\frac{\varrho}{\sqrt{2}}(-i a \sin \theta,0,1,\frac{-i}{\sin \theta}). 
\end{align*}
where 
$\Sigma \equiv r^2 + a^2 \cos^2 \theta,\Delta \equiv r^2 - 2m_2 r + a^2$.The non-zero spin-coefficients which appear in Sec.~\ref{Sec:RgMPs} are
\begin{align*}
\varrho=\frac{-1}{r-i a \cos \theta},
\qquad\tau=\frac{-i a \sin \theta}{\sqrt{2} \Sigma} ,
\qquad\beta=-\frac{\bar{\varrho}\cot \theta}{2\sqrt{2}},\\
\gamma=\mu+\frac{r-m_2}{2 \Sigma},
\qquad \mu=\frac{\Delta \varrho}{2 \Sigma},
\qquad \varpi=\frac{i a\varrho^2 \sin \theta }{\sqrt{2}},
\end{align*}
\begin{align*}
\alpha=\varpi-\bar{\beta} .
\end{align*}
We give these in Kerr spacetime, our situation is recovered by setting $a$ to zero in the above.

\bibliography{SpinPrecSchw}

\begin{thebibliography}{78}%
\makeatletter
\providecommand \@ifxundefined [1]{%
 \@ifx{#1\undefined}
}%
\providecommand \@ifnum [1]{%
 \ifnum #1\expandafter \@firstoftwo
 \else \expandafter \@secondoftwo
 \fi
}%
\providecommand \@ifx [1]{%
 \ifx #1\expandafter \@firstoftwo
 \else \expandafter \@secondoftwo
 \fi
}%
\providecommand \natexlab [1]{#1}%
\providecommand \enquote  [1]{``#1''}%
\providecommand \bibnamefont  [1]{#1}%
\providecommand \bibfnamefont [1]{#1}%
\providecommand \citenamefont [1]{#1}%
\providecommand \href@noop [0]{\@secondoftwo}%
\providecommand \href [0]{\begingroup \@sanitize@url \@href}%
\providecommand \@href[1]{\@@startlink{#1}\@@href}%
\providecommand \@@href[1]{\endgroup#1\@@endlink}%
\providecommand \@sanitize@url [0]{\catcode `\\12\catcode `\$12\catcode
  `\&12\catcode `\#12\catcode `\^12\catcode `\_12\catcode `\%12\relax}%
\providecommand \@@startlink[1]{}%
\providecommand \@@endlink[0]{}%
\providecommand \url  [0]{\begingroup\@sanitize@url \@url }%
\providecommand \@url [1]{\endgroup\@href {#1}{\urlprefix }}%
\providecommand \urlprefix  [0]{URL }%
\providecommand \Eprint [0]{\href }%
\providecommand \doibase [0]{http://dx.doi.org/}%
\providecommand \selectlanguage [0]{\@gobble}%
\providecommand \bibinfo  [0]{\@secondoftwo}%
\providecommand \bibfield  [0]{\@secondoftwo}%
\providecommand \translation [1]{[#1]}%
\providecommand \BibitemOpen [0]{}%
\providecommand \bibitemStop [0]{}%
\providecommand \bibitemNoStop [0]{.\EOS\space}%
\providecommand \EOS [0]{\spacefactor3000\relax}%
\providecommand \BibitemShut  [1]{\csname bibitem#1\endcsname}%
\let\auto@bib@innerbib\@empty
\bibitem [{\citenamefont {Abbott}\ \emph
  {et~al.}(2016{\natexlab{a}})\citenamefont {Abbott} \emph
  {et~al.}}]{Abbott:2016blz}%
  \BibitemOpen
  \bibfield  {author} {\bibinfo {author} {\bibfnamefont {B.~P.}\ \bibnamefont
  {Abbott}} \emph {et~al.} (\bibinfo {collaboration} {Virgo, LIGO
  Scientific}),\ }\bibfield  {title} {\enquote {\bibinfo {title} {{Observation
  of Gravitational Waves from a Binary Black Hole Merger}},}\ }\href {\doibase
  10.1103/PhysRevLett.116.061102} {\bibfield  {journal} {\bibinfo  {journal}
  {Phys. Rev. Lett.}\ }\textbf {\bibinfo {volume} {116}},\ \bibinfo {pages}
  {061102} (\bibinfo {year} {2016}{\natexlab{a}})},\ \Eprint
  {http://arxiv.org/abs/1602.03837} {arXiv:1602.03837 [gr-qc]} \BibitemShut
  {NoStop}%
\bibitem [{\citenamefont {Abbott}\ \emph
  {et~al.}(2016{\natexlab{b}})\citenamefont {Abbott} \emph
  {et~al.}}]{Abbott:2016nmj}%
  \BibitemOpen
  \bibfield  {author} {\bibinfo {author} {\bibfnamefont {B.~P.}\ \bibnamefont
  {Abbott}} \emph {et~al.} (\bibinfo {collaboration} {Virgo, LIGO
  Scientific}),\ }\bibfield  {title} {\enquote {\bibinfo {title} {{GW151226:
  Observation of Gravitational Waves from a 22-Solar-Mass Binary Black Hole
  Coalescence}},}\ }\href {\doibase 10.1103/PhysRevLett.116.241103} {\bibfield
  {journal} {\bibinfo  {journal} {Phys. Rev. Lett.}\ }\textbf {\bibinfo
  {volume} {116}},\ \bibinfo {pages} {241103} (\bibinfo {year}
  {2016}{\natexlab{b}})},\ \Eprint {http://arxiv.org/abs/1606.04855}
  {arXiv:1606.04855 [gr-qc]} \BibitemShut {NoStop}%
\bibitem [{\citenamefont {Taracchini}\ \emph {et~al.}(2014)\citenamefont
  {Taracchini} \emph {et~al.}}]{Taracchini:2013rva}%
  \BibitemOpen
  \bibfield  {author} {\bibinfo {author} {\bibfnamefont {Andrea}\ \bibnamefont
  {Taracchini}} \emph {et~al.},\ }\bibfield  {title} {\enquote {\bibinfo
  {title} {{Effective-one-body model for black-hole binaries with generic mass
  ratios and spins}},}\ }\href {\doibase 10.1103/PhysRevD.89.061502} {\bibfield
   {journal} {\bibinfo  {journal} {Phys. Rev.}\ }\textbf {\bibinfo {volume}
  {D89}},\ \bibinfo {pages} {061502} (\bibinfo {year} {2014})},\ \Eprint
  {http://arxiv.org/abs/1311.2544} {arXiv:1311.2544 [gr-qc]} \BibitemShut
  {NoStop}%
\bibitem [{\citenamefont {Pürrer}(2016)}]{Purrer:2015tud}%
  \BibitemOpen
  \bibfield  {author} {\bibinfo {author} {\bibfnamefont {Michael}\ \bibnamefont
  {Pürrer}},\ }\bibfield  {title} {\enquote {\bibinfo {title} {{Frequency
  domain reduced order model of aligned-spin effective-one-body waveforms with
  generic mass-ratios and spins}},}\ }\href {\doibase
  10.1103/PhysRevD.93.064041} {\bibfield  {journal} {\bibinfo  {journal} {Phys.
  Rev.}\ }\textbf {\bibinfo {volume} {D93}},\ \bibinfo {pages} {064041}
  (\bibinfo {year} {2016})},\ \Eprint {http://arxiv.org/abs/1512.02248}
  {arXiv:1512.02248 [gr-qc]} \BibitemShut {NoStop}%
\bibitem [{\citenamefont {Buonanno}\ and\ \citenamefont
  {Damour}(1999)}]{Buonanno:1998gg}%
  \BibitemOpen
  \bibfield  {author} {\bibinfo {author} {\bibfnamefont {A.}~\bibnamefont
  {Buonanno}}\ and\ \bibinfo {author} {\bibfnamefont {T.}~\bibnamefont
  {Damour}},\ }\bibfield  {title} {\enquote {\bibinfo {title} {{Effective
  one-body approach to general relativistic two-body dynamics}},}\ }\href
  {\doibase 10.1103/PhysRevD.59.084006} {\bibfield  {journal} {\bibinfo
  {journal} {Phys. Rev.}\ }\textbf {\bibinfo {volume} {D59}},\ \bibinfo {pages}
  {084006} (\bibinfo {year} {1999})},\ \Eprint
  {http://arxiv.org/abs/gr-qc/9811091} {arXiv:gr-qc/9811091 [gr-qc]}
  \BibitemShut {NoStop}%
\bibitem [{\citenamefont {Buonanno}\ and\ \citenamefont
  {Damour}(2000)}]{Buonanno:2000ef}%
  \BibitemOpen
  \bibfield  {author} {\bibinfo {author} {\bibfnamefont {Alessandra}\
  \bibnamefont {Buonanno}}\ and\ \bibinfo {author} {\bibfnamefont {Thibault}\
  \bibnamefont {Damour}},\ }\bibfield  {title} {\enquote {\bibinfo {title}
  {{Transition from inspiral to plunge in binary black hole coalescences}},}\
  }\href {\doibase 10.1103/PhysRevD.62.064015} {\bibfield  {journal} {\bibinfo
  {journal} {Phys. Rev.}\ }\textbf {\bibinfo {volume} {D62}},\ \bibinfo {pages}
  {064015} (\bibinfo {year} {2000})},\ \Eprint
  {http://arxiv.org/abs/gr-qc/0001013} {arXiv:gr-qc/0001013 [gr-qc]}
  \BibitemShut {NoStop}%
\bibitem [{\citenamefont {Damour}\ \emph {et~al.}(2000)\citenamefont {Damour},
  \citenamefont {Jaranowski},\ and\ \citenamefont {Schaefer}}]{Damour:2000we}%
  \BibitemOpen
  \bibfield  {author} {\bibinfo {author} {\bibfnamefont {Thibault}\
  \bibnamefont {Damour}}, \bibinfo {author} {\bibfnamefont {Piotr}\
  \bibnamefont {Jaranowski}}, \ and\ \bibinfo {author} {\bibfnamefont
  {Gerhard}\ \bibnamefont {Schaefer}},\ }\bibfield  {title} {\enquote {\bibinfo
  {title} {{On the determination of the last stable orbit for circular general
  relativistic binaries at the third postNewtonian approximation}},}\ }\href
  {\doibase 10.1103/PhysRevD.62.084011} {\bibfield  {journal} {\bibinfo
  {journal} {Phys. Rev.}\ }\textbf {\bibinfo {volume} {D62}},\ \bibinfo {pages}
  {084011} (\bibinfo {year} {2000})},\ \Eprint
  {http://arxiv.org/abs/gr-qc/0005034} {arXiv:gr-qc/0005034 [gr-qc]}
  \BibitemShut {NoStop}%
\bibitem [{\citenamefont {Damour}(2001)}]{Damour:2001tu}%
  \BibitemOpen
  \bibfield  {author} {\bibinfo {author} {\bibfnamefont {Thibault}\
  \bibnamefont {Damour}},\ }\bibfield  {title} {\enquote {\bibinfo {title}
  {{Coalescence of two spinning black holes: an effective one-body
  approach}},}\ }\href {\doibase 10.1103/PhysRevD.64.124013} {\bibfield
  {journal} {\bibinfo  {journal} {Phys. Rev.}\ }\textbf {\bibinfo {volume}
  {D64}},\ \bibinfo {pages} {124013} (\bibinfo {year} {2001})},\ \Eprint
  {http://arxiv.org/abs/gr-qc/0103018} {arXiv:gr-qc/0103018 [gr-qc]}
  \BibitemShut {NoStop}%
\bibitem [{\citenamefont {Damour}\ \emph {et~al.}(2015)\citenamefont {Damour},
  \citenamefont {Jaranowski},\ and\ \citenamefont {Schäfer}}]{Damour:2015isa}%
  \BibitemOpen
  \bibfield  {author} {\bibinfo {author} {\bibfnamefont {Thibault}\
  \bibnamefont {Damour}}, \bibinfo {author} {\bibfnamefont {Piotr}\
  \bibnamefont {Jaranowski}}, \ and\ \bibinfo {author} {\bibfnamefont
  {Gerhard}\ \bibnamefont {Schäfer}},\ }\bibfield  {title} {\enquote {\bibinfo
  {title} {{Fourth post-Newtonian effective one-body dynamics}},}\ }\href
  {\doibase 10.1103/PhysRevD.91.084024} {\bibfield  {journal} {\bibinfo
  {journal} {Phys. Rev.}\ }\textbf {\bibinfo {volume} {D91}},\ \bibinfo {pages}
  {084024} (\bibinfo {year} {2015})},\ \Eprint
  {http://arxiv.org/abs/1502.07245} {arXiv:1502.07245 [gr-qc]} \BibitemShut
  {NoStop}%
\bibitem [{\citenamefont {Damour}\ \emph {et~al.}(2016)\citenamefont {Damour},
  \citenamefont {Jaranowski},\ and\ \citenamefont {Schäfer}}]{Damour:2016abl}%
  \BibitemOpen
  \bibfield  {author} {\bibinfo {author} {\bibfnamefont {Thibault}\
  \bibnamefont {Damour}}, \bibinfo {author} {\bibfnamefont {Piotr}\
  \bibnamefont {Jaranowski}}, \ and\ \bibinfo {author} {\bibfnamefont
  {Gerhard}\ \bibnamefont {Schäfer}},\ }\bibfield  {title} {\enquote {\bibinfo
  {title} {{Conservative dynamics of two-body systems at the fourth
  post-Newtonian approximation of general relativity}},}\ }\href {\doibase
  10.1103/PhysRevD.93.084014} {\bibfield  {journal} {\bibinfo  {journal} {Phys.
  Rev.}\ }\textbf {\bibinfo {volume} {D93}},\ \bibinfo {pages} {084014}
  (\bibinfo {year} {2016})},\ \Eprint {http://arxiv.org/abs/1601.01283}
  {arXiv:1601.01283 [gr-qc]} \BibitemShut {NoStop}%
\bibitem [{\citenamefont {Julié}\ and\ \citenamefont
  {Deruelle}(2017)}]{Julie:2017pkb}%
  \BibitemOpen
  \bibfield  {author} {\bibinfo {author} {\bibfnamefont {Félix-Louis}\
  \bibnamefont {Julié}}\ and\ \bibinfo {author} {\bibfnamefont {Nathalie}\
  \bibnamefont {Deruelle}},\ }\bibfield  {title} {\enquote {\bibinfo {title}
  {{Two-body problem in Scalar-Tensor theories as a deformation of General
  Relativity : an Effective-One-Body approach}},}\ }\href@noop {} {\  (\bibinfo
  {year} {2017})},\ \Eprint {http://arxiv.org/abs/1703.05360} {arXiv:1703.05360
  [gr-qc]} \BibitemShut {NoStop}%
\bibitem [{\citenamefont {Messina}\ and\ \citenamefont
  {Nagar}(2017)}]{Messina:2017yjg}%
  \BibitemOpen
  \bibfield  {author} {\bibinfo {author} {\bibfnamefont {Francesco}\
  \bibnamefont {Messina}}\ and\ \bibinfo {author} {\bibfnamefont {Alessandro}\
  \bibnamefont {Nagar}},\ }\bibfield  {title} {\enquote {\bibinfo {title}
  {{Back to post-Newtonian: parametrized-4.5PN TaylorF2 approximant(s) and tail
  effects to quartic nonlinear order from the effective one body formalism}},}\
  }\href@noop {} {\  (\bibinfo {year} {2017})},\ \Eprint
  {http://arxiv.org/abs/1703.08107} {arXiv:1703.08107 [gr-qc]} \BibitemShut
  {NoStop}%
\bibitem [{\citenamefont {Damour}(2016)}]{Damour:2016gwp}%
  \BibitemOpen
  \bibfield  {author} {\bibinfo {author} {\bibfnamefont {Thibault}\
  \bibnamefont {Damour}},\ }\bibfield  {title} {\enquote {\bibinfo {title}
  {{Gravitational scattering, post-Minkowskian approximation and Effective
  One-Body theory}},}\ }\href {\doibase 10.1103/PhysRevD.94.104015} {\bibfield
  {journal} {\bibinfo  {journal} {Phys. Rev.}\ }\textbf {\bibinfo {volume}
  {D94}},\ \bibinfo {pages} {104015} (\bibinfo {year} {2016})},\ \Eprint
  {http://arxiv.org/abs/1609.00354} {arXiv:1609.00354 [gr-qc]} \BibitemShut
  {NoStop}%
\bibitem [{\citenamefont {Bini}\ \emph
  {et~al.}(2016{\natexlab{a}})\citenamefont {Bini}, \citenamefont {Damour},\
  and\ \citenamefont {Geralico}}]{Bini:2015bfb}%
  \BibitemOpen
  \bibfield  {author} {\bibinfo {author} {\bibfnamefont {Donato}\ \bibnamefont
  {Bini}}, \bibinfo {author} {\bibfnamefont {Thibault}\ \bibnamefont {Damour}},
  \ and\ \bibinfo {author} {\bibfnamefont {Andrea}\ \bibnamefont {Geralico}},\
  }\bibfield  {title} {\enquote {\bibinfo {title} {{Confirming and improving
  post-Newtonian and effective-one-body results from self-force computations
  along eccentric orbits around a Schwarzschild black hole}},}\ }\href
  {\doibase 10.1103/PhysRevD.93.064023} {\bibfield  {journal} {\bibinfo
  {journal} {Phys. Rev.}\ }\textbf {\bibinfo {volume} {D93}},\ \bibinfo {pages}
  {064023} (\bibinfo {year} {2016}{\natexlab{a}})},\ \Eprint
  {http://arxiv.org/abs/1511.04533} {arXiv:1511.04533 [gr-qc]} \BibitemShut
  {NoStop}%
\bibitem [{\citenamefont {Hopper}\ \emph {et~al.}(2016)\citenamefont {Hopper},
  \citenamefont {Kavanagh},\ and\ \citenamefont {Ottewill}}]{Hopper:2015icj}%
  \BibitemOpen
  \bibfield  {author} {\bibinfo {author} {\bibfnamefont {Seth}\ \bibnamefont
  {Hopper}}, \bibinfo {author} {\bibfnamefont {Chris}\ \bibnamefont
  {Kavanagh}}, \ and\ \bibinfo {author} {\bibfnamefont {Adrian~C.}\
  \bibnamefont {Ottewill}},\ }\bibfield  {title} {\enquote {\bibinfo {title}
  {{Analytic self-force calculations in the post-Newtonian regime: eccentric
  orbits on a Schwarzschild background}},}\ }\href {\doibase
  10.1103/PhysRevD.93.044010} {\bibfield  {journal} {\bibinfo  {journal} {Phys.
  Rev.}\ }\textbf {\bibinfo {volume} {D93}},\ \bibinfo {pages} {044010}
  (\bibinfo {year} {2016})},\ \Eprint {http://arxiv.org/abs/1512.01556}
  {arXiv:1512.01556 [gr-qc]} \BibitemShut {NoStop}%
\bibitem [{\citenamefont {Akcay}\ and\ \citenamefont {van~de
  Meent}(2016)}]{Akcay:2015pjz}%
  \BibitemOpen
  \bibfield  {author} {\bibinfo {author} {\bibfnamefont {Sarp}\ \bibnamefont
  {Akcay}}\ and\ \bibinfo {author} {\bibfnamefont {Maarten}\ \bibnamefont
  {van~de Meent}},\ }\bibfield  {title} {\enquote {\bibinfo {title} {{Numerical
  computation of the effective-one-body potential $q$ using self-force
  results}},}\ }\href {\doibase 10.1103/PhysRevD.93.064063} {\bibfield
  {journal} {\bibinfo  {journal} {Phys. Rev.}\ }\textbf {\bibinfo {volume}
  {D93}},\ \bibinfo {pages} {064063} (\bibinfo {year} {2016})},\ \Eprint
  {http://arxiv.org/abs/1512.03392} {arXiv:1512.03392 [gr-qc]} \BibitemShut
  {NoStop}%
\bibitem [{\citenamefont {Bini}\ and\ \citenamefont
  {Damour}(2016)}]{Bini:2016cje}%
  \BibitemOpen
  \bibfield  {author} {\bibinfo {author} {\bibfnamefont {Donato}\ \bibnamefont
  {Bini}}\ and\ \bibinfo {author} {\bibfnamefont {Thibault}\ \bibnamefont
  {Damour}},\ }\bibfield  {title} {\enquote {\bibinfo {title} {{Conservative
  second-order gravitational self-force on circular orbits and the effective
  one-body formalism}},}\ }\href {\doibase 10.1103/PhysRevD.93.104040}
  {\bibfield  {journal} {\bibinfo  {journal} {Phys. Rev.}\ }\textbf {\bibinfo
  {volume} {D93}},\ \bibinfo {pages} {104040} (\bibinfo {year} {2016})},\
  \Eprint {http://arxiv.org/abs/1603.09175} {arXiv:1603.09175 [gr-qc]}
  \BibitemShut {NoStop}%
\bibitem [{\citenamefont {Nagar}\ \emph {et~al.}(2016)\citenamefont {Nagar},
  \citenamefont {Damour}, \citenamefont {Reisswig},\ and\ \citenamefont
  {Pollney}}]{Nagar:2015xqa}%
  \BibitemOpen
  \bibfield  {author} {\bibinfo {author} {\bibfnamefont {Alessandro}\
  \bibnamefont {Nagar}}, \bibinfo {author} {\bibfnamefont {Thibault}\
  \bibnamefont {Damour}}, \bibinfo {author} {\bibfnamefont {Christian}\
  \bibnamefont {Reisswig}}, \ and\ \bibinfo {author} {\bibfnamefont {Denis}\
  \bibnamefont {Pollney}},\ }\bibfield  {title} {\enquote {\bibinfo {title}
  {{Energetics and phasing of nonprecessing spinning coalescing black hole
  binaries}},}\ }\href {\doibase 10.1103/PhysRevD.93.044046} {\bibfield
  {journal} {\bibinfo  {journal} {Phys. Rev.}\ }\textbf {\bibinfo {volume}
  {D93}},\ \bibinfo {pages} {044046} (\bibinfo {year} {2016})},\ \Eprint
  {http://arxiv.org/abs/1506.08457} {arXiv:1506.08457 [gr-qc]} \BibitemShut
  {NoStop}%
\bibitem [{\citenamefont {Bohé}\ \emph {et~al.}(2017)\citenamefont {Bohé}
  \emph {et~al.}}]{Bohe:2016gbl}%
  \BibitemOpen
  \bibfield  {author} {\bibinfo {author} {\bibfnamefont {Alejandro}\
  \bibnamefont {Bohé}} \emph {et~al.},\ }\bibfield  {title} {\enquote
  {\bibinfo {title} {{Improved effective-one-body model of spinning,
  nonprecessing binary black holes for the era of gravitational-wave
  astrophysics with advanced detectors}},}\ }\href {\doibase
  10.1103/PhysRevD.95.044028} {\bibfield  {journal} {\bibinfo  {journal} {Phys.
  Rev.}\ }\textbf {\bibinfo {volume} {D95}},\ \bibinfo {pages} {044028}
  (\bibinfo {year} {2017})},\ \Eprint {http://arxiv.org/abs/1611.03703}
  {arXiv:1611.03703 [gr-qc]} \BibitemShut {NoStop}%
\bibitem [{\citenamefont {Babak}\ \emph {et~al.}(2017)\citenamefont {Babak},
  \citenamefont {Taracchini},\ and\ \citenamefont {Buonanno}}]{Babak:2016tgq}%
  \BibitemOpen
  \bibfield  {author} {\bibinfo {author} {\bibfnamefont {Stanislav}\
  \bibnamefont {Babak}}, \bibinfo {author} {\bibfnamefont {Andrea}\
  \bibnamefont {Taracchini}}, \ and\ \bibinfo {author} {\bibfnamefont
  {Alessandra}\ \bibnamefont {Buonanno}},\ }\bibfield  {title} {\enquote
  {\bibinfo {title} {{Validating the effective-one-body model of spinning,
  precessing binary black holes against numerical relativity}},}\ }\href
  {\doibase 10.1103/PhysRevD.95.024010} {\bibfield  {journal} {\bibinfo
  {journal} {Phys. Rev.}\ }\textbf {\bibinfo {volume} {D95}},\ \bibinfo {pages}
  {024010} (\bibinfo {year} {2017})},\ \Eprint
  {http://arxiv.org/abs/1607.05661} {arXiv:1607.05661 [gr-qc]} \BibitemShut
  {NoStop}%
\bibitem [{\citenamefont {Dietrich}\ and\ \citenamefont
  {Hinderer}(2017)}]{Dietrich:2017feu}%
  \BibitemOpen
  \bibfield  {author} {\bibinfo {author} {\bibfnamefont {Tim}\ \bibnamefont
  {Dietrich}}\ and\ \bibinfo {author} {\bibfnamefont {Tanja}\ \bibnamefont
  {Hinderer}},\ }\bibfield  {title} {\enquote {\bibinfo {title} {{Comprehensive
  numerical relativity -- effective-one-body comparison to inform improvements
  in waveform models for binary neutron star systems}},}\ }\href@noop {} {\
  (\bibinfo {year} {2017})},\ \Eprint {http://arxiv.org/abs/1702.02053}
  {arXiv:1702.02053 [gr-qc]} \BibitemShut {NoStop}%
\bibitem [{\citenamefont {Nagar}\ \emph {et~al.}(2017)\citenamefont {Nagar},
  \citenamefont {Riemenschneider},\ and\ \citenamefont
  {Pratten}}]{Nagar:2017jdw}%
  \BibitemOpen
  \bibfield  {author} {\bibinfo {author} {\bibfnamefont {Alessandro}\
  \bibnamefont {Nagar}}, \bibinfo {author} {\bibfnamefont {Gunnar}\
  \bibnamefont {Riemenschneider}}, \ and\ \bibinfo {author} {\bibfnamefont
  {Geraint}\ \bibnamefont {Pratten}},\ }\bibfield  {title} {\enquote {\bibinfo
  {title} {{Impact of Numerical Relativity information on effective-one-body
  waveform models}},}\ }\href@noop {} {\  (\bibinfo {year} {2017})},\ \Eprint
  {http://arxiv.org/abs/1703.06814} {arXiv:1703.06814 [gr-qc]} \BibitemShut
  {NoStop}%
\bibitem [{\citenamefont {Detweiler}(2008)}]{Detweiler:2008ft}%
  \BibitemOpen
  \bibfield  {author} {\bibinfo {author} {\bibfnamefont {Steven~L.}\
  \bibnamefont {Detweiler}},\ }\bibfield  {title} {\enquote {\bibinfo {title}
  {{A Consequence of the gravitational self-force for circular orbits of the
  Schwarzschild geometry}},}\ }\href {\doibase 10.1103/PhysRevD.77.124026}
  {\bibfield  {journal} {\bibinfo  {journal} {Phys. Rev.}\ }\textbf {\bibinfo
  {volume} {D77}},\ \bibinfo {pages} {124026} (\bibinfo {year} {2008})},\
  \Eprint {http://arxiv.org/abs/0804.3529} {arXiv:0804.3529 [gr-qc]}
  \BibitemShut {NoStop}%
\bibitem [{\citenamefont {Barack}\ and\ \citenamefont
  {Sago}(2009)}]{Barack:2009ey}%
  \BibitemOpen
  \bibfield  {author} {\bibinfo {author} {\bibfnamefont {Leor}\ \bibnamefont
  {Barack}}\ and\ \bibinfo {author} {\bibfnamefont {Norichika}\ \bibnamefont
  {Sago}},\ }\bibfield  {title} {\enquote {\bibinfo {title} {{Gravitational
  self-force correction to the innermost stable circular orbit of a
  Schwarzschild black hole}},}\ }\href {\doibase
  10.1103/PhysRevLett.102.191101} {\bibfield  {journal} {\bibinfo  {journal}
  {Phys. Rev. Lett.}\ }\textbf {\bibinfo {volume} {102}},\ \bibinfo {pages}
  {191101} (\bibinfo {year} {2009})},\ \Eprint {http://arxiv.org/abs/0902.0573}
  {arXiv:0902.0573 [gr-qc]} \BibitemShut {NoStop}%
\bibitem [{\citenamefont {Damour}(2010)}]{Damour:2009sm}%
  \BibitemOpen
  \bibfield  {author} {\bibinfo {author} {\bibfnamefont {Thibault}\
  \bibnamefont {Damour}},\ }\bibfield  {title} {\enquote {\bibinfo {title}
  {{Gravitational Self Force in a Schwarzschild Background and the Effective
  One Body Formalism}},}\ }\href {\doibase 10.1103/PhysRevD.81.024017}
  {\bibfield  {journal} {\bibinfo  {journal} {Phys. Rev.}\ }\textbf {\bibinfo
  {volume} {D81}},\ \bibinfo {pages} {024017} (\bibinfo {year} {2010})},\
  \Eprint {http://arxiv.org/abs/0910.5533} {arXiv:0910.5533 [gr-qc]}
  \BibitemShut {NoStop}%
\bibitem [{\citenamefont {Barausse}\ \emph {et~al.}(2012)\citenamefont
  {Barausse}, \citenamefont {Buonanno},\ and\ \citenamefont
  {Le~Tiec}}]{Barausse:2011dq}%
  \BibitemOpen
  \bibfield  {author} {\bibinfo {author} {\bibfnamefont {Enrico}\ \bibnamefont
  {Barausse}}, \bibinfo {author} {\bibfnamefont {Alessandra}\ \bibnamefont
  {Buonanno}}, \ and\ \bibinfo {author} {\bibfnamefont {Alexandre}\
  \bibnamefont {Le~Tiec}},\ }\bibfield  {title} {\enquote {\bibinfo {title}
  {{The complete non-spinning effective-one-body metric at linear order in the
  mass ratio}},}\ }\href {\doibase 10.1103/PhysRevD.85.064010} {\bibfield
  {journal} {\bibinfo  {journal} {Phys. Rev.}\ }\textbf {\bibinfo {volume}
  {D85}},\ \bibinfo {pages} {064010} (\bibinfo {year} {2012})},\ \Eprint
  {http://arxiv.org/abs/1111.5610} {arXiv:1111.5610 [gr-qc]} \BibitemShut
  {NoStop}%
\bibitem [{\citenamefont {Colleoni}\ \emph {et~al.}(2015)\citenamefont
  {Colleoni}, \citenamefont {Barack}, \citenamefont {Shah},\ and\ \citenamefont
  {van~de Meent}}]{Colleoni:2015ena}%
  \BibitemOpen
  \bibfield  {author} {\bibinfo {author} {\bibfnamefont {Marta}\ \bibnamefont
  {Colleoni}}, \bibinfo {author} {\bibfnamefont {Leor}\ \bibnamefont {Barack}},
  \bibinfo {author} {\bibfnamefont {Abhay~G.}\ \bibnamefont {Shah}}, \ and\
  \bibinfo {author} {\bibfnamefont {Maarten}\ \bibnamefont {van~de Meent}},\
  }\bibfield  {title} {\enquote {\bibinfo {title} {{Self-force as a cosmic
  censor in the Kerr overspinning problem}},}\ }\href {\doibase
  10.1103/PhysRevD.92.084044} {\bibfield  {journal} {\bibinfo  {journal} {Phys.
  Rev.}\ }\textbf {\bibinfo {volume} {D92}},\ \bibinfo {pages} {084044}
  (\bibinfo {year} {2015})},\ \Eprint {http://arxiv.org/abs/1508.04031}
  {arXiv:1508.04031 [gr-qc]} \BibitemShut {NoStop}%
\bibitem [{\citenamefont {Sago}\ \emph {et~al.}(2008)\citenamefont {Sago},
  \citenamefont {Barack},\ and\ \citenamefont {Detweiler}}]{Sago:2008id}%
  \BibitemOpen
  \bibfield  {author} {\bibinfo {author} {\bibfnamefont {Norichika}\
  \bibnamefont {Sago}}, \bibinfo {author} {\bibfnamefont {Leor}\ \bibnamefont
  {Barack}}, \ and\ \bibinfo {author} {\bibfnamefont {Steven~L.}\ \bibnamefont
  {Detweiler}},\ }\bibfield  {title} {\enquote {\bibinfo {title} {{Two
  approaches for the gravitational self force in black hole spacetime:
  Comparison of numerical results}},}\ }\href {\doibase
  10.1103/PhysRevD.78.124024} {\bibfield  {journal} {\bibinfo  {journal} {Phys.
  Rev.}\ }\textbf {\bibinfo {volume} {D78}},\ \bibinfo {pages} {124024}
  (\bibinfo {year} {2008})},\ \Eprint {http://arxiv.org/abs/0810.2530}
  {arXiv:0810.2530 [gr-qc]} \BibitemShut {NoStop}%
\bibitem [{\citenamefont {Blanchet}\ \emph
  {et~al.}(2010{\natexlab{a}})\citenamefont {Blanchet}, \citenamefont
  {Detweiler}, \citenamefont {Le~Tiec},\ and\ \citenamefont
  {Whiting}}]{Blanchet:2009sd}%
  \BibitemOpen
  \bibfield  {author} {\bibinfo {author} {\bibfnamefont {Luc}\ \bibnamefont
  {Blanchet}}, \bibinfo {author} {\bibfnamefont {Steven~L.}\ \bibnamefont
  {Detweiler}}, \bibinfo {author} {\bibfnamefont {Alexandre}\ \bibnamefont
  {Le~Tiec}}, \ and\ \bibinfo {author} {\bibfnamefont {Bernard~F.}\
  \bibnamefont {Whiting}},\ }\bibfield  {title} {\enquote {\bibinfo {title}
  {{Post-Newtonian and Numerical Calculations of the Gravitational Self-Force
  for Circular Orbits in the Schwarzschild Geometry}},}\ }\href {\doibase
  10.1103/PhysRevD.81.064004} {\bibfield  {journal} {\bibinfo  {journal} {Phys.
  Rev.}\ }\textbf {\bibinfo {volume} {D81}},\ \bibinfo {pages} {064004}
  (\bibinfo {year} {2010}{\natexlab{a}})},\ \Eprint
  {http://arxiv.org/abs/0910.0207} {arXiv:0910.0207 [gr-qc]} \BibitemShut
  {NoStop}%
\bibitem [{\citenamefont {Akcay}\ \emph {et~al.}(2015)\citenamefont {Akcay},
  \citenamefont {Le~Tiec}, \citenamefont {Barack}, \citenamefont {Sago},\ and\
  \citenamefont {Warburton}}]{Akcay:2015pza}%
  \BibitemOpen
  \bibfield  {author} {\bibinfo {author} {\bibfnamefont {Sarp}\ \bibnamefont
  {Akcay}}, \bibinfo {author} {\bibfnamefont {Alexandre}\ \bibnamefont
  {Le~Tiec}}, \bibinfo {author} {\bibfnamefont {Leor}\ \bibnamefont {Barack}},
  \bibinfo {author} {\bibfnamefont {Norichika}\ \bibnamefont {Sago}}, \ and\
  \bibinfo {author} {\bibfnamefont {Niels}\ \bibnamefont {Warburton}},\
  }\bibfield  {title} {\enquote {\bibinfo {title} {{Comparison Between
  Self-Force and Post-Newtonian Dynamics: Beyond Circular Orbits}},}\ }\href
  {\doibase 10.1103/PhysRevD.91.124014} {\bibfield  {journal} {\bibinfo
  {journal} {Phys. Rev.}\ }\textbf {\bibinfo {volume} {D91}},\ \bibinfo {pages}
  {124014} (\bibinfo {year} {2015})},\ \Eprint
  {http://arxiv.org/abs/1503.01374} {arXiv:1503.01374 [gr-qc]} \BibitemShut
  {NoStop}%
\bibitem [{\citenamefont {van~de Meent}(2017)}]{vandeMeent:2016hel}%
  \BibitemOpen
  \bibfield  {author} {\bibinfo {author} {\bibfnamefont {Maarten}\ \bibnamefont
  {van~de Meent}},\ }\bibfield  {title} {\enquote {\bibinfo {title}
  {{Self-force corrections to the periapsis advance around a spinning black
  hole}},}\ }\href {\doibase 10.1103/PhysRevLett.118.011101} {\bibfield
  {journal} {\bibinfo  {journal} {Phys. Rev. Lett.}\ }\textbf {\bibinfo
  {volume} {118}},\ \bibinfo {pages} {011101} (\bibinfo {year} {2017})},\
  \Eprint {http://arxiv.org/abs/1610.03497} {arXiv:1610.03497 [gr-qc]}
  \BibitemShut {NoStop}%
\bibitem [{\citenamefont {Le~Tiec}\ \emph {et~al.}(2011)\citenamefont
  {Le~Tiec}, \citenamefont {Mroue}, \citenamefont {Barack}, \citenamefont
  {Buonanno}, \citenamefont {Pfeiffer}, \citenamefont {Sago},\ and\
  \citenamefont {Taracchini}}]{LeTiec:2011bk}%
  \BibitemOpen
  \bibfield  {author} {\bibinfo {author} {\bibfnamefont {Alexandre}\
  \bibnamefont {Le~Tiec}}, \bibinfo {author} {\bibfnamefont {Abdul~H.}\
  \bibnamefont {Mroue}}, \bibinfo {author} {\bibfnamefont {Leor}\ \bibnamefont
  {Barack}}, \bibinfo {author} {\bibfnamefont {Alessandra}\ \bibnamefont
  {Buonanno}}, \bibinfo {author} {\bibfnamefont {Harald~P.}\ \bibnamefont
  {Pfeiffer}}, \bibinfo {author} {\bibfnamefont {Norichika}\ \bibnamefont
  {Sago}}, \ and\ \bibinfo {author} {\bibfnamefont {Andrea}\ \bibnamefont
  {Taracchini}},\ }\bibfield  {title} {\enquote {\bibinfo {title} {{Periastron
  Advance in Black Hole Binaries}},}\ }\href {\doibase
  10.1103/PhysRevLett.107.141101} {\bibfield  {journal} {\bibinfo  {journal}
  {Phys. Rev. Lett.}\ }\textbf {\bibinfo {volume} {107}},\ \bibinfo {pages}
  {141101} (\bibinfo {year} {2011})},\ \Eprint {http://arxiv.org/abs/1106.3278}
  {arXiv:1106.3278 [gr-qc]} \BibitemShut {NoStop}%
\bibitem [{\citenamefont {Bini}\ and\ \citenamefont
  {Damour}(2013)}]{Bini:2013zaa}%
  \BibitemOpen
  \bibfield  {author} {\bibinfo {author} {\bibfnamefont {Donato}\ \bibnamefont
  {Bini}}\ and\ \bibinfo {author} {\bibfnamefont {Thibault}\ \bibnamefont
  {Damour}},\ }\bibfield  {title} {\enquote {\bibinfo {title} {{Analytical
  determination of the two-body gravitational interaction potential at the
  fourth post-Newtonian approximation}},}\ }\href {\doibase
  10.1103/PhysRevD.87.121501} {\bibfield  {journal} {\bibinfo  {journal} {Phys.
  Rev.}\ }\textbf {\bibinfo {volume} {D87}},\ \bibinfo {pages} {121501}
  (\bibinfo {year} {2013})},\ \Eprint {http://arxiv.org/abs/1305.4884}
  {arXiv:1305.4884 [gr-qc]} \BibitemShut {NoStop}%
\bibitem [{\citenamefont {Bini}\ and\ \citenamefont
  {Damour}(2014{\natexlab{a}})}]{Bini:2014ica}%
  \BibitemOpen
  \bibfield  {author} {\bibinfo {author} {\bibfnamefont {Donato}\ \bibnamefont
  {Bini}}\ and\ \bibinfo {author} {\bibfnamefont {Thibault}\ \bibnamefont
  {Damour}},\ }\bibfield  {title} {\enquote {\bibinfo {title} {{Two-body
  gravitational spin-orbit interaction at linear order in the mass ratio}},}\
  }\href {\doibase 10.1103/PhysRevD.90.024039} {\bibfield  {journal} {\bibinfo
  {journal} {Phys. Rev.}\ }\textbf {\bibinfo {volume} {D90}},\ \bibinfo {pages}
  {024039} (\bibinfo {year} {2014}{\natexlab{a}})},\ \Eprint
  {http://arxiv.org/abs/1404.2747} {arXiv:1404.2747 [gr-qc]} \BibitemShut
  {NoStop}%
\bibitem [{\citenamefont {Kavanagh}\ \emph {et~al.}(2015)\citenamefont
  {Kavanagh}, \citenamefont {Ottewill},\ and\ \citenamefont
  {Wardell}}]{Kavanagh:2015lva}%
  \BibitemOpen
  \bibfield  {author} {\bibinfo {author} {\bibfnamefont {Chris}\ \bibnamefont
  {Kavanagh}}, \bibinfo {author} {\bibfnamefont {Adrian~C.}\ \bibnamefont
  {Ottewill}}, \ and\ \bibinfo {author} {\bibfnamefont {Barry}\ \bibnamefont
  {Wardell}},\ }\bibfield  {title} {\enquote {\bibinfo {title} {{Analytical
  high-order post-Newtonian expansions for extreme mass ratio binaries}},}\
  }\href {\doibase 10.1103/PhysRevD.92.084025} {\bibfield  {journal} {\bibinfo
  {journal} {Phys. Rev.}\ }\textbf {\bibinfo {volume} {D92}},\ \bibinfo {pages}
  {084025} (\bibinfo {year} {2015})},\ \Eprint
  {http://arxiv.org/abs/1503.02334} {arXiv:1503.02334 [gr-qc]} \BibitemShut
  {NoStop}%
\bibitem [{\citenamefont {Akcay}\ \emph {et~al.}(2017)\citenamefont {Akcay},
  \citenamefont {Dempsey},\ and\ \citenamefont {Dolan}}]{Akcay:2016dku}%
  \BibitemOpen
  \bibfield  {author} {\bibinfo {author} {\bibfnamefont {Sarp}\ \bibnamefont
  {Akcay}}, \bibinfo {author} {\bibfnamefont {David}\ \bibnamefont {Dempsey}},
  \ and\ \bibinfo {author} {\bibfnamefont {Sam}\ \bibnamefont {Dolan}},\
  }\bibfield  {title} {\enquote {\bibinfo {title} {{Spin-orbit precession for
  eccentric black hole binaries at first order in the mass ratio}},}\ }\href
  {\doibase 10.1088/1361-6382/aa61d6} {\bibfield  {journal} {\bibinfo
  {journal} {Class. Quant. Grav.}\ }\textbf {\bibinfo {volume} {34}},\ \bibinfo
  {pages} {084001} (\bibinfo {year} {2017})},\ \Eprint
  {http://arxiv.org/abs/1608.04811} {arXiv:1608.04811 [gr-qc]} \BibitemShut
  {NoStop}%
\bibitem [{\citenamefont {Detweiler}\ and\ \citenamefont
  {Whiting}(2003)}]{Detweiler:2002mi}%
  \BibitemOpen
  \bibfield  {author} {\bibinfo {author} {\bibfnamefont {Steven~L.}\
  \bibnamefont {Detweiler}}\ and\ \bibinfo {author} {\bibfnamefont
  {Bernard~F.}\ \bibnamefont {Whiting}},\ }\bibfield  {title} {\enquote
  {\bibinfo {title} {{Selfforce via a Green's function decomposition}},}\
  }\href {\doibase 10.1103/PhysRevD.67.024025} {\bibfield  {journal} {\bibinfo
  {journal} {Phys. Rev.}\ }\textbf {\bibinfo {volume} {D67}},\ \bibinfo {pages}
  {024025} (\bibinfo {year} {2003})},\ \Eprint
  {http://arxiv.org/abs/gr-qc/0202086} {arXiv:gr-qc/0202086 [gr-qc]}
  \BibitemShut {NoStop}%
\bibitem [{\citenamefont {{Darwin}}(1959)}]{Darw59}%
  \BibitemOpen
  \bibfield  {author} {\bibinfo {author} {\bibfnamefont {C.}~\bibnamefont
  {{Darwin}}},\ }\bibfield  {title} {\enquote {\bibinfo {title} {{The Gravity
  Field of a Particle}},}\ }\href {\doibase 10.1098/rspa.1959.0015} {\bibfield
  {journal} {\bibinfo  {journal} {Proc. R. Soc. Lond. A}\ }\textbf {\bibinfo
  {volume} {249}},\ \bibinfo {pages} {180--194} (\bibinfo {year}
  {1959})}\BibitemShut {NoStop}%
\bibitem [{\citenamefont {Marck}(1983)}]{Marck431}%
  \BibitemOpen
  \bibfield  {author} {\bibinfo {author} {\bibfnamefont {J.-A.}\ \bibnamefont
  {Marck}},\ }\bibfield  {title} {\enquote {\bibinfo {title} {Solution to the
  equations of parallel transport in kerr geometry; tidal tensor},}\ }\href
  {\doibase 10.1098/rspa.1983.0021} {\bibfield  {journal} {\bibinfo  {journal}
  {Proceedings of the Royal Society of London A: Mathematical, Physical and
  Engineering Sciences}\ }\textbf {\bibinfo {volume} {385}},\ \bibinfo {pages}
  {431--438} (\bibinfo {year} {1983})},\ \Eprint
  {http://arxiv.org/abs/http://rspa.royalsocietypublishing.org/content/385/178\\9/431.full.pdf}
  {http://rspa.royalsocietypublishing.org/content/385/178\\9/431.full.pdf}
  \BibitemShut {NoStop}%
\bibitem [{\citenamefont {Barack}\ and\ \citenamefont
  {Sago}(2011)}]{Barack:2011ed}%
  \BibitemOpen
  \bibfield  {author} {\bibinfo {author} {\bibfnamefont {Leor}\ \bibnamefont
  {Barack}}\ and\ \bibinfo {author} {\bibfnamefont {Norichika}\ \bibnamefont
  {Sago}},\ }\bibfield  {title} {\enquote {\bibinfo {title} {{Beyond the
  geodesic approximation: conservative effects of the gravitational self-force
  in eccentric orbits around a Schwarzschild black hole}},}\ }\href {\doibase
  10.1103/PhysRevD.83.084023} {\bibfield  {journal} {\bibinfo  {journal} {Phys.
  Rev.}\ }\textbf {\bibinfo {volume} {D83}},\ \bibinfo {pages} {084023}
  (\bibinfo {year} {2011})},\ \Eprint {http://arxiv.org/abs/1101.3331}
  {arXiv:1101.3331 [gr-qc]} \BibitemShut {NoStop}%
\bibitem [{\citenamefont {Chrzanowski}(1975)}]{Chrzanowski:1975wv}%
  \BibitemOpen
  \bibfield  {author} {\bibinfo {author} {\bibfnamefont {P.~L.}\ \bibnamefont
  {Chrzanowski}},\ }\bibfield  {title} {\enquote {\bibinfo {title} {{Vector
  Potential and Metric Perturbations of a Rotating Black Hole}},}\ }\href
  {\doibase 10.1103/PhysRevD.11.2042} {\bibfield  {journal} {\bibinfo
  {journal} {Phys. Rev.}\ }\textbf {\bibinfo {volume} {D11}},\ \bibinfo {pages}
  {2042--2062} (\bibinfo {year} {1975})}\BibitemShut {NoStop}%
\bibitem [{\citenamefont {Cohen}\ and\ \citenamefont
  {Kegeles}(1975)}]{COHEN19755}%
  \BibitemOpen
  \bibfield  {author} {\bibinfo {author} {\bibfnamefont {J.M.}\ \bibnamefont
  {Cohen}}\ and\ \bibinfo {author} {\bibfnamefont {L.S.}\ \bibnamefont
  {Kegeles}},\ }\bibfield  {title} {\enquote {\bibinfo {title} {Space-time
  perturbations},}\ }\href {\doibase
  http://dx.doi.org/10.1016/0375-9601(75)90583-6} {\bibfield  {journal}
  {\bibinfo  {journal} {Physics Letters A}\ }\textbf {\bibinfo {volume} {54}},\
  \bibinfo {pages} {5 -- 7} (\bibinfo {year} {1975})}\BibitemShut {NoStop}%
\bibitem [{\citenamefont {Kegeles}\ and\ \citenamefont
  {Cohen}(1979)}]{Kegeles:1979an}%
  \BibitemOpen
  \bibfield  {author} {\bibinfo {author} {\bibfnamefont {L.~S.}\ \bibnamefont
  {Kegeles}}\ and\ \bibinfo {author} {\bibfnamefont {J.~M.}\ \bibnamefont
  {Cohen}},\ }\bibfield  {title} {\enquote {\bibinfo {title} {{Constructive
  procedure for perturbations of space-times}},}\ }\href {\doibase
  10.1103/PhysRevD.19.1641} {\bibfield  {journal} {\bibinfo  {journal} {Phys.
  Rev.}\ }\textbf {\bibinfo {volume} {D19}},\ \bibinfo {pages} {1641--1664}
  (\bibinfo {year} {1979})}\BibitemShut {NoStop}%
\bibitem [{\citenamefont {Lousto}\ and\ \citenamefont
  {Whiting}(2002)}]{Lousto:2002em}%
  \BibitemOpen
  \bibfield  {author} {\bibinfo {author} {\bibfnamefont {Carlos~O.}\
  \bibnamefont {Lousto}}\ and\ \bibinfo {author} {\bibfnamefont {Bernard~F.}\
  \bibnamefont {Whiting}},\ }\bibfield  {title} {\enquote {\bibinfo {title}
  {{Reconstruction of black hole metric perturbations from Weyl curvature}},}\
  }\href {\doibase 10.1103/PhysRevD.66.024026} {\bibfield  {journal} {\bibinfo
  {journal} {Phys. Rev.}\ }\textbf {\bibinfo {volume} {D66}},\ \bibinfo {pages}
  {024026} (\bibinfo {year} {2002})},\ \Eprint
  {http://arxiv.org/abs/gr-qc/0203061} {arXiv:gr-qc/0203061 [gr-qc]}
  \BibitemShut {NoStop}%
\bibitem [{\citenamefont {Keidl}\ \emph {et~al.}(2010)\citenamefont {Keidl},
  \citenamefont {Shah}, \citenamefont {Friedman}, \citenamefont {Kim},\ and\
  \citenamefont {Price}}]{Keidl:2010pm}%
  \BibitemOpen
  \bibfield  {author} {\bibinfo {author} {\bibfnamefont {Tobias~S.}\
  \bibnamefont {Keidl}}, \bibinfo {author} {\bibfnamefont {Abhay~G.}\
  \bibnamefont {Shah}}, \bibinfo {author} {\bibfnamefont {John~L.}\
  \bibnamefont {Friedman}}, \bibinfo {author} {\bibfnamefont {Dong-Hoon}\
  \bibnamefont {Kim}}, \ and\ \bibinfo {author} {\bibfnamefont {Larry~R.}\
  \bibnamefont {Price}},\ }\bibfield  {title} {\enquote {\bibinfo {title}
  {{Gravitational Self-force in a Radiation Gauge}},}\ }\href {\doibase
  10.1103/PhysRevD.82.124012, 10.1103/PhysRevD.90.109902} {\bibfield  {journal}
  {\bibinfo  {journal} {Phys. Rev.}\ }\textbf {\bibinfo {volume} {D82}},\
  \bibinfo {pages} {124012} (\bibinfo {year} {2010})},\ \bibinfo {note}
  {[Erratum: Phys. Rev.D90,no.10,109902(2014)]},\ \Eprint
  {http://arxiv.org/abs/1004.2276} {arXiv:1004.2276 [gr-qc]} \BibitemShut
  {NoStop}%
\bibitem [{\citenamefont {Shah}\ \emph {et~al.}(2011)\citenamefont {Shah},
  \citenamefont {Keidl}, \citenamefont {Friedman}, \citenamefont {Kim},\ and\
  \citenamefont {Price}}]{Shah:2010bi}%
  \BibitemOpen
  \bibfield  {author} {\bibinfo {author} {\bibfnamefont {Abhay~G.}\
  \bibnamefont {Shah}}, \bibinfo {author} {\bibfnamefont {Tobias~S.}\
  \bibnamefont {Keidl}}, \bibinfo {author} {\bibfnamefont {John~L.}\
  \bibnamefont {Friedman}}, \bibinfo {author} {\bibfnamefont {Dong-Hoon}\
  \bibnamefont {Kim}}, \ and\ \bibinfo {author} {\bibfnamefont {Larry~R.}\
  \bibnamefont {Price}},\ }\bibfield  {title} {\enquote {\bibinfo {title}
  {{Conservative, gravitational self-force for a particle in circular orbit
  around a Schwarzschild black hole in a Radiation Gauge}},}\ }\href {\doibase
  10.1103/PhysRevD.83.064018} {\bibfield  {journal} {\bibinfo  {journal} {Phys.
  Rev.}\ }\textbf {\bibinfo {volume} {D83}},\ \bibinfo {pages} {064018}
  (\bibinfo {year} {2011})},\ \Eprint {http://arxiv.org/abs/1009.4876}
  {arXiv:1009.4876 [gr-qc]} \BibitemShut {NoStop}%
\bibitem [{\citenamefont {Kavanagh}\ \emph {et~al.}(2016)\citenamefont
  {Kavanagh}, \citenamefont {Ottewill},\ and\ \citenamefont
  {Wardell}}]{Kavanagh:2016idg}%
  \BibitemOpen
  \bibfield  {author} {\bibinfo {author} {\bibfnamefont {Chris}\ \bibnamefont
  {Kavanagh}}, \bibinfo {author} {\bibfnamefont {Adrian~C.}\ \bibnamefont
  {Ottewill}}, \ and\ \bibinfo {author} {\bibfnamefont {Barry}\ \bibnamefont
  {Wardell}},\ }\bibfield  {title} {\enquote {\bibinfo {title} {{Analytical
  high-order post-Newtonian expansions for spinning extreme mass ratio
  binaries}},}\ }\href {\doibase 10.1103/PhysRevD.93.124038} {\bibfield
  {journal} {\bibinfo  {journal} {Phys. Rev.}\ }\textbf {\bibinfo {volume}
  {D93}},\ \bibinfo {pages} {124038} (\bibinfo {year} {2016})},\ \Eprint
  {http://arxiv.org/abs/1601.03394} {arXiv:1601.03394 [gr-qc]} \BibitemShut
  {NoStop}%
\bibitem [{\citenamefont {van~de Meent}\ and\ \citenamefont
  {Shah}(2015)}]{vandeMeent:2015lxa}%
  \BibitemOpen
  \bibfield  {author} {\bibinfo {author} {\bibfnamefont {Maarten}\ \bibnamefont
  {van~de Meent}}\ and\ \bibinfo {author} {\bibfnamefont {Abhay~G.}\
  \bibnamefont {Shah}},\ }\bibfield  {title} {\enquote {\bibinfo {title}
  {{Metric perturbations produced by eccentric equatorial orbits around a Kerr
  black hole}},}\ }\href {\doibase 10.1103/PhysRevD.92.064025} {\bibfield
  {journal} {\bibinfo  {journal} {Phys. Rev.}\ }\textbf {\bibinfo {volume}
  {D92}},\ \bibinfo {pages} {064025} (\bibinfo {year} {2015})},\ \Eprint
  {http://arxiv.org/abs/1506.04755} {arXiv:1506.04755 [gr-qc]} \BibitemShut
  {NoStop}%
\bibitem [{\citenamefont {Bini}\ and\ \citenamefont
  {Damour}(2014{\natexlab{b}})}]{Bini:2013rfa}%
  \BibitemOpen
  \bibfield  {author} {\bibinfo {author} {\bibfnamefont {Donato}\ \bibnamefont
  {Bini}}\ and\ \bibinfo {author} {\bibfnamefont {Thibault}\ \bibnamefont
  {Damour}},\ }\bibfield  {title} {\enquote {\bibinfo {title} {{High-order
  post-Newtonian contributions to the two-body gravitational interaction
  potential from analytical gravitational self-force calculations}},}\ }\href
  {\doibase 10.1103/PhysRevD.89.064063} {\bibfield  {journal} {\bibinfo
  {journal} {Phys. Rev.}\ }\textbf {\bibinfo {volume} {D89}},\ \bibinfo {pages}
  {064063} (\bibinfo {year} {2014}{\natexlab{b}})},\ \Eprint
  {http://arxiv.org/abs/1312.2503} {arXiv:1312.2503 [gr-qc]} \BibitemShut
  {NoStop}%
\bibitem [{\citenamefont {Wardell}\ and\ \citenamefont
  {Gopakumar}(2015)}]{Wardell:2015kea}%
  \BibitemOpen
  \bibfield  {author} {\bibinfo {author} {\bibfnamefont {Barry}\ \bibnamefont
  {Wardell}}\ and\ \bibinfo {author} {\bibfnamefont {Achamveedu}\ \bibnamefont
  {Gopakumar}},\ }\bibfield  {title} {\enquote {\bibinfo {title} {{Self-force:
  Computational Strategies}},}\ }\bibfield  {booktitle} {\emph {\bibinfo
  {booktitle} {{Proceedings, 524th WE-Heraeus-Seminar: Equations of Motion in
  Relativistic Gravity (EOM 2013): Bad Honnef, Germany, February 17-23,
  2013}}},\ }\href {\doibase 10.1007/978-3-319-18335-0_14} {\bibfield
  {journal} {\bibinfo  {journal} {Fund. Theor. Phys.}\ }\textbf {\bibinfo
  {volume} {179}},\ \bibinfo {pages} {487--522} (\bibinfo {year} {2015})},\
  \Eprint {http://arxiv.org/abs/1501.07322} {arXiv:1501.07322 [gr-qc]}
  \BibitemShut {NoStop}%
\bibitem [{\citenamefont {Hinderer}\ and\ \citenamefont
  {Flanagan}(2008)}]{Hinderer:2008dm}%
  \BibitemOpen
  \bibfield  {author} {\bibinfo {author} {\bibfnamefont {Tanja}\ \bibnamefont
  {Hinderer}}\ and\ \bibinfo {author} {\bibfnamefont {Eanna~E.}\ \bibnamefont
  {Flanagan}},\ }\bibfield  {title} {\enquote {\bibinfo {title} {{Two timescale
  analysis of extreme mass ratio inspirals in Kerr. I. Orbital Motion}},}\
  }\href {\doibase 10.1103/PhysRevD.78.064028} {\bibfield  {journal} {\bibinfo
  {journal} {Phys. Rev.}\ }\textbf {\bibinfo {volume} {D78}},\ \bibinfo {pages}
  {064028} (\bibinfo {year} {2008})},\ \Eprint {http://arxiv.org/abs/0805.3337}
  {arXiv:0805.3337 [gr-qc]} \BibitemShut {NoStop}%
\bibitem [{\citenamefont {Barack}\ and\ \citenamefont
  {Sago}(2010)}]{Barack:2010tm}%
  \BibitemOpen
  \bibfield  {author} {\bibinfo {author} {\bibfnamefont {Leor}\ \bibnamefont
  {Barack}}\ and\ \bibinfo {author} {\bibfnamefont {Norichika}\ \bibnamefont
  {Sago}},\ }\bibfield  {title} {\enquote {\bibinfo {title} {{Gravitational
  self-force on a particle in eccentric orbit around a Schwarzschild black
  hole}},}\ }\href {\doibase 10.1103/PhysRevD.81.084021} {\bibfield  {journal}
  {\bibinfo  {journal} {Phys. Rev.}\ }\textbf {\bibinfo {volume} {D81}},\
  \bibinfo {pages} {084021} (\bibinfo {year} {2010})},\ \Eprint
  {http://arxiv.org/abs/1002.2386} {arXiv:1002.2386 [gr-qc]} \BibitemShut
  {NoStop}%
\bibitem [{\citenamefont {Wardell}\ and\ \citenamefont
  {Warburton}(2015)}]{Wardell:2015ada}%
  \BibitemOpen
  \bibfield  {author} {\bibinfo {author} {\bibfnamefont {Barry}\ \bibnamefont
  {Wardell}}\ and\ \bibinfo {author} {\bibfnamefont {Niels}\ \bibnamefont
  {Warburton}},\ }\bibfield  {title} {\enquote {\bibinfo {title} {{Applying the
  effective-source approach to frequency-domain self-force calculations:
  Lorenz-gauge gravitational perturbations}},}\ }\href {\doibase
  10.1103/PhysRevD.92.084019} {\bibfield  {journal} {\bibinfo  {journal} {Phys.
  Rev.}\ }\textbf {\bibinfo {volume} {D92}},\ \bibinfo {pages} {084019}
  (\bibinfo {year} {2015})},\ \Eprint {http://arxiv.org/abs/1505.07841}
  {arXiv:1505.07841 [gr-qc]} \BibitemShut {NoStop}%
\bibitem [{\citenamefont {Barack}\ and\ \citenamefont
  {Ori}(2001)}]{Barack:2001ph}%
  \BibitemOpen
  \bibfield  {author} {\bibinfo {author} {\bibfnamefont {Leor}\ \bibnamefont
  {Barack}}\ and\ \bibinfo {author} {\bibfnamefont {Amos}\ \bibnamefont
  {Ori}},\ }\bibfield  {title} {\enquote {\bibinfo {title} {{Gravitational
  selfforce and gauge transformations}},}\ }\href {\doibase
  10.1103/PhysRevD.64.124003} {\bibfield  {journal} {\bibinfo  {journal} {Phys.
  Rev.}\ }\textbf {\bibinfo {volume} {D64}},\ \bibinfo {pages} {124003}
  (\bibinfo {year} {2001})},\ \Eprint {http://arxiv.org/abs/gr-qc/0107056}
  {arXiv:gr-qc/0107056 [gr-qc]} \BibitemShut {NoStop}%
\bibitem [{\citenamefont {Pound}\ \emph {et~al.}(2014)\citenamefont {Pound},
  \citenamefont {Merlin},\ and\ \citenamefont {Barack}}]{Pound:2013faa}%
  \BibitemOpen
  \bibfield  {author} {\bibinfo {author} {\bibfnamefont {Adam}\ \bibnamefont
  {Pound}}, \bibinfo {author} {\bibfnamefont {Cesar}\ \bibnamefont {Merlin}}, \
  and\ \bibinfo {author} {\bibfnamefont {Leor}\ \bibnamefont {Barack}},\
  }\bibfield  {title} {\enquote {\bibinfo {title} {{Gravitational self-force
  from radiation-gauge metric perturbations}},}\ }\href {\doibase
  10.1103/PhysRevD.89.024009} {\bibfield  {journal} {\bibinfo  {journal} {Phys.
  Rev.}\ }\textbf {\bibinfo {volume} {D89}},\ \bibinfo {pages} {024009}
  (\bibinfo {year} {2014})},\ \Eprint {http://arxiv.org/abs/1310.1513}
  {arXiv:1310.1513 [gr-qc]} \BibitemShut {NoStop}%
\bibitem [{\citenamefont {Barack}\ and\ \citenamefont
  {Ori}(2000)}]{Barack:1999wf}%
  \BibitemOpen
  \bibfield  {author} {\bibinfo {author} {\bibfnamefont {Leor}\ \bibnamefont
  {Barack}}\ and\ \bibinfo {author} {\bibfnamefont {Amos}\ \bibnamefont
  {Ori}},\ }\bibfield  {title} {\enquote {\bibinfo {title} {{Mode sum
  regularization approach for the selfforce in black hole space-time}},}\
  }\href {\doibase 10.1103/PhysRevD.61.061502} {\bibfield  {journal} {\bibinfo
  {journal} {Phys. Rev.}\ }\textbf {\bibinfo {volume} {D61}},\ \bibinfo {pages}
  {061502} (\bibinfo {year} {2000})},\ \Eprint
  {http://arxiv.org/abs/gr-qc/9912010} {arXiv:gr-qc/9912010 [gr-qc]}
  \BibitemShut {NoStop}%
\bibitem [{\citenamefont {Heffernan}\ \emph {et~al.}(2012)\citenamefont
  {Heffernan}, \citenamefont {Ottewill},\ and\ \citenamefont
  {Wardell}}]{Heffernan:2012su}%
  \BibitemOpen
  \bibfield  {author} {\bibinfo {author} {\bibfnamefont {Anna}\ \bibnamefont
  {Heffernan}}, \bibinfo {author} {\bibfnamefont {Adrian}\ \bibnamefont
  {Ottewill}}, \ and\ \bibinfo {author} {\bibfnamefont {Barry}\ \bibnamefont
  {Wardell}},\ }\bibfield  {title} {\enquote {\bibinfo {title} {{High-order
  expansions of the Detweiler-Whiting singular field in Schwarzschild
  spacetime}},}\ }\href {\doibase 10.1103/PhysRevD.86.104023} {\bibfield
  {journal} {\bibinfo  {journal} {Phys. Rev.}\ }\textbf {\bibinfo {volume}
  {D86}},\ \bibinfo {pages} {104023} (\bibinfo {year} {2012})},\ \Eprint
  {http://arxiv.org/abs/1204.0794} {arXiv:1204.0794 [gr-qc]} \BibitemShut
  {NoStop}%
\bibitem [{onl()}]{online}%
  \BibitemOpen
  \href@noop {} {\enquote {\bibinfo {title} {Electronic archive of
  post-{N}ewtonian coefficients},}\ }\bibinfo {howpublished}
  {\url{http://www.barrywardell.net/research/code}}\BibitemShut {NoStop}%
\bibitem [{\citenamefont {Bini}\ \emph
  {et~al.}(2016{\natexlab{b}})\citenamefont {Bini}, \citenamefont {Damour},\
  and\ \citenamefont {Geralico}}]{Bini:2016qtx}%
  \BibitemOpen
  \bibfield  {author} {\bibinfo {author} {\bibfnamefont {Donato}\ \bibnamefont
  {Bini}}, \bibinfo {author} {\bibfnamefont {Thibault}\ \bibnamefont {Damour}},
  \ and\ \bibinfo {author} {\bibfnamefont {Andrea}\ \bibnamefont {Geralico}},\
  }\bibfield  {title} {\enquote {\bibinfo {title} {{New gravitational
  self-force analytical results for eccentric orbits around a Schwarzschild
  black hole}},}\ }\href {\doibase 10.1103/PhysRevD.93.104017} {\bibfield
  {journal} {\bibinfo  {journal} {Phys. Rev.}\ }\textbf {\bibinfo {volume}
  {D93}},\ \bibinfo {pages} {104017} (\bibinfo {year} {2016}{\natexlab{b}})},\
  \Eprint {http://arxiv.org/abs/1601.02988} {arXiv:1601.02988 [gr-qc]}
  \BibitemShut {NoStop}%
\bibitem [{\citenamefont {Damour}\ \emph
  {et~al.}(2008{\natexlab{a}})\citenamefont {Damour}, \citenamefont
  {Jaranowski},\ and\ \citenamefont {Schaefer}}]{Damour:2008qf}%
  \BibitemOpen
  \bibfield  {author} {\bibinfo {author} {\bibfnamefont {Thibault}\
  \bibnamefont {Damour}}, \bibinfo {author} {\bibfnamefont {Piotr}\
  \bibnamefont {Jaranowski}}, \ and\ \bibinfo {author} {\bibfnamefont
  {Gerhard}\ \bibnamefont {Schaefer}},\ }\bibfield  {title} {\enquote {\bibinfo
  {title} {{Effective one body approach to the dynamics of two spinning black
  holes with next-to-leading order spin-orbit coupling}},}\ }\href {\doibase
  10.1103/PhysRevD.78.024009} {\bibfield  {journal} {\bibinfo  {journal} {Phys.
  Rev.}\ }\textbf {\bibinfo {volume} {D78}},\ \bibinfo {pages} {024009}
  (\bibinfo {year} {2008}{\natexlab{a}})},\ \Eprint
  {http://arxiv.org/abs/0803.0915} {arXiv:0803.0915 [gr-qc]} \BibitemShut
  {NoStop}%
\bibitem [{\citenamefont {Barausse}\ \emph {et~al.}(2009)\citenamefont
  {Barausse}, \citenamefont {Racine},\ and\ \citenamefont
  {Buonanno}}]{Barausse:2009aa}%
  \BibitemOpen
  \bibfield  {author} {\bibinfo {author} {\bibfnamefont {Enrico}\ \bibnamefont
  {Barausse}}, \bibinfo {author} {\bibfnamefont {Etienne}\ \bibnamefont
  {Racine}}, \ and\ \bibinfo {author} {\bibfnamefont {Alessandra}\ \bibnamefont
  {Buonanno}},\ }\bibfield  {title} {\enquote {\bibinfo {title} {{Hamiltonian
  of a spinning test-particle in curved spacetime}},}\ }\href {\doibase
  10.1103/PhysRevD.85.069904, 10.1103/PhysRevD.80.104025} {\bibfield  {journal}
  {\bibinfo  {journal} {Phys. Rev.}\ }\textbf {\bibinfo {volume} {D80}},\
  \bibinfo {pages} {104025} (\bibinfo {year} {2009})},\ \bibinfo {note}
  {[Erratum: Phys. Rev.D85,069904(2012)]},\ \Eprint
  {http://arxiv.org/abs/0907.4745} {arXiv:0907.4745 [gr-qc]} \BibitemShut
  {NoStop}%
\bibitem [{\citenamefont {Barausse}\ and\ \citenamefont
  {Buonanno}(2010)}]{Barausse:2009xi}%
  \BibitemOpen
  \bibfield  {author} {\bibinfo {author} {\bibfnamefont {Enrico}\ \bibnamefont
  {Barausse}}\ and\ \bibinfo {author} {\bibfnamefont {Alessandra}\ \bibnamefont
  {Buonanno}},\ }\bibfield  {title} {\enquote {\bibinfo {title} {{An Improved
  effective-one-body Hamiltonian for spinning black-hole binaries}},}\ }\href
  {\doibase 10.1103/PhysRevD.81.084024} {\bibfield  {journal} {\bibinfo
  {journal} {Phys. Rev.}\ }\textbf {\bibinfo {volume} {D81}},\ \bibinfo {pages}
  {084024} (\bibinfo {year} {2010})},\ \Eprint {http://arxiv.org/abs/0912.3517}
  {arXiv:0912.3517 [gr-qc]} \BibitemShut {NoStop}%
\bibitem [{\citenamefont {Nagar}(2011)}]{Nagar:2011fx}%
  \BibitemOpen
  \bibfield  {author} {\bibinfo {author} {\bibfnamefont {Alessandro}\
  \bibnamefont {Nagar}},\ }\bibfield  {title} {\enquote {\bibinfo {title}
  {{Effective one body Hamiltonian of two spinning black-holes with
  next-to-next-to-leading order spin-orbit coupling}},}\ }\href {\doibase
  10.1103/PhysRevD.84.084028, 10.1103/PhysRevD.88.089901} {\bibfield  {journal}
  {\bibinfo  {journal} {Phys. Rev.}\ }\textbf {\bibinfo {volume} {D84}},\
  \bibinfo {pages} {084028} (\bibinfo {year} {2011})},\ \bibinfo {note}
  {[Erratum: Phys. Rev.D88,no.8,089901(2013)]},\ \Eprint
  {http://arxiv.org/abs/1106.4349} {arXiv:1106.4349 [gr-qc]} \BibitemShut
  {NoStop}%
\bibitem [{\citenamefont {Barausse}\ and\ \citenamefont
  {Buonanno}(2011)}]{Barausse:2011ys}%
  \BibitemOpen
  \bibfield  {author} {\bibinfo {author} {\bibfnamefont {Enrico}\ \bibnamefont
  {Barausse}}\ and\ \bibinfo {author} {\bibfnamefont {Alessandra}\ \bibnamefont
  {Buonanno}},\ }\bibfield  {title} {\enquote {\bibinfo {title} {{Extending the
  effective-one-body Hamiltonian of black-hole binaries to include
  next-to-next-to-leading spin-orbit couplings}},}\ }\href {\doibase
  10.1103/PhysRevD.84.104027} {\bibfield  {journal} {\bibinfo  {journal} {Phys.
  Rev.}\ }\textbf {\bibinfo {volume} {D84}},\ \bibinfo {pages} {104027}
  (\bibinfo {year} {2011})},\ \Eprint {http://arxiv.org/abs/1107.2904}
  {arXiv:1107.2904 [gr-qc]} \BibitemShut {NoStop}%
\bibitem [{\citenamefont {Taracchini}\ \emph {et~al.}(2012)\citenamefont
  {Taracchini}, \citenamefont {Pan}, \citenamefont {Buonanno}, \citenamefont
  {Barausse}, \citenamefont {Boyle}, \citenamefont {Chu}, \citenamefont
  {Lovelace}, \citenamefont {Pfeiffer},\ and\ \citenamefont
  {Scheel}}]{Taracchini:2012ig}%
  \BibitemOpen
  \bibfield  {author} {\bibinfo {author} {\bibfnamefont {Andrea}\ \bibnamefont
  {Taracchini}}, \bibinfo {author} {\bibfnamefont {Yi}~\bibnamefont {Pan}},
  \bibinfo {author} {\bibfnamefont {Alessandra}\ \bibnamefont {Buonanno}},
  \bibinfo {author} {\bibfnamefont {Enrico}\ \bibnamefont {Barausse}}, \bibinfo
  {author} {\bibfnamefont {Michael}\ \bibnamefont {Boyle}}, \bibinfo {author}
  {\bibfnamefont {Tony}\ \bibnamefont {Chu}}, \bibinfo {author} {\bibfnamefont
  {Geoffrey}\ \bibnamefont {Lovelace}}, \bibinfo {author} {\bibfnamefont
  {Harald~P.}\ \bibnamefont {Pfeiffer}}, \ and\ \bibinfo {author}
  {\bibfnamefont {Mark~A.}\ \bibnamefont {Scheel}},\ }\bibfield  {title}
  {\enquote {\bibinfo {title} {{Prototype effective-one-body model for
  nonprecessing spinning inspiral-merger-ringdown waveforms}},}\ }\href
  {\doibase 10.1103/PhysRevD.86.024011} {\bibfield  {journal} {\bibinfo
  {journal} {Phys. Rev.}\ }\textbf {\bibinfo {volume} {D86}},\ \bibinfo {pages}
  {024011} (\bibinfo {year} {2012})},\ \Eprint {http://arxiv.org/abs/1202.0790}
  {arXiv:1202.0790 [gr-qc]} \BibitemShut {NoStop}%
\bibitem [{\citenamefont {Damour}\ and\ \citenamefont
  {Nagar}(2014)}]{Damour:2014sva}%
  \BibitemOpen
  \bibfield  {author} {\bibinfo {author} {\bibfnamefont {Thibault}\
  \bibnamefont {Damour}}\ and\ \bibinfo {author} {\bibfnamefont {Alessandro}\
  \bibnamefont {Nagar}},\ }\bibfield  {title} {\enquote {\bibinfo {title} {{New
  effective-one-body description of coalescing nonprecessing spinning
  black-hole binaries}},}\ }\href {\doibase 10.1103/PhysRevD.90.044018}
  {\bibfield  {journal} {\bibinfo  {journal} {Phys. Rev.}\ }\textbf {\bibinfo
  {volume} {D90}},\ \bibinfo {pages} {044018} (\bibinfo {year} {2014})},\
  \Eprint {http://arxiv.org/abs/1406.6913} {arXiv:1406.6913 [gr-qc]}
  \BibitemShut {NoStop}%
\bibitem [{\citenamefont {Balmelli}\ and\ \citenamefont
  {Jetzer}(2015)}]{Balmelli:2015lva}%
  \BibitemOpen
  \bibfield  {author} {\bibinfo {author} {\bibfnamefont {Simone}\ \bibnamefont
  {Balmelli}}\ and\ \bibinfo {author} {\bibfnamefont {Philippe}\ \bibnamefont
  {Jetzer}},\ }\bibfield  {title} {\enquote {\bibinfo {title}
  {{Effective-one-body Hamiltonian with next-to-leading order spin-spin
  coupling}},}\ }\href {\doibase 10.1103/PhysRevD.91.064011} {\bibfield
  {journal} {\bibinfo  {journal} {Phys. Rev.}\ }\textbf {\bibinfo {volume}
  {D91}},\ \bibinfo {pages} {064011} (\bibinfo {year} {2015})},\ \Eprint
  {http://arxiv.org/abs/1502.01343} {arXiv:1502.01343 [gr-qc]} \BibitemShut
  {NoStop}%
\bibitem [{\citenamefont {Bini}\ and\ \citenamefont
  {Damour}(2015{\natexlab{a}})}]{Bini:2015mza}%
  \BibitemOpen
  \bibfield  {author} {\bibinfo {author} {\bibfnamefont {Donato}\ \bibnamefont
  {Bini}}\ and\ \bibinfo {author} {\bibfnamefont {Thibault}\ \bibnamefont
  {Damour}},\ }\bibfield  {title} {\enquote {\bibinfo {title} {{Analytic
  determination of high-order post-Newtonian self-force contributions to
  gravitational spin precession}},}\ }\href {\doibase
  10.1103/PhysRevD.91.064064} {\bibfield  {journal} {\bibinfo  {journal} {Phys.
  Rev.}\ }\textbf {\bibinfo {volume} {D91}},\ \bibinfo {pages} {064064}
  (\bibinfo {year} {2015}{\natexlab{a}})},\ \Eprint
  {http://arxiv.org/abs/1503.01272} {arXiv:1503.01272 [gr-qc]} \BibitemShut
  {NoStop}%
\bibitem [{\citenamefont {Balmelli}\ and\ \citenamefont
  {Damour}(2015)}]{Balmelli:2015zsa}%
  \BibitemOpen
  \bibfield  {author} {\bibinfo {author} {\bibfnamefont {Simone}\ \bibnamefont
  {Balmelli}}\ and\ \bibinfo {author} {\bibfnamefont {Thibault}\ \bibnamefont
  {Damour}},\ }\bibfield  {title} {\enquote {\bibinfo {title} {{New
  effective-one-body Hamiltonian with next-to-leading order spin-spin
  coupling}},}\ }\href {\doibase 10.1103/PhysRevD.92.124022} {\bibfield
  {journal} {\bibinfo  {journal} {Phys. Rev.}\ }\textbf {\bibinfo {volume}
  {D92}},\ \bibinfo {pages} {124022} (\bibinfo {year} {2015})},\ \Eprint
  {http://arxiv.org/abs/1509.08135} {arXiv:1509.08135 [gr-qc]} \BibitemShut
  {NoStop}%
\bibitem [{\citenamefont {Bini}\ \emph {et~al.}(2015)\citenamefont {Bini},
  \citenamefont {Damour},\ and\ \citenamefont {Geralico}}]{Bini:2015xua}%
  \BibitemOpen
  \bibfield  {author} {\bibinfo {author} {\bibfnamefont {Donato}\ \bibnamefont
  {Bini}}, \bibinfo {author} {\bibfnamefont {Thibault}\ \bibnamefont {Damour}},
  \ and\ \bibinfo {author} {\bibfnamefont {Andrea}\ \bibnamefont {Geralico}},\
  }\bibfield  {title} {\enquote {\bibinfo {title} {{Spin-dependent two-body
  interactions from gravitational self-force computations}},}\ }\href {\doibase
  10.1103/PhysRevD.93.109902, 10.1103/PhysRevD.92.124058} {\bibfield  {journal}
  {\bibinfo  {journal} {Phys. Rev.}\ }\textbf {\bibinfo {volume} {D92}},\
  \bibinfo {pages} {124058} (\bibinfo {year} {2015})},\ \bibinfo {note}
  {[Erratum: Phys. Rev.D93,no.10,109902(2016)]},\ \Eprint
  {http://arxiv.org/abs/1510.06230} {arXiv:1510.06230 [gr-qc]} \BibitemShut
  {NoStop}%
\bibitem [{\citenamefont {Damour}\ \emph
  {et~al.}(2008{\natexlab{b}})\citenamefont {Damour}, \citenamefont
  {Jaranowski},\ and\ \citenamefont {Schaefer}}]{Damour:2007nc}%
  \BibitemOpen
  \bibfield  {author} {\bibinfo {author} {\bibfnamefont {Thibault}\
  \bibnamefont {Damour}}, \bibinfo {author} {\bibfnamefont {Piotr}\
  \bibnamefont {Jaranowski}}, \ and\ \bibinfo {author} {\bibfnamefont
  {Gerhard}\ \bibnamefont {Schaefer}},\ }\bibfield  {title} {\enquote {\bibinfo
  {title} {{Hamiltonian of two spinning compact bodies with next-to-leading
  order gravitational spin-orbit coupling}},}\ }\href {\doibase
  10.1103/PhysRevD.77.064032} {\bibfield  {journal} {\bibinfo  {journal} {Phys.
  Rev.}\ }\textbf {\bibinfo {volume} {D77}},\ \bibinfo {pages} {064032}
  (\bibinfo {year} {2008}{\natexlab{b}})},\ \Eprint
  {http://arxiv.org/abs/0711.1048} {arXiv:0711.1048 [gr-qc]} \BibitemShut
  {NoStop}%
\bibitem [{\citenamefont {Dolan}\ \emph {et~al.}(2014)\citenamefont {Dolan},
  \citenamefont {Warburton}, \citenamefont {Harte}, \citenamefont {Le~Tiec},
  \citenamefont {Wardell},\ and\ \citenamefont {Barack}}]{Dolan:2013roa}%
  \BibitemOpen
  \bibfield  {author} {\bibinfo {author} {\bibfnamefont {Sam~R.}\ \bibnamefont
  {Dolan}}, \bibinfo {author} {\bibfnamefont {Niels}\ \bibnamefont
  {Warburton}}, \bibinfo {author} {\bibfnamefont {Abraham~I.}\ \bibnamefont
  {Harte}}, \bibinfo {author} {\bibfnamefont {Alexandre}\ \bibnamefont
  {Le~Tiec}}, \bibinfo {author} {\bibfnamefont {Barry}\ \bibnamefont
  {Wardell}}, \ and\ \bibinfo {author} {\bibfnamefont {Leor}\ \bibnamefont
  {Barack}},\ }\bibfield  {title} {\enquote {\bibinfo {title} {{Gravitational
  self-torque and spin precession in compact binaries}},}\ }\href {\doibase
  10.1103/PhysRevD.89.064011} {\bibfield  {journal} {\bibinfo  {journal} {Phys.
  Rev.}\ }\textbf {\bibinfo {volume} {D89}},\ \bibinfo {pages} {064011}
  (\bibinfo {year} {2014})},\ \Eprint {http://arxiv.org/abs/1312.0775}
  {arXiv:1312.0775 [gr-qc]} \BibitemShut {NoStop}%
\bibitem [{\citenamefont {Akcay}\ \emph {et~al.}(2012)\citenamefont {Akcay},
  \citenamefont {Barack}, \citenamefont {Damour},\ and\ \citenamefont
  {Sago}}]{Akcay:2012ea}%
  \BibitemOpen
  \bibfield  {author} {\bibinfo {author} {\bibfnamefont {Sarp}\ \bibnamefont
  {Akcay}}, \bibinfo {author} {\bibfnamefont {Leor}\ \bibnamefont {Barack}},
  \bibinfo {author} {\bibfnamefont {Thibault}\ \bibnamefont {Damour}}, \ and\
  \bibinfo {author} {\bibfnamefont {Norichika}\ \bibnamefont {Sago}},\
  }\bibfield  {title} {\enquote {\bibinfo {title} {{Gravitational self-force
  and the effective-one-body formalism between the innermost stable circular
  orbit and the light ring}},}\ }\href {\doibase 10.1103/PhysRevD.86.104041}
  {\bibfield  {journal} {\bibinfo  {journal} {Phys. Rev.}\ }\textbf {\bibinfo
  {volume} {D86}},\ \bibinfo {pages} {104041} (\bibinfo {year} {2012})},\
  \Eprint {http://arxiv.org/abs/1209.0964} {arXiv:1209.0964 [gr-qc]}
  \BibitemShut {NoStop}%
\bibitem [{\citenamefont {Blanchet}\ \emph
  {et~al.}(2010{\natexlab{b}})\citenamefont {Blanchet}, \citenamefont
  {Detweiler}, \citenamefont {Le~Tiec},\ and\ \citenamefont
  {Whiting}}]{Blanchet:2010zd}%
  \BibitemOpen
  \bibfield  {author} {\bibinfo {author} {\bibfnamefont {Luc}\ \bibnamefont
  {Blanchet}}, \bibinfo {author} {\bibfnamefont {Steven~L.}\ \bibnamefont
  {Detweiler}}, \bibinfo {author} {\bibfnamefont {Alexandre}\ \bibnamefont
  {Le~Tiec}}, \ and\ \bibinfo {author} {\bibfnamefont {Bernard~F.}\
  \bibnamefont {Whiting}},\ }\bibfield  {title} {\enquote {\bibinfo {title}
  {{High-Order Post-Newtonian Fit of the Gravitational Self-Force for Circular
  Orbits in the Schwarzschild Geometry}},}\ }\href {\doibase
  10.1103/PhysRevD.81.084033} {\bibfield  {journal} {\bibinfo  {journal} {Phys.
  Rev.}\ }\textbf {\bibinfo {volume} {D81}},\ \bibinfo {pages} {084033}
  (\bibinfo {year} {2010}{\natexlab{b}})},\ \Eprint
  {http://arxiv.org/abs/1002.0726} {arXiv:1002.0726 [gr-qc]} \BibitemShut
  {NoStop}%
\bibitem [{\citenamefont {Barack}\ \emph {et~al.}(2010)\citenamefont {Barack},
  \citenamefont {Damour},\ and\ \citenamefont {Sago}}]{Barack:2010ny}%
  \BibitemOpen
  \bibfield  {author} {\bibinfo {author} {\bibfnamefont {Leor}\ \bibnamefont
  {Barack}}, \bibinfo {author} {\bibfnamefont {Thibault}\ \bibnamefont
  {Damour}}, \ and\ \bibinfo {author} {\bibfnamefont {Norichika}\ \bibnamefont
  {Sago}},\ }\bibfield  {title} {\enquote {\bibinfo {title} {{Precession effect
  of the gravitational self-force in a Schwarzschild spacetime and the
  effective one-body formalism}},}\ }\href {\doibase
  10.1103/PhysRevD.82.084036} {\bibfield  {journal} {\bibinfo  {journal} {Phys.
  Rev.}\ }\textbf {\bibinfo {volume} {D82}},\ \bibinfo {pages} {084036}
  (\bibinfo {year} {2010})},\ \Eprint {http://arxiv.org/abs/1008.0935}
  {arXiv:1008.0935 [gr-qc]} \BibitemShut {NoStop}%
\bibitem [{\citenamefont {Bini}\ and\ \citenamefont
  {Damour}(2014{\natexlab{c}})}]{Bini:2014nfa}%
  \BibitemOpen
  \bibfield  {author} {\bibinfo {author} {\bibfnamefont {Donato}\ \bibnamefont
  {Bini}}\ and\ \bibinfo {author} {\bibfnamefont {Thibault}\ \bibnamefont
  {Damour}},\ }\bibfield  {title} {\enquote {\bibinfo {title} {{Analytic
  determination of the eight-and-a-half post-Newtonian self-force contributions
  to the two-body gravitational interaction potential}},}\ }\href {\doibase
  10.1103/PhysRevD.89.104047} {\bibfield  {journal} {\bibinfo  {journal} {Phys.
  Rev.}\ }\textbf {\bibinfo {volume} {D89}},\ \bibinfo {pages} {104047}
  (\bibinfo {year} {2014}{\natexlab{c}})},\ \Eprint
  {http://arxiv.org/abs/1403.2366} {arXiv:1403.2366 [gr-qc]} \BibitemShut
  {NoStop}%
\bibitem [{\citenamefont {Bini}\ and\ \citenamefont
  {Damour}(2015{\natexlab{b}})}]{Bini:2015bla}%
  \BibitemOpen
  \bibfield  {author} {\bibinfo {author} {\bibfnamefont {Donato}\ \bibnamefont
  {Bini}}\ and\ \bibinfo {author} {\bibfnamefont {Thibault}\ \bibnamefont
  {Damour}},\ }\bibfield  {title} {\enquote {\bibinfo {title} {{Detweiler’s
  gauge-invariant redshift variable: Analytic determination of the nine and
  nine-and-a-half post-Newtonian self-force contributions}},}\ }\href {\doibase
  10.1103/PhysRevD.91.064050} {\bibfield  {journal} {\bibinfo  {journal} {Phys.
  Rev.}\ }\textbf {\bibinfo {volume} {D91}},\ \bibinfo {pages} {064050}
  (\bibinfo {year} {2015}{\natexlab{b}})},\ \Eprint
  {http://arxiv.org/abs/1502.02450} {arXiv:1502.02450 [gr-qc]} \BibitemShut
  {NoStop}%
\bibitem [{\citenamefont {Damour}\ and\ \citenamefont
  {Schaefer}(1988)}]{Damour:1988mr}%
  \BibitemOpen
  \bibfield  {author} {\bibinfo {author} {\bibfnamefont {T.}~\bibnamefont
  {Damour}}\ and\ \bibinfo {author} {\bibfnamefont {Gerhard}\ \bibnamefont
  {Schaefer}},\ }\bibfield  {title} {\enquote {\bibinfo {title} {{Higher Order
  Relativistic Periastron Advances and Binary Pulsars}},}\ }\href {\doibase
  10.1007/BF02828697} {\bibfield  {journal} {\bibinfo  {journal} {Nuovo Cim.}\
  }\textbf {\bibinfo {volume} {B101}},\ \bibinfo {pages} {127} (\bibinfo {year}
  {1988})}\BibitemShut {NoStop}%
\end{thebibliography}%

\end{document}